\documentclass[10pt,aps,prd,twocolumn,showpacs,amsmath,nofootinbib]{revtex4-1}

\usepackage[english]{babel}

\usepackage[caption=false]{subfig}
\usepackage{slashed}
\usepackage{dcolumn}
\usepackage{slashbox}
\usepackage{graphicx}

\usepackage[unicode=true]{hyperref}
\hypersetup{pdftitle=Radiative corrections to the \textbackslash eta(') Dalitz decays}
\hypersetup{pdfauthor=Tom\'{a}\v{s} Husek}

\def\cuteps{\epsilon}
\def\dimeps{\varepsilon}

\makeatletter
\providecommand*{\diff}%
  {\@ifnextchar^{\DIfF}{\DIfF^{}}}
\def\DIfF^#1{%
  \mathop{\mathrm{\mathstrut d}}%
    \nolimits^{#1}\gobblespace}
\def\gobblespace{%
    \futurelet\diffarg\opspace}
\def\opspace{%
    \let\DiffSpace\!%
    \ifx\diffarg(%
      \let\DiffSpace\relax
     \else
      \ifx\diffarg[%
	\let\DiffSpace\relax
      \else
	\ifx\diffarg\{%
	  \let\DiffSpace\relax
	\fi\fi\fi\DiffSpace}

\allowdisplaybreaks

\begin{document}

\title{Radiative corrections to the \texorpdfstring{$\eta^{(\prime)}$}{\textbackslash eta(')} Dalitz decays}
%
\author{Tom\'{a}\v{s} Husek}
\thanks{Present address: School of Physics and Astronomy, University of Birmingham, Edgbaston, Birmingham, B15 2TT, UK}
\email{tomas.husek@mff.cuni.cz}
\author{Karol Kampf}
\email{karol.kampf@mff.cuni.cz}
\author{Ji\v{r}\'{i} Novotn\'{y}}
\email{jiri.novotny@mff.cuni.cz}
\affiliation{Institute of Particle and Nuclear Physics, 
Faculty of Mathematics and Physics,
Charles University, V Hole\v{s}ovi\v{c}k\'{a}ch 2, Praha 8, Czech Republic}
\author{Stefan Leupold}
\email{stefan.leupold@physics.uu.se}
\affiliation{Institutionen f\"or Fysik och Astronomi, Uppsala Universitet, Box 516, S--75120 Uppsala, Sweden}

\begin{abstract}
We provide the complete set of radiative corrections to the Dalitz decays $\eta^{(\prime)}\to \ell^+\ell^-\gamma$ beyond the soft-photon approximation, i.e.\ over the whole range of the Dalitz plot and with no restrictions on the energy of a radiative photon.
The corrections inevitably depend on the $\eta^{(\prime)}\to\gamma^*\gamma^{(*)}$ transition form factors.
For the singly virtual transition form factor appearing e.g.\ in the bremsstrahlung correction, recent dispersive calculations are used.
For the one-photon-irreducible contribution at the one-loop level (for the doubly virtual form factor), we use a vector-meson-dominance-inspired model while taking into account the $\eta$-$\eta^\prime$ mixing.
\end{abstract}


\maketitle


\section{Introduction and summary}
\label{sec:intro}

Looking for effects of new physics is certainly one of the major contemporary goals of particle physics.
The discovery and quantification of new phenomena at this frontier is a very complicated task: it goes along with having our present knowledge under control.
Due to the fact that experimental devices and techniques are getting more and more precise, theorists should provide sufficiently low uncertainties together with their predictions.
Only in this way an eventual discrepancy can be clearly and correctly revealed.
New physics has therefore no meaning without the `old one' being fully explored.
Unfortunately, the low-energy sector of strong interactions remains a significant challenge.
The correct incorporation of radiative corrections in the QED sector might help to extract information about the strong sector for the chosen processes.
It is the subject of this paper to study the next-to-leading-order (NLO) radiative corrections to the Dalitz decays $\eta^{(\prime)}\to\ell^+\ell^-\gamma$.

Unlike in the neutral pion case, the $\eta^{(\prime)}$ Dalitz decays do not belong to those with the highest branching ratio, since due to the higher $\eta^{(\prime)}$ rest masses hadronic decay channels are open.
Nevertheless, studying these decays provides a way to access the electromagnetic transition form factors and consequently information about the structure of related mesons.
The form factors in turn represent a valuable input for precision predictions of some other quantities like the anomalous magnetic moment ($g-2$) of a muon.
Moreover, these decays are used as normalization channels in rare decay ($\eta^{(\prime)}\to\ell^+\ell^-$) searches.

In~\cite{Husek:2015sma}, motivated by contemporary experimental needs of the NA62 experiment~\cite{TheNA62:2016fhr}, we revisited the classical paper of Mikaelian and Smith (M\&S)~\cite{Mikaelian:1972yg}.
We concentrated on the detailed recalculation and completion of the full set of NLO QED radiative corrections to the Dalitz decay of a neutral pion, i.e.\ to the process $\pi^0\to e^+e^-\gamma$.
Doing so, we have avoided simplifications connected to neglecting higher orders in the final-state lepton mass and thus retaining generality for a future use --- in particular for Dalitz decays including muon pairs.
The one-photon-irreducible (1$\gamma$IR) contribution at the one-loop level, which turned out to be indeed non-negligible in view of the two-fold differential NLO decay width, was also included.
Finally, we have discussed contributions, which were numerically irrelevant for the pion case but were expected to became important e.g.\ for $\eta$ decays.
Here, muon loops as a part of the virtual radiative corrections need to be taken into account.
Most importantly, we have also touched the fact that the deviations due to the slope of the $\eta$ transition form factor cannot be overlooked and provided expressions covering a related correction.

It is the aim of the present paper to discuss in detail not only the case of $\eta$ decays, but also the $\eta^\prime$ decays.
In the former case, we could conveniently and extensively draw from the previous work~\cite{Husek:2015sma}, which governed the neutral pion Dalitz decay, since it was already written in a sufficiently general way.
Consequently, we would proceed along very similar lines and after properly treating the $\eta$-$\eta^\prime$ mixing and generalizing the one-photon-irreducible contribution beyond the effective approach, we could immediately provide the relevant tables and plots.
What really brings the current topic to a different level of difficulty is a desire to tackle the radiative corrections for the $\eta^\prime$ decays.
The resulting framework is, of course, directly applicable for the $\eta$ case and one can check that the results are (backward) compatible with the previous, although somewhat generalized, approach used in~\cite{Husek:2015sma}.
Needless to say, the same framework introduced here applies for the pion case as well and it provides some minor corrections compared to the numerical results given in~\cite{Husek:2015sma}.
This stems from the fact that therein we have intentionally neglected the form-factor dependence of the bremsstrahlung correction.
Let us only mention that the numerical results obtained for the pion decay using the new framework are indeed compatible with the form-factor slope correction suggested at the end of Section V of~\cite{Husek:2015sma}.
There is thus no particular need to use the presented framework for the pion case: one gains a correction to the correction at the level of 1\,\% and it would be rather an overkill.
For the Dalitz decay of a neutral pion, the approach shown in~\cite{Husek:2015sma} is sufficient.

Let us briefly discuss the subtleties and difficulties which one encounters and needs to deal with when facing the Dalitz decays of $\eta^{(\prime)}$ mesons and associated NLO radiative corrections and which are mainly driven by the properties of the $\eta^\prime$ meson.
The main differences compared to the pion case stem from the following facts.
First, it is the higher rest mass, which in the case of $\eta$ is above the production of a muon pair and in the case of $\eta^\prime$ even above the lowest-lying resonances $\rho$ and $\omega$, the former of which is a broad resonance in $\pi\pi$ scattering.
This is connected to the fact that the form-factor slope parameter is not negligible as it was in the pion case: the form factor cannot be scaled out anymore and its particular model is required to be taken into account.
We then need to distinguish between two separate cases.
Similarly to the leading-order (LO) decay width, in the case of the bremsstrahlung correction the singly virtual transition form factor appears.
The calculation of this contribution includes integration over angles and energies of the bremsstrahlung photon.
For these integrals to be well-defined in order to obtain reasonable results, including the width of the lowest-lying vector-meson resonances becomes necessary.
Due to the fact that such a calculation will be unavoidably sensitive to the width of the broad $\rho$ resonance, we have decided to incorporate the recent dispersive calculations~\cite{Hanhart:2013vba,Hanhart:2016pcd}.
In the case of the 1$\gamma$IR correction, one needs to take into account the doubly virtual transition form factor.
Here we do not expect any substantial dependence of the result on the vector-meson decay widths and we use a simple model, which incorporates the strange-flavor content of $\eta^{(\prime)}$ mesons and the $\eta$-$\eta^\prime$ mixing.

Na\"ive radiative corrections for the $\eta\to e^+e^-\gamma$ process were already published~\cite{Mikaelian:1972jn} soon after the work~\cite{Mikaelian:1972yg}: compared to~\cite{Mikaelian:1972yg}, the numerical results presented in~\cite{Mikaelian:1972jn} correspond to the case in which only the numerical value of the physical mass of the decaying pseudoscalar was changed.
Other work related to this paper is~\cite{Silagadze:2006rt}, where the two-photon exchange contributions to the cross sections of $e^+e^-\to\eta^{(\prime)}\gamma$ processes were calculated.
In the current work, we provide a complete systematic study of the NLO radiative corrections to the differential decay widths of the four processes under consideration: $\eta\to e^+e^-\gamma$, $\eta\to \mu^+\mu^-\gamma$, $\eta^\prime\to e^+e^-\gamma$ and $\eta^\prime\to \mu^+\mu^-\gamma$.

As it was in the pion case, radiative corrections for these processes are crucial in order to extract relevant information from the data.
This goes together with the fact that currently an ambitious experimental $\eta^{(\prime)}$ program aiming for an accuracy never reached before is running, for instance, at experiments BES-III~\cite{Fang:2017qgz}, A2~\cite{Adlarson:2016hpp} or GlueX~\cite{AlekSejevs:2013mkl}.
Note that in \cite{Husek:2015sma} and also throughout the present work we study fully inclusive radiative corrections, i.e.\ no momentum or angular cuts on the additional 
bremsstrahlung photon(s) are applied.
Consequently we are free of the collinear divergences (sensitivity to the smallness of the electron mass) that would otherwise appear; see, e.g., the corresponding discussion in \cite{Kubis:2010mp}.

The main message of the present work is the completion of the list of the NLO corrections in the QED sector and improving the previous approach~\cite{Mikaelian:1972jn}.
Compared to~\cite{Mikaelian:1972jn}, which relates to the case of the $\eta\to e^+e^-\gamma$ decay, we took into account muon loops and hadronic corrections as a part of the vacuum polarization contribution, 1$\gamma$IR contribution, higher-order final-state lepton mass correction and form-factor effects.
Moreover, we treat three additional processes including $\eta^\prime$ decays: $\eta\to \mu^+\mu^-\gamma$, $\eta^\prime\to e^+e^-\gamma$ and $\eta^\prime\to \mu^+\mu^-\gamma$.
All the formulae necessary for the calculation of the considered correction are listed in the present paper or, whenever a repetition should occur, the reader is referred to the previous work~\cite{Husek:2015sma}.
Let us mention that in contrary to the pion case, the eventual NLO Monte Carlo event generator would not be able to profit from the real-time code calculating the desired correction at a given kinematical point.
This is due to the much higher CPU time consumption caused by higher complexity of the involved integrals.
On the other hand, sufficiently dense tables of values suitable for interpolation are submitted together with this text in a form of ancillary files.

Our paper is organized as follows.
We recapitulate first some basic facts about the LO differential decay width calculation in Section~\ref{sec:LO}. 
Then we proceed to the review of the NLO radiative corrections in the QED sector in Sections~\ref{sec:virt}, \ref{sec:1gIR} and \ref{sec:BS}.
In particular, in Section~\ref{sec:virt} we discuss the virtual corrections, in Section~\ref{sec:1gIR} we introduce the one-photon irreducible contribution and in Section~\ref{sec:BS} we describe the bremsstrahlung correction calculation.
Some technical details together with extensive results concerning the bremsstrahlung and 1$\gamma$IR contributions to the NLO correction have been moved to appendices.
The building blocks for the 1$\gamma$IR matrix element in terms of scalar form factors can be found in Appendix~\ref{app:M1gIR}.
In Appendix~\ref{app:JTr} we present explicit results regarding the bremsstrahlung matrix element squared.
This is related to Appendix~\ref{app:J}, where new basic integrals are listed which are necessary to append to the previously used basis presented in~\cite{Husek:2015sma} due to a necessary generalization.
Appendix~\ref{app:eps} shows the partial fraction decompositions used to simplify the bremsstrahlung matrix element squared.
In Appendix~\ref{app:etaVMD} we briefly describe how a simple vector-meson-dominance (VMD)-inspired model for the $\eta$ and $\eta^\prime$ electromagnetic transition form factors is derived and provide its phenomenological test.
Finally, we make use of the doubly virtual transition form factor from Appendix~\ref{app:etaVMD} and show in Appendix~\ref{app:Pll} a simple example of the approach discussed in Section~\ref{sec:1gIR} on the case of the $P\to\ell^+\ell^-$ decays.


\section{Leading order}
\label{sec:LO}

In what follows we will stick to the notation used in~\cite{Husek:2015sma}. Let us briefly recapitulate it for completeness.
Throughout the text we denote the four-momenta of the decaying pseudoscalar meson (of mass $M_P$), lepton (mass $m_\ell$), antilepton and photon by $P$, $p$, $q$ and $k$, respectively.
For a parent meson we have in mind $\eta$ or $\eta^\prime$.
Traditionally, we introduce kinematical variables $x$ and $y$ defined as
\begin{equation}
x=\frac{(p+q)^2}{M_P^2}\,,\quad y=-\frac{2}{M_P^2}\frac{P\cdot(p-q)}{(1-x)}\,,
\label{eq:defxy}
\end{equation}
where $x$ is a normalized lepton-antilepton pair invariant mass squared and $y$ has the meaning of the rescaled cosine of the angle between the directions of the outgoing photon and antilepton in the $\ell^+\ell^-$ center-of-mass system (CMS).
If we introduce a lepton-specific constant $\nu_\ell=2m_\ell/M_P$ and associated CMS lepton speed
\begin{equation}
\beta_\ell=\beta_\ell(x)\equiv\sqrt{1-\frac{\nu_\ell^2}{x}}\,,
\label{eq:beta}
\end{equation}
we can write the limits on $x$ and $y$ as
\begin{equation}
x\in [\nu_\ell^2,1]\,,\quad
y\in [-\beta_\ell,\beta_\ell]\,.
\label{eq:xylimits}
\end{equation}
Consequently, these depend on the final-state lepton mass.

\begin{figure}[t!]
\includegraphics[width=0.5\columnwidth]{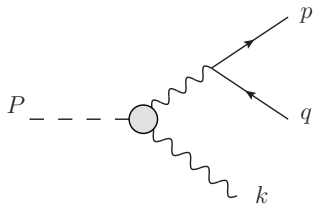}
\caption{
\label{fig:LO}
Leading-order diagram of the decay $P\to \ell^+\ell^-\gamma$ in the QED expansion.
}
\end{figure}

The leading-order diagram of the decay $P\to \ell^+\ell^-\gamma$ is shown in Fig.~\ref{fig:LO}.
The shaded blob corresponds to the singly virtual electromagnetic transition form factor $\mathcal{F}((p+q)^2)$, which is closely related to the doubly virtual transition form factor by
\begin{equation}
\mathcal{F}\big((p+q)^2\big)
\equiv\mathcal{F}_{P\gamma^*\gamma^*}\big(0,(p+q)^2\big)
=\mathcal{F}_{P\gamma^*\gamma^*}\big((p+q)^2,0\big)\,.
\label{eq:F}
\end{equation}
The two-fold differential decay rate reads
\begin{equation}
\begin{split}
&\frac{{\mathrm d}^2\Gamma_{P\to\ell\bar\ell\gamma}^\text{LO}}{{\mathrm d} x{\mathrm d} y}\\
&=\frac{\alpha}{\pi}\,\Gamma_{P\to\gamma\gamma}^\text{LO}\bigg|\frac{\mathcal{F}\big(M_P^2x\big)}{\mathcal{F}(0)}\bigg|^2\frac{(1-x)^3}{4x}\left[1+y^2+\frac{\nu_\ell^2}{x}\right].
\end{split}
\label{eq:dLOxy}
\end{equation}
Above, we have used the LO expression for the most prominent (electromagnetic) decay rate of a neutral pseudoscalar meson $P$:
\begin{equation}
\Gamma_{P\to\gamma\gamma}^\text{LO}
=\frac{e^4M_P^3}{64\pi}|\mathcal{F}(0)|^2\,.
\end{equation}
Integrating (\ref{eq:dLOxy}) over $y$, we find the one-fold differential decay width
\begin{equation}
\begin{split}
\frac{{\mathrm d}\Gamma_{P\to\ell\bar\ell\gamma}^\text{LO}}{{\mathrm d} x}
&=\frac{\alpha}{\pi}\,\Gamma_{P\to\gamma\gamma}^\text{LO}\bigg|\frac{\mathcal{F}\big(M_P^2x\big)}{\mathcal{F}(0)}\bigg|^2\frac{8\beta_\ell}{3}\frac{(1-x)^3}{4x}\bigg[1+\frac{\nu_\ell^2}{2x}\bigg].
\end{split}
\end{equation}

Moving beyond the leading order, it is convenient to introduce the NLO correction $\delta$ to the LO differential decay width, which allows us to write schematically $\text{d}\Gamma=(1+\delta+\dots)\,\text{d}\Gamma^\text{LO}$.
In particular, in the case of the two-fold differential decay width we define
\begin{equation}
\delta(x,y)
=\frac{{\mathrm d}^2\Gamma^\text{NLO}}{{\mathrm d} x{\mathrm d} y}\bigg/\frac{{\mathrm d}^2\Gamma^\text{LO}}{{\mathrm d} x{\mathrm d} y}\,,
\label{eq:dxy}
\end{equation}
and in the one-fold differential case we have
\begin{equation}
\delta(x)
=\frac{{\mathrm d}\Gamma^\text{NLO}}{{\mathrm d} x}\bigg/\frac{{\mathrm d}\Gamma^\text{LO}}{{\mathrm d} x}\,.
\end{equation}
Concluding the definitions closely related to the previous work~\cite{Husek:2015sma}, such a correction can be divided into three parts emphasizing its origin
\begin{equation}
\delta
=\delta^\text{virt}+\delta^{1\gamma\text{IR}}+\delta^\text{BS}\,.
\label{eq:delta_origins}
\end{equation}
Here, $\delta^\text{virt}$ stands for the virtual radiative corrections, $\delta^{1\gamma\text{IR}}$ for the one-photon-irreducible contribution, which is treated separately from $\delta^\text{virt}$ in our approach due to the reasons of historical development, and $\delta^\text{BS}$ for the bremsstrahlung.
As a trivial consequence of previous equations, having knowledge of $\delta(x,y)$ allows for obtaining $\delta(x)$ using the following prescription:
\begin{equation}
\delta(x)=
\frac 3{8\beta_\ell}\frac1{\big(1+\frac{\nu_\ell^2}{2x}\big)}\int_{-\beta_\ell}^{\beta_\ell}{\mathrm d} y\,\delta(x,y)\left[1+y^2+\frac{\nu_\ell^2}{x}\right].
\label{eq:dx}
\end{equation}
In the following sections we discuss the individual contributions one by one.
We mainly point out the differences in comparison to the pion case.

\begin{figure}[!ht]
\subfloat[][]{
\includegraphics[width=0.4\columnwidth]{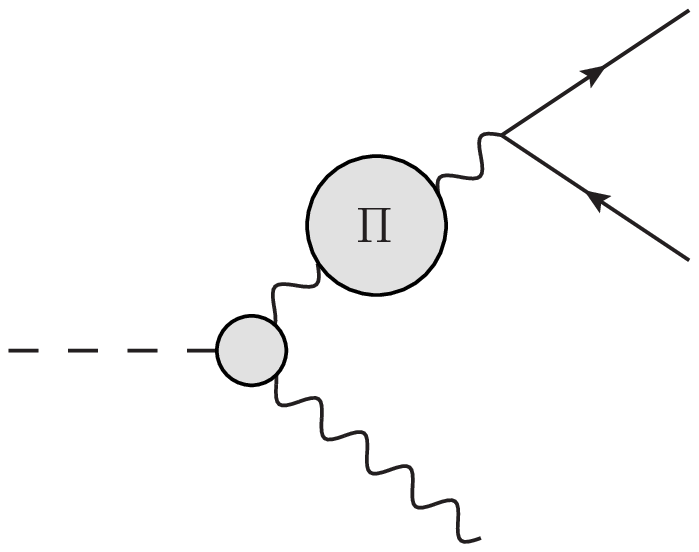}
\label{fig:virta}
}
\subfloat[][]{
\includegraphics[width=0.4\columnwidth]{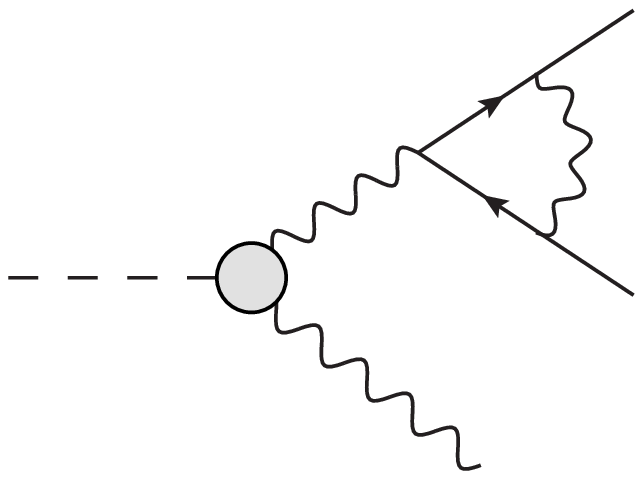}
\label{fig:virtb}
}

\subfloat[][]{
\includegraphics[width=0.4\columnwidth]{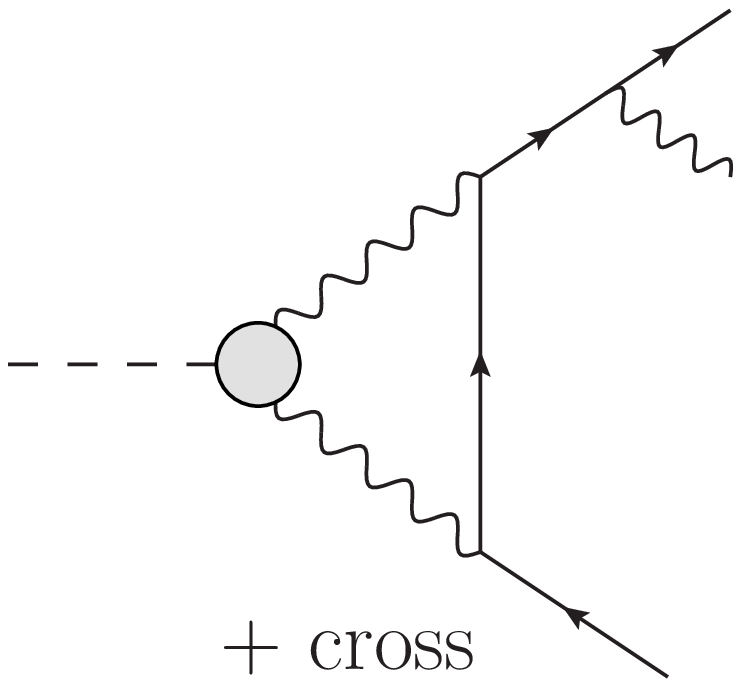}
\label{fig:1gIRa}
}
\subfloat[][]{
\includegraphics[width=0.4\columnwidth]{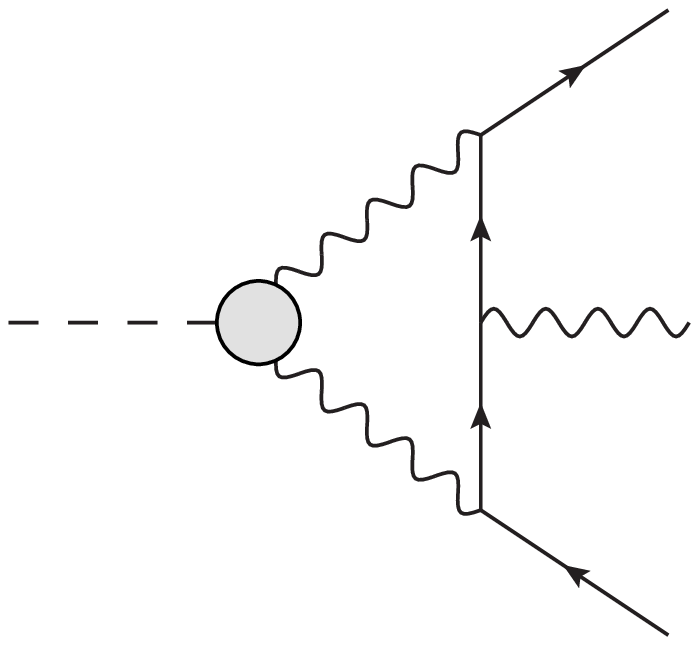}
\label{fig:1gIRb}
}

\subfloat[][]{
\includegraphics[width=0.85\columnwidth]{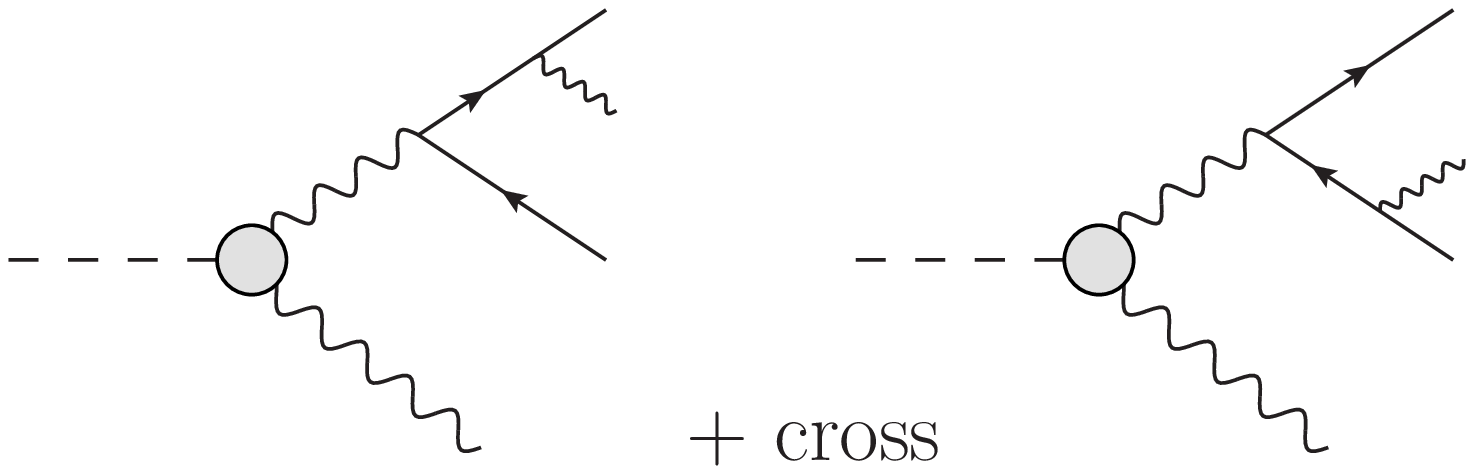}
\label{fig:BS}
}
\caption{
\label{fig:diagrams}
NLO QED radiative corrections to the Dalitz decay $P\to \ell^+\ell^-\gamma$: a) vacuum polarization insertion, b) correction to the QED vertex, c) \& d) one-loop one-photon-irreducible contributions, e) bremsstrahlung.
Note that `cross' in figure (c) corresponds to a diagram where the photon is emitted from the outgoing positron line.
Needless to say, `cross' in figure (e) stands for the diagrams with outgoing photons interchanged.
}
\end{figure}


\section{Virtual radiative corrections}
\label{sec:virt}

What we call the virtual radiative corrections is obtained from the interference terms of the LO diagram shown in Fig.~\ref{fig:LO} and the diagrams in Fig.~\ref{fig:virta} and \ref{fig:virtb}.
In the pion case, the vacuum polarization was dominated by the electron loop.
It turns out though that in the high-invariant-mass region of the photon propagator the hadronic effects become significant, which should be taken into account for the $\eta^{(\prime)}$ decays.
Thus, in general, we shall deal with the photon self-energy in the form
\begin{equation}
\Pi(s)=\Pi_\text{L}(s)+\Pi_\text{H}(s)\,,
\label{eq:Pi}
\end{equation}
where $\Pi_\text{L}$ and $\Pi_\text{H}$ stand for the leptonic and hadronic parts, respectively.
We discuss these contributions separately in the following. 
At NLO, the contribution of the vacuum polarization (\ref{eq:Pi}) to the virtual radiative correction then reads
\begin{equation}
\delta_{\Pi}^\text{virt}(x,y)
=2\operatorname{Re}\left\{-\Pi(M_P^2x)\right\}\,.
\label{eq:deltaPi}
\end{equation}
Note that $\delta_\Pi^\text{virt}(x,y)$ does not depend on $y$ and consistently with (\ref{eq:dx}) we have $\delta_\Pi^\text{virt}(x)=\delta_\Pi^\text{virt}(x,y)$.


Regarding the lepton part, all the necessary formulae connected with the contributions of lepton loops to the vacuum polarization are stated in Section III of~\cite{Husek:2015sma} and hold also in the current cases.
However, we shall point out some interesting details.
It was already discussed in~\cite{Husek:2015sma} that in the case of the $\eta^{(\prime)}$-meson decays not only the electron loop but also the muon loop should be taken into account as a part of the vacuum polarization contribution.
The contribution of the leptonic vacuum polarization insertion --- i.e.\ the contribution of the lepton loops to the photon propagator --- to the correction $\delta^\text{virt}$ can be expressed as (cf.\ the formulae in Section III of~\cite{Husek:2015sma})
\begin{equation}
\begin{split}
&\delta_{\Pi_\text{L}}^\text{virt}(x)
=2\operatorname{Re}\left\{-\Pi_\text{L}(M_P^2x)\right\}\\
&\equiv-2\,\frac{\alpha}{\pi}\sum_{\ell^\prime=e,\mu}
\bigg\{
\frac89-\frac{\beta_{\ell^\prime}^2}3
-\left(1-\frac{\beta_{\ell^\prime}^2}3\right)\\
&\times|\beta_{\ell^\prime}|\left[\theta(\beta_{\ell^\prime}^2)\operatorname{arctanh}\beta_{\ell^\prime}+\theta(-\beta_{\ell^\prime}^2)\operatorname{arctan}\frac1{|\beta_{\ell^\prime}|}\right]\bigg\}\,.
\label{eq:L}
\end{split}
\end{equation}
Here, $M_P^2x$ is the invariant mass of the final-state lepton pair and $\ell^\prime$ stands for the leptons circulating in the loop.
Consistently with definition (\ref{eq:beta}) we have $\beta_{\ell^\prime}\equiv\beta(M_P^2x;m_{\ell^\prime}^2)$ and $|\beta_{\ell^\prime}|=\sqrt{|\beta_{\ell^\prime}^2|}$\,.
Let us note that $\beta_{\ell^\prime}$ no longer stands for the CMS speed of the final-state leptons as was the case for $\beta_\ell$ in (\ref{eq:beta}).
Since $\beta$ depends on $x$, which for the given process of a decay to a lepton pair of flavor $\ell$ satisfies the limit $x\in[\nu_\ell^2,1]$, we can run into different kinematical regimes, which are explicitly covered by formula (\ref{eq:L}); for this purpose the Heaviside step function $\theta$ was used.
Whenever $\beta_{\ell^\prime}^2<0$, i.e.\ $\beta_{\ell^\prime}$ itself becomes imaginary, no on-shell lepton-antilepton pair can be created in the loop and the diagram in Fig.~\ref{fig:virta} lacks the imaginary part.
This happens if the invariant mass of the final-state lepton pair is under the production threshold of the two loop leptons of flavor $\ell^\prime$: $\nu_\ell^2\le x<\nu_{\ell^\prime}^2$.
In turn, this condition can be, of course, only realized in the case of the muon loop contribution to the NLO decay widths of the $\pi^0\to e^+e^-\gamma$ decay and in the process $\eta^{(\prime)}\to e^+e^-\gamma$, where the kinematical condition $\nu_e^2\le x<\nu_\mu^2$ is met within a part of the kinematically allowed region.


The hadronic contribution to the photon self-energy can be expressed via a dispersive integral~\cite{Jegerlehner:1985gq}
\begin{equation}
\Pi_\text{H}(s)
=-\frac{s}{4\pi^2\alpha}\int_{4m_\pi^2}^{\infty}\frac{\sigma_\text{H}(s^\prime)\diff s^\prime}{s-s^\prime+i\cuteps}\,.
\label{eq:PiH}
\end{equation}
Here, $\sigma_\text{H}$ is the total cross section of the $e^+e^-$ annihilation into hadrons and is related to the ratio $R(s)$ as $\sigma_\text{H}(s)=(4\pi\alpha^2/3s)R(s)$.
In order to obtain a smooth curve from the data~\cite{Olive:2016xmw}, we fit the experimental values in a similar way as it was done in~\cite{Jegerlehner:1985gq}.
Concerning the upper bound of the integral in (\ref{eq:PiH}), in practice it is sufficient to use the scale $\Lambda_\text{cut}\simeq4\,\text{GeV}$.
It is set in such a way that all the important resonances are covered while the higher-energy region does not contribute significantly for $s\lesssim M_{\eta^\prime}$.
The resulting contribution to the radiative corrections is plotted as one of the constituent curves in Fig.~\ref{fig:deltax0}.


Instead of (\ref{eq:deltaPi}), one might take into account the whole geometric series of vacuum polarization insertions and sum it in order to find
\begin{equation}
\begin{split}
\delta_\Pi^\text{virt}(x)
&=\frac1{|1+\Pi(M_P^2x)|^2}-1\\
&=-\frac{2\operatorname{Re}\Pi(M_P^2x)+|\Pi(M_P^2x)|^2}{|1+\Pi(M_P^2x)|^2}\,.
\label{eq:VPIalt}
\end{split}
\end{equation}
Eq.~(\ref{eq:deltaPi}) is then recovered by taking the linear expansion of (\ref{eq:VPIalt}), since $|\Pi(s)|\ll1$.
Note that in (\ref{eq:VPIalt}) one should use a full form of the vacuum polarization including its imaginary part, similarly as it was e.g.\ defined in (20) of~\cite{Husek:2015sma} for the lepton case:
\begin{equation}
\begin{split}
&\Pi_\text{L}(M_P^2x)\\
&=\frac{\alpha}{\pi}\sum_{\ell^\prime=e,\mu}
\left\{\frac89-\frac{\beta_{\ell^\prime}^2}3+\left(1-\frac{\beta_{\ell^\prime}^2}3\right)\frac{\beta_{\ell^\prime}}2\log\left[-\gamma_{\ell^\prime}+i\cuteps\right]\right\},
\label{eq:PiL}
\end{split}
\end{equation}
with $\gamma_{\ell}\equiv({1-\beta_{\ell}})/({1+\beta_{\ell}}).$
But since $|\Pi(s)|\ll1$, also $|\operatorname{Im}\Pi(s)|\ll1+\operatorname{Re}\Pi(s)$ and $2\operatorname{Re}\Pi(s)\gg|\Pi(s)|^2$.
One could thus safely use only the real part of $\Pi(s)$, as it was defined in (\ref{eq:deltaPi}), for the purpose of the numerical evaluation of (\ref{eq:VPIalt}).
This is equivalent to effectively using $\operatorname{Im}\Pi(s)\to0$.
On the other hand, the numerical difference between the definitions (\ref{eq:deltaPi}) and (\ref{eq:VPIalt}) is $\Delta\delta_\Pi^\text{virt}(x)\simeq0.25\,\%$ at $s\simeq M_\omega^2$.
Therefore we take into account the precise formula (\ref{eq:VPIalt}) together with (\ref{eq:PiH}) and (\ref{eq:PiL}).

For completeness, let us revise at this point what we mean by the virtual correction.
In agreement with Eq.~(16) in~\cite{Husek:2015sma}, we use
\begin{equation}
\begin{split}
&\delta^\text{virt}(x,y)\\
&=\frac1{|1+\Pi(M_P^2x)|^2}-1+2\operatorname{Re}\left\{F_1(x)+\frac{2F_2(x)}{1+y^2+\frac{\nu^2}{x}}\right\},
\label{eq:dvirt}
\end{split}
\end{equation}
where $\Pi(s)$ contains not only single electron and muon loops, but also the whole hadronic contribution to the photon self-energy.
For the form factors $F_1$ and $F_2$, which arise from the vertex correction diagram in Fig.~\ref{fig:diagrams}b, the reader is referred to seek the expressions in~\cite{Husek:2015sma}.


\section{One-photon-irreducible virtual radiative correction}
\label{sec:1gIR}

Considering the QED expansion, the NLO diagrams contributing to the one-loop one-photon-irreducible (1$\gamma$IR) correction to the $P\to \ell^+\ell^-\gamma$ process are shown in Figs.~\ref{fig:1gIRa} and~\ref{fig:1gIRb}.
We treat this contribution separately from the virtual correction to emphasize the fact that it was not included in the original approach~\cite{Mikaelian:1972yg}.
Therein, it was considered to be negligible already for the pion case.
This statement had been corrected in Ref.~\cite{Tupper:1983uw} many years before the debate about this issue was finally closed.
Note also that in the pion case, the 1$\gamma$IR contribution was studied in~\cite{Husek:2014tna} in connection with the bremsstrahlung correction to the $\pi^0\to e^+e^-$ rare decay.

In the $1\gamma$IR contribution one cannot factorize out the electromagnetic transition form factor.
This correction becomes therefore unavoidably model-dependent already at the two-fold differential level and it is necessary to choose a particular model to evaluate the correction numerically.
Compared to the previous calculations of the LO diagram and NLO virtual radiative corrections, a doubly virtual transition form factor $\mathcal{F}_{P\gamma^*\gamma^*}(l^2,(P-l)^2)$ is needed to be used.
This form factor enters the loop and the integration over the unconstrained momentum $l$ is then performed.
Note that $P$ stands for both the decaying pseudoscalar as well as for its four-momentum.

Let us now proceed further and consider that during the calculation of the 1$\gamma$IR loop diagrams the following structure inevitably appears:
\begin{equation}
\frac{\mathcal{F}_{P\gamma^*\gamma^*}(l^2,(P-l)^2)}{[l^2+i\cuteps][(P-l)^2+i\cuteps]}\,.
\label{eq:FFloop}
\end{equation}
By construction, the arguments of the form factor coincide with the photon propagators in the loop; $l$ denotes the loop momentum as is usual.
In what follows we consider a family of large-$N_\text{c}$ motivated analytic resonance-saturation models, which are in detail discussed in~\cite{Husek:2015wta} ($N_\text{c}$ denotes the number of colors).
Here $\mathcal{F}_{P\gamma^*\gamma^*}$ is a rational function with vector-meson poles.
We then realize that due to a use of the partial fraction decomposition one can perform --- within the loop integrals appearing during the evaluation of the diagrams --- the following substitution:
\begin{equation}
\frac{\mathcal{F}_{P\gamma^*\gamma^*}(p^2,q^2)}{p^2q^2}
=-\frac{N_\text{c}}{12\pi^2 F_\pi}\sum_i \alpha_i\,h(c_{1,i},c_{2,i},M_{1,i}^{2},M_{2,i}^{2})\,.\\
\label{eq:FFsubsth}
\end{equation}
Note that from now on we will suppress the explicit writing of the `+$i\cuteps$' part.
Above,
\begin{equation}
\begin{split}
&h(c_1,c_2,M_1^2,M_2^2)
\equiv
\frac{1}{p^2q^2}
+\frac{c_2}{(p^2-M_1^2)(q^2-M_2^2)}\\
&-c_1\left[\frac1{p^2(q^2-M_1^2)}+\frac1{q^2(p^2-M_1^2)}\right].
\label{eq:hc1c2}
\end{split}
\end{equation}
In this way we come from the matrix element $\mathcal{M}_{1\gamma\text{IR}}$ to $\sum\alpha_i\mathcal{M}_{1\gamma\text{IR}}^h[h_i]$; note that the normalization constant $\mathcal{F}_{\pi^0\gamma^*\gamma^*}(0,0)=-{N_\text{c}}/{(12\pi^2 F_\pi)}$ from decomposition (\ref{eq:FFsubsth}) is already included in $\mathcal{M}_{1\gamma\text{IR}}^h[h_i]$ and that we have used a shorthand notation $h_i\equiv h(c_{1,i},c_{2,i},M_{1,i}^{2},M_{2,i}^{2})$.
In order to get results for the whole family of models it is necessary to analytically integrate over the loop momentum just once; this is the main advantage of this approach.
At the end one can choose the particular model of the form factor by setting the parameters in the final matrix element appropriately.
We can find the above used constants $c_1$ and $c_2$ from projecting on the product of the normalized form factor and the photon propagators: for instance in the case $M_{V_1}=M_{V_2}$ we have
\begin{equation}
c_2
=\lim_{p^2,q^2\to M_{V_1}^2}\hspace{-1mm}\frac{\mathcal{F}_{P\gamma^*\gamma^*}(p^2,q^2)}{\mathcal{F}_{\pi^0\gamma^*\gamma^*}(0,0)}\,\frac{(p^2-M_{V_1}^2)(q^2-M_{V_1}^2)}{p^2q^2}\,.
\label{eq:c2}
\end{equation}

This little trick is highly convenient when it is necessary to create a universal code for calculating radiative corrections within different models.
Let us also note that substitution (\ref{eq:hc1c2}) does obviously not conserve term by term the desired property of the doubly virtual form factors --- the symmetry in their arguments.
This ansatz though works generally for the rational models mentioned above, which might have a rather complicated structure.
The symmetry in question is then always restored in the final result after including all the pieces $\mathcal{M}_{1\gamma\text{IR}}^h[h_i]$\,.
Moreover, we realize that if we define
\begin{equation}
g(M_1^2,M_2^2)
\equiv\frac{1}{(p^2-M_1^2)(q^2-M_2^2)}\,,
\end{equation}
we can immediately write
\begin{equation}
\begin{split}
&h(c_1,c_2,M_1^2,M_2^2)\\
&=g(0,0)+c_2\,g(M_1^2,M_2^2)
-c_1\big[g(0,M_1^2)+g(M_1^2,0)\big]\,.
\end{split}
\label{eq:htog}
\end{equation}
This trivial observation simplifies the loop integration even further.
Instead of (\ref{eq:FFsubsth}), we can then write
\begin{equation}
\frac{\mathcal{F}_{P\gamma^*\gamma^*}(p^2,q^2)}{p^2q^2}
=-\frac{N_\text{c}}{12\pi^2 F_\pi}\sum_i \beta_i\,g(M_{1,i}^{2},M_{2,i}^{2})\,.\\
\label{eq:FFsubstg}
\end{equation}
During the calculation of the amplitude it is only necessary to perform the following substitution:
\begin{equation}
\frac{\mathcal{F}_{P\gamma^*\gamma^*}(p^2,q^2)}{p^2q^2}
\rightarrow-\frac{N_\text{c}}{12\pi^2 F_\pi}\,g(M_1^2,M_2^2)\,.\\
\label{eq:FFsubstgBB}
\end{equation}
The desired final amplitude for the particular model is then obtained by writing a suitable combination (in spirit of (\ref{eq:FFsubstg})) of such amplitudes which are calculated using substitution (\ref{eq:FFsubstgBB}).
One only needs to insert the correct masses and coefficients into this combination, which goes along the lines of (\ref{eq:htog}), (\ref{eq:c2}) and (\ref{eq:hc1c2}).

Lets discuss the previous procedure on a particular case and consider for a while the eta-meson decays: $P=\eta$.
The simplest physically relevant model we can imagine is based on the VMD scenario, i.e.\ by an ansatz which assumes that the form factor is saturated by the lowest-lying multiplet of vector mesons.
It has the following form:
\begin{equation}
\begin{split}
&e^2\mathcal{F}_{\eta\gamma^*\gamma^*}^\text{VMD}(p^2,q^2)\\
&=-\frac{N_\text{c}}{8\pi^2 F_\pi}
\frac{2e^2}{3}
\bigg[
\frac53\frac{\cos\phi}{f_\ell}\,\frac{M_{\omega/\rho}^4}{(p^2-M_{\omega/\rho}^2)(q^2-M_{\omega/\rho}^2)}\\
&-\frac{\sqrt2}3\frac{\sin\phi}{f_\text{s}}\,\frac{M_\phi^4}{(p^2-M_\phi^2)(q^2-M_\phi^2)}
\bigg]\,.
\end{split}
\label{eq:FFVMD}
\end{equation}
Above, $\phi$ is the $\eta$-$\eta^\prime$ mixing angle and $f_\ell$ together with $f_s$ are the associated decay constants in the quark-flavor basis of the quark currents~\cite{Feldmann:1998sh,Escribano:2005qq}; for further details concerning derivation of this model see Appendix~\ref{app:etaVMD}.
It is then clear after counting of the loop-momenta powers that such a form factor guarantees the UV convergence of the loop integrals.
The matrix element for such a form factor can be schematically written as
\begin{equation}
\begin{split}
\mathcal{M}_{1\gamma\text{IR}}^\text{VMD}
&=\frac53\frac{\cos\phi}{f_\ell}\mathcal{M}_{1\gamma\text{IR}}^h\big[h(1,1,M_{\omega/\rho}^2,M_{\omega/\rho}^2)\big]\\
&-\frac{\sqrt2}3\frac{\sin\phi}{f_\text{s}}\mathcal{M}_{1\gamma\text{IR}}^h\big[h(1,1,M_\phi^2,M_\phi^2)\big]\,.
\end{split}
\label{eq:hVMDeta}
\end{equation}
Following the subsequent decomposition (\ref{eq:htog}) and using linearity, one then finds
\begin{equation}
\begin{split}
&\mathcal{M}_{1\gamma\text{IR}}^h\big[h(1,1,M_\mathcal{V}^2,M_\mathcal{V}^2)\big]\\
&=\mathcal{M}_{1\gamma\text{IR}}^h\big[g(0,0)\big]
+\mathcal{M}_{1\gamma\text{IR}}^h\big[g(M_\mathcal{V}^2,M_\mathcal{V}^2)\big]\\
&-\mathcal{M}_{1\gamma\text{IR}}^h\big[g(0,M_\mathcal{V}^2)\big]
-\mathcal{M}_{1\gamma\text{IR}}^h\big[g(M_\mathcal{V}^2,0)\big]\,.
\end{split}
\label{eq:Mhing}
\end{equation}
As the only building block, one needs to calculate $\mathcal{M}_{1\gamma\text{IR}}^h\big[g(M_1^2,M_2^2)\big]$ obtained in terms of substitution (\ref{eq:FFsubstgBB}) in the original matrix element.
For the purpose of the 1$\gamma$IR contribution to the correction $\delta(x,y)$ at the one loop level, the same decomposition will work.
Indeed, an interference of $\mathcal{M}_{1\gamma\text{IR}}$ with the LO matrix element needs to be considered and thus linearity is preserved.
Therefore it is necessary to take into account a linear combination of $\delta_{1\gamma\text{IR}}^h$ defined according to the prescription (26) in~\cite{Husek:2015sma}:
\begin{equation}
\begin{split}
&\delta_{1\gamma\text{IR}}^h\big[g(M_1^2,M_2^2)\big]\\
&=2\operatorname{Re}\left\{-\frac{\alpha}{\pi}\frac{\mathcal{F}_{\pi^0\gamma^*\gamma^*}(0,0)}{\mathcal{F}(M_P^2x)}\frac{i\pi^2M_P}{\left[1+y^2+\frac{\nu_\ell^2}{x}\right]}\right.\\
&\times\Big\{4\nu_\ell T[g(M_1^2,M_2^2)](x,y)\\
&\qquad+\Big[A[g(M_1^2,M_2^2)](x,y)M_P[x(1-y)^2-\nu_\ell^2]\\
&\qquad\qquad+(y\to-y)\Big]\Big\}\bigg\}\,.
\end{split}
\label{eq:d1gIRh}
\end{equation}
The explicit expressions for the building-block form factors $A$ and $T$ are shown in Appendix~\ref{app:M1gIR}.
Let us have a look at the ratio of the electromagnetic form factors in (\ref{eq:d1gIRh}).
Clearly, irrespective of the model that we use for the doubly virtual form factor, it is convenient that the normalization of the form factor in such a model equals to $\mathcal{F}(0)$ appearing in the LO expression; see also (\ref{eq:F}).
This leads to the fact that the correction will be independent of any overall normalization effects.
In the next section we introduce a spectral representation of a normalized form factor; see~(\ref{eq:KLrepre}).
Following what we have just assumed, we can write
\begin{equation}
\frac{\mathcal{F}_{\pi^0\gamma^*\gamma^*}(0,0)}{\mathcal{F}(M_P^2x)}
=\frac{\mathcal{F}(0)}{\mathcal{F}(M_P^2x)}\frac{\mathcal{F}_{\pi^0\gamma^*\gamma^*}(0,0)}{\mathcal{F}_{P\gamma^*\gamma^*}(0,0)}\,,
\end{equation}
and using the VMD-based scenario for instance for $\eta$ as in (\ref{eq:FFVMD}), we find
\begin{equation}
\frac{\mathcal{F}_{\pi^0\gamma^*\gamma^*}(0,0)}{\mathcal{F}_{\eta\gamma^*\gamma^*}^\text{VMD}(0,0)}
=\left[\frac53\frac{\cos\phi}{f_\ell}-\frac{\sqrt2}3\frac{\sin\phi}{f_\text{s}}\right]^{-1}.
\end{equation}

A simple example of the presented approach applied on the $P\to\ell^+\ell^-$ decays is included as Appendix~\ref{app:Pll}.
There is also a complementary technique how to cover a whole set of form factors under consideration.
Instead of putting the particular form factor into our diagrams, we can use the local Wess--Zumino--Witten (WZW) term.
In other words we trade the form factor for the constant given by the chiral anomaly.
It is then clear from simple considerations, that the contributions from Fig.~\ref{fig:1gIRa} need counter terms to compensate UV divergences.
The convergent part of such a counter term carries an undetermined constant $\chi^\text{(r)}(\mu)$ renormalized at scale $\mu$, which can effectively mimic the high-energy behavior of the would-be complete form factor.
Using a proper matching procedure, it is possible to acquire a numerical value of this constant $\chi^\text{(r)}(\mu)$ for a given form-factor model.
Up to mass corrections we can use an approximate formula to estimate this effective parameter (see Eq.~(46) in~\cite{Husek:2015wta}).
The question is, if this procedure can be used also for a box diagram in Fig.~\ref{fig:1gIRb} which is already convergent for the local WZW form factor.
It turns out that the corrections are of order $m_\ell^2/M_V^2$ and $M_P^2/M_V^2$.
Hence, for the pion case this assumption works well.
On the other hand, for $\eta^{(\prime)}$ it does not and one would need to introduce additional effective parameters in a consistent way.
This is though not a trivial task and is beyond the scope of this work.
Let us also mention that $\chi^\text{(r)}(\mu)$ enters the corrections being multiplied by $\nu_\ell^2$ and its effect is thus negligible for the decays with electrons in the final state.

In the result section (Section~\ref{sec:res}) we will use the simple VMD-inspired model for the doubly virtual transition form factor to estimate the importance of the 1$\gamma$IR contributions; for model details see Appendix~\ref{app:etaVMD}.
For completeness, let us then present the correction $\delta^{1\gamma\text{IR}}(x,y)$ within the model under consideration:
\begin{equation}
\begin{split}
&\delta_{\text{VMD}}^{1\gamma\text{IR}}(x,y)\\
&=\sum_{A\in\{\ell,\text{s}\}}\kappa_A^P
\Big\{
\delta_{1\gamma\text{IR}}^h\big[g(0,0)\big]
+\delta_{1\gamma\text{IR}}^h\big[g(M_A^2,M_A^2)\big]\\
&\qquad\qquad\;-\delta_{1\gamma\text{IR}}^h\big[g(0,M_A^2)\big]
-\delta_{1\gamma\text{IR}}^h\big[g(M_A^2,0)\big]\Big\}\,.
\end{split}
\label{eq:d1gIRVMD}
\end{equation}
For the $\eta$ case, we have $\kappa_\ell^\eta\equiv\frac53\frac{\cos\phi}{f_\ell}$, $\kappa_\text{s}^\eta\equiv-\frac{\sqrt2}3\frac{\sin\phi}{f_\text{s}}$.
The $\eta^{\prime}$ case then corresponds to the simultaneous interchange \{$\cos\phi\to\sin\phi$, $\sin\phi\to-\cos\phi$\}, of course followed by specifying the right $M_P$ within the building blocks $\delta_{1\gamma\text{IR}}^h$\,.
For the resonance masses we put $M_\ell\equiv M_{\rho/\omega}$ (the average physical mass of $\rho$ and $\omega$ mesons) for the light sector and $M_\text{s}\equiv M_\phi$ (the physical mass of the $\phi$ meson) for the strange sector.
However, we would like to stress again that the framework presented in the present section is capable of dealing also with more sophisticated form-factor models.
In general, the model dependence of part of the radiative corrections is, of course, a nuisance.
But for the doubly virtual transition form factors of $\eta$ and $\eta^\prime$ there is at present no alternative.
In the next section the singly virtual transition form factors are required.
Here the model dependence can be reduced by the use of dispersive methods~\cite{Hanhart:2013vba,Hanhart:2016pcd}.


\section{Bremsstrahlung}
\label{sec:BS}

In this section we briefly build on~\cite{Husek:2015sma} --- which among others discussed the bremsstrahlung correction calculation for the pion case --- and show the differences which come into play due to the fact that we are now interested in $\eta^{(\prime)}$ mesons.
We also show some techniques how to deal with the obstacles which arise.
Concerning the notation, we will restrict ourselves to the one used in previous works.

The diagrams which contribute to the Dalitz decay bremsstrahlung are shown in Fig.~\ref{fig:BS}.
Their contribution is (among others) important to cancel IR divergences stemming from the virtual corrections depicted in Fig.~\ref{fig:virtb}.
The corresponding invariant matrix element (including cross terms) can be written in the form
\begin{equation}
i\mathcal{M}_\text{BS}
=\frac{\mathcal{F}\big((l+p+q)^2\big)}{(l+p+q)^2+i\epsilon}\,I(k,l)
+(k\leftrightarrow l)\,,
\label{eq:MBS}
\end{equation}
where
\begin{equation}
I(k,l)
=\bar u(p)I^{\rho\sigma}(k,l)v(q)\epsilon_\rho^*(k)\epsilon_\sigma^*(l)
\label{eq:M}
\end{equation}
with
\begin{equation}
\begin{split}
&I^{\alpha\beta}(k,l)
=-i^5e^4{\varepsilon^{(l+p+q)(k)\mu\alpha}}\\
&\times\bigg[\gamma^\beta\frac{(\slashed l+\slashed p+m)}{2l\cdot p+i\epsilon}\gamma^\mu-\gamma^\mu\frac{(\slashed l+\slashed q-m)}{2l\cdot q+i\epsilon}\gamma^\beta\bigg]\,.
\end{split}
\end{equation}
Here, we use $l$ and $k$ for the photons and $p$ and $q$ for the electron and positron four-momenta, respectively, and $\mathcal{F}((p+q)^2)\equiv\mathcal{F}_{P\gamma^*\gamma^*}(0,(p+q)^2)$.
Note that we use the shorthand notation for the product of the Levi-Civita tensor and four-momenta in which $\varepsilon^{(k)\dots}=\varepsilon^{\mu\dots} k_\mu$.
Inasmuch as an additional photon comes into play it is convenient to introduce a new kinematical variable which stands for the normalized invariant mass squared of the two photons
\begin{equation}
x_\gamma=\frac{(k+l)^2}{M_P^2}\,.
\end{equation}
It has the similar meaning as $x$ in the case of the lepton-antilepton pair.
The contribution of the bremsstrahlung to the next-to-leading-order two-fold differential decay width can be written as
\begin{equation}
\frac{\diff^2\Gamma_\text{BS}^\text{NLO}}{\diff x\diff y}
=\frac{(1-x)}{4M_P(2\pi)^8}\frac{\pi^3M_P^4}{16}\int J\Big\{\overline{|\mathcal{M}_\text{BS}|^2}\Big\}\diff x_\gamma\,.
\label{eq:dBS}
\end{equation}
The above used operator $J$ is defined for an arbitrary invariant $f(k,l)$ of the momenta $k$ and $l$ as follows:
\begin{equation}
J\{f(k,l)\}
=\frac 1{2\pi}\int\frac{\diff^3k}{k_0}\frac{\diff^3l}{l_0}f(k,l)\,\delta^{(4)}(P-p-q-k-l)\,.
\end{equation}

In the case of the pion Dalitz decay, the value of the slope parameter of the form factor is small: $a_\pi\simeq0.03$.
Consequently, the form factor $\mathcal{F}((l+p+q)^2)$ which enters (\ref{eq:MBS}) can be conveniently expanded in the following way:
\begin{equation}
\mathcal{F}\big((l+p+q)^2\big)
\simeq\mathcal{F}(M_P^2x)\bigg[1+a\,\frac{2l\cdot(p+q)}{M_P^2}\bigg]\,.
\label{eq:BSFFexpand}
\end{equation}
Therefore, $\mathcal{F}({(l+p+q)^2})$ can be approximated by $\mathcal{F}(M_P^2x)$ for the process $\pi^0\to e^+e^-\gamma$.
This squared leads to the direct cancellation with $|\mathcal{F}(M_P^2x)|^2$ appearing in the leading-order expression.
The bremsstrahlung contribution to the radiative corrections then becomes effectively independent of the particular model of the pion transition form factor.
However, in the case of an $\eta$ meson, the slope parameter is $a_\eta\simeq0.5$, which is definitely not negligible anymore.
One would need to include higher-order corrections in expansion (\ref{eq:BSFFexpand}), the convergence would be slower and things would become in general more complicated since additional terms would need to be treated; see also the discussion at the end of Section V of~\cite{Husek:2015sma}.
The real obstacles though appear with the $\eta^\prime$-meson case.
Due to the fact that $a_{\eta^\prime}\simeq1.4$, expansion (\ref{eq:BSFFexpand}) is not applicable at all.
One thus needs to calculate with the full form factor.

Although in such a case the situation is somewhat different compared to the one in which the form factor cancels out, in general it is possible to use a similar framework as in the pion case --- at least in the sense of treating the kinematical integrals.
Accordingly, one needs to rewrite the bremsstrahlung correction in terms of integrals which are known from the pion case.
These need to be somewhat generalized due to the presence of poles in the form factor; for results see Appendix~\ref{app:J}.
This becomes more important in the $\eta$ case compared to the pion decay and needs to be taken into account explicitly.
For the $\eta^\prime$ case, this procedure then becomes absolutely crucial since in the hadron spectrum the mass of the $\eta^\prime$ meson lies above the masses of the lightest vector-meson resonances $\rho$ and $\omega$ and close to the $\phi$-meson mass.
The subsequent (numerical) integrations over all the relevant kinematical configurations of the bremsstrahlung photon become significantly nontrivial due to the running over these poles, which are regulated by incorporating physical widths of the resonances.
The narrow resonances like $\omega$ and $\phi$ are somewhat straightforward to include.
However, the width of the broad $\rho$ resonance is sensitive to the $\pi$-$\pi$ scattering.
This can be also taken into account by the use of recent dispersive approaches~\cite{Hanhart:2013vba,Hanhart:2016pcd}.
To this extent, it is convenient to use the K\"all\'en--Lehmann spectral representation of the Feynman propagator, which allows for the use of a common spectral density function for all the mentioned resonances.
The facts stated above make the bremsstrahlung contribution --- especially in the case of the $\eta^\prime$ meson --- significantly sensitive to the form-factor model.
To minimize the model dependence, whenever possible, data and general principles of QFT, analyticity and unitarity, are used that give rise to a dispersive framework.

In the K\"all\'en--Lehmann spectral representation, the form factor has the following form:
\begin{equation}
\frac{\mathcal{F}(q^2)}{\mathcal{F}(0)}
\simeq1+q^2\int_{4m_{\pi}^2}^{\Lambda^2}
\frac{\mathcal{A}(s)\diff s}{q^2-s+i\cuteps}
\equiv1+q^2\mathcal{I}(q^2)\,.
\label{eq:KLrepre}
\end{equation}
Concerning the integration range, we restrict ourselves to $(4m_\pi^2,\Lambda^2)$.
This is sufficient for our purpose and covers all the important physics.
In the lower bound, $m_\pi$ is the mass of a charged pion $\pi^{\pm}$ and $4m_\pi^2$ then constitutes the squared mass of the lowest hadronic state that can couple to a photon.
The upper bound is governed by the cut-off $\Lambda\simeq1.05$\,GeV chosen in such a way just to cover the peak and width of the $\phi$ resonance.
The spectral function has in the chosen energy range two main contributions distinguished by isospin, $\mathcal{A}(s)=\mathcal{A}_0(s)+\mathcal{A}_1(s)$.

The narrow resonances $\omega$ and $\phi$ contribute to the isospin-zero part
\begin{equation}
\mathcal{A}_0(s)
=w_\omega\mathcal{A}_\omega(s)
+w_\phi\mathcal{A}_\phi(s)\,,
\label{eq:A0}
\end{equation}
with the narrow-resonance spectral function
\begin{equation}
\mathcal{A_V}(s)
\equiv\mathcal{A}(s;M_\mathcal{V},\Gamma_\mathcal{V})
\equiv
-\frac1\pi\frac{M_\mathcal{V}\Gamma_\mathcal{V}}{(s-M_\mathcal{V}^2)^2+(M_\mathcal{V}\Gamma_\mathcal{V})^2}\,.
\label{eq:A_V}
\end{equation}
Note that after a full integration over $s$ one indeed gets (up to a sign) the resonance propagator:
\begin{equation}
\int_{-\infty}^\infty\frac{\mathcal{A_V}(s)\diff s}{q^2-s+i\cuteps}
=-\frac1{q^2-M_\mathcal{V}^2+iM_\mathcal{V}\Gamma_\mathcal{V}}\,.
\end{equation}
Finally, let us mention, that a one-narrow-resonance VMD model for the form factor can be written in terms of (\ref{eq:KLrepre}) with $\mathcal{A}(s)=
\mathcal{A_V}(s)$.

The isospin-one part is governed by the broad $\rho$ meson.
In order to include the important effect of $\pi\pi$ scattering, we use the dispersive approach of~\cite{Hanhart:2013vba,Hanhart:2016pcd}.
There the spectral function has the form
\begin{equation}
\mathcal{A}_1(s)
=-\frac\kappa{96\pi^2F_\pi^2}
\left[1-\frac{4m_\pi^2}s\right]^{3/2}P(s)\,R(s)\,|\Omega(s)|^2\,,
\label{eq:A1Omnes}
\end{equation}
where $F_\pi=92.2$\,MeV is the pion decay constant, $\Omega(s)$ is the Omn\`es function, a dispersive tool incorporating pion rescattering, and $P(s)$ and $R(s)=(1+\alpha_\text{V}s)$ are polynomials related to the $\eta\to\pi\pi\gamma$ reaction amplitude and the pion vector form factor $F_\text{V}(s)=R(s)\Omega(s)$, respectively.
In the case of $\eta^\prime$, we use $P_{\eta^\prime}(s)=(1+\alpha_{\eta^\prime}s+\beta_{\eta^\prime}s^2)$ with $\alpha_{\eta^\prime}=0.99(4)\,\text{GeV}^{-2}$ and $\beta_{\eta^\prime}=-0.55(4)\,\text{GeV}^{-4}$~\cite{Hanhart:2016pcd}.
For the $\eta$ spectral function, we take (based on~\cite{Hanhart:2013vba}) $P_\eta(s)=(1+\alpha_\eta s)R(s)$ with $\alpha_\eta=1.32(13)\,\text{GeV}^{-2}$.
Note that in view of (\ref{eq:A1Omnes}) and taking into account only terms linear in $s$, $P_\eta(s)=1+(\alpha_\eta+\alpha_V)s$.
The rest of the parameters from (\ref{eq:A0}) and (\ref{eq:A1Omnes}) are given in Table~\ref{tab:Aparams}.
\begin{table}[ht!]
\begin{ruledtabular}
\begin{tabular}{c | c c c c c c}
$P$ & $\kappa$ & $\alpha_\text{V}$\,[GeV$^{-2}$] & $w_\omega$ & $w_\phi$\\
\hline
$\vphantom{\Big(}\eta$ & 0.56 & 
0.115 & 0.78(4)$\times\frac18$ & 0.75(3)$\times\left(-\frac28\right)$\\
$\eta^\prime$ & 0.415 & 
0.09 & 1.27(7)$\times\frac1{14}$ & 0.54(2)$\times\frac4{14}$\\
\end{tabular}
\end{ruledtabular}
\caption{
Values for $w$ are taken from~\cite{Hanhart:2013vba} and values of $\kappa$ were calculated using the following prescription therein: $\kappa_{\eta^{(\prime)}}=eA_{\pi\pi\gamma}^{\eta^{(\prime)}}F_\pi^2/A_{\gamma\gamma}^{\eta^{(\prime)}}$.
The numerical value of $\alpha_\text{V}^\eta$ was estimated from the fit in the upper panel of Fig.~1 in~\cite{Hanhart:2013vba} and the value of $\alpha_\text{V}^{\eta^\prime}$ is based on values tabulated in~\cite{Hanhart:2016pcd}.
}
\label{tab:Aparams}
\end{table}

\begin{figure}[th!]
\resizebox{\columnwidth}{!}{\includegraphics{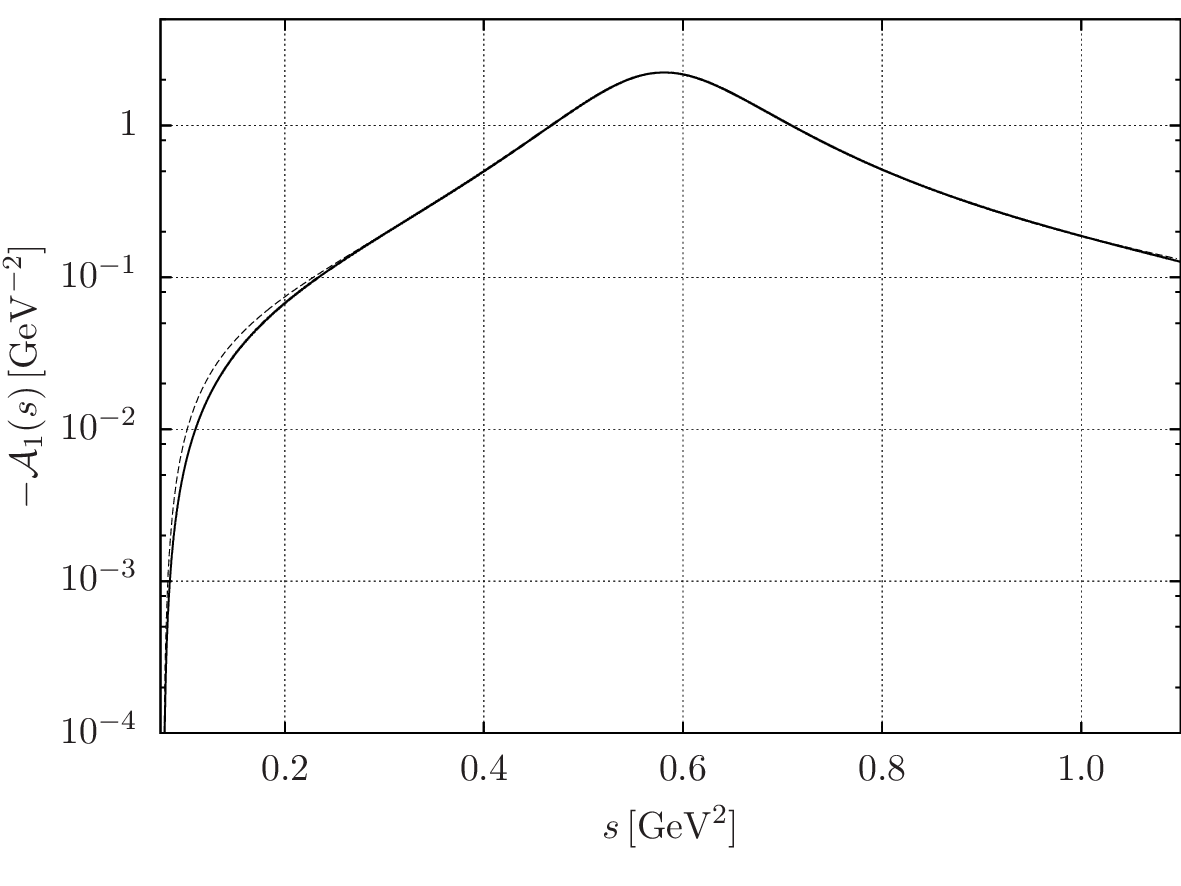}}
\resizebox{\columnwidth}{!}{\includegraphics{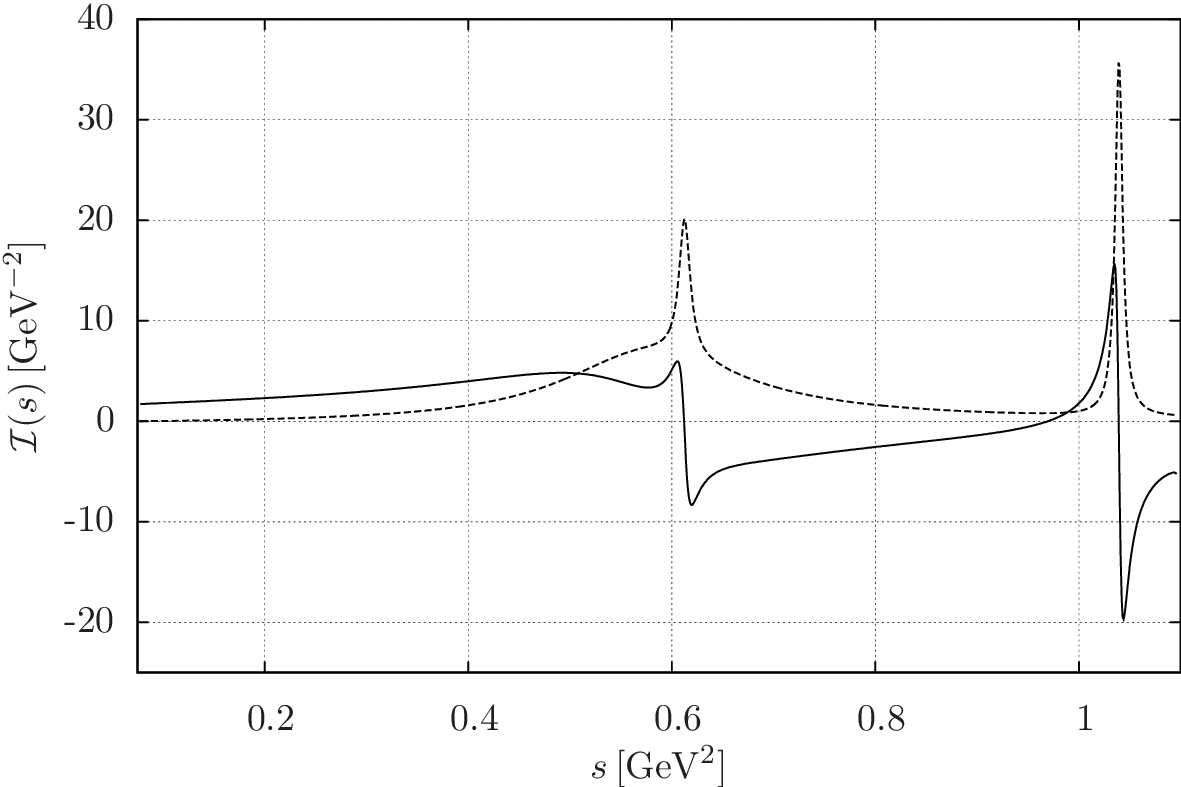}}
\caption{
The fit and the dispersive integral of the spectral function for the $\eta^\prime$ case.
In the first panel we display the fit of the spectral function $\mathcal{A}_1(s)$.
The curve based on Eq.~(\ref{eq:A1Omnes}) (shown as a dashed line) is compared to the fit based on the ansatz (\ref{eq:A1fit}) (shown as a solid line).
In the second panel one can find the dispersive integral $\mathcal{I}(s)$ as it is defined in (\ref{eq:KLrepre}): $\operatorname{Re}\mathcal{I}(s)$ is shown as a solid line and $\operatorname{Im}\mathcal{I}(s)$ as a dashed line.
}
\label{fig:A1fit}
\vspace{-5mm}
\end{figure}

For numerical reasons, the spectral function for the broad $\rho$ resonance may be fitted, since it is much faster to numerically integrate the analytical expression compared to the dispersive data interpolation.
To this extent, it is necessary to model the resonance peak behavior.
The following function copies the dispersive shape of $\mathcal{A}_1$ satisfactorily:
\begin{equation}
\mathcal{A}_1(s)
\simeq\left[a_0+a_1s+(a_2s)^2+(a_3s)^3\right]\mathcal{A}(s;M_\rho,\Gamma_\rho(s))\,.
\label{eq:A1fit}
\end{equation}
The energy dependence of the width of the $\rho$ meson --- assuming the main contribution comes from the 2$\pi$-decay --- can be expressed as
\begin{equation}
\begin{split}
\Gamma_\rho(s)
&\equiv\Gamma_\rho\frac{M_\rho}{\sqrt{s}}\left[\frac{s-4m_\pi^2}{M_\rho^2-4m_\pi^2}\right]^{3/2}\\
&\simeq\Gamma_\rho{\frac{M_\rho}{\sqrt{M_\rho^2-4m_\pi^2}}}\left(1-\frac{4m_\pi^2}{2s}\right)\left[\frac{s-4m_\pi^2}{M_\rho^2-4m_\pi^2}\right].
\end{split}
\label{eq:Gamma_rho}
\end{equation}
The advantage of the latter approximation is the fact, that when we insert it into (\ref{eq:A1fit}) and subsequently into (\ref{eq:KLrepre}), the form factor $\mathcal{F}(q^2)$ can be evaluated analytically.
This speeds up the numerics even further.
The values of the fit are shown in Table~\ref{tab:A1fit}.
\begin{table}[ht!]
\begin{ruledtabular}
\begin{tabular}{c | c c c c}
$P$ & $a_0$ & $a_1$\,[GeV$^{-2}$] & $a_2$\,[GeV$^{-2}$] & $a_3$\,[GeV$^{-2}$]\\
\hline
$\vphantom{\Big(}\eta$ & 3.336 & -6.364 & 2.622 & -1.342\\
$\eta^\prime$ & 2.333 & -4.969 & 2.261 & -1.240\\
\end{tabular}
\end{ruledtabular}
\caption{
The fitted values of ansatz (\ref{eq:A1fit}) for the spectral function $\mathcal{A}_1$.
}
\label{tab:A1fit}
\end{table}
For the $\eta^\prime$ case, the formula (\ref{eq:A1Omnes}) is compared to the fit in the upper panel of Fig.~\ref{fig:A1fit}.
The dispersive integral $\mathcal{I}(s)$ of the final spectral function $\mathcal{A}(s)$ is then shown in the second panel of Fig.~\ref{fig:A1fit}.

Let us now show, how the matrix element squared looks like in the spectral representation.
For the sake of writing down its structure, we recast (\ref{eq:MBS}) as
\begin{equation}
i\mathcal{M}_\text{BS}
=\frac{\mathcal{F}(E)}{E}\,I_\text{E}
+\frac{\mathcal{F}(F)}{F}\,I_\text{F}
\label{eq:MBSEF}
\end{equation}
where we have used the shorthand notation $E=(l+p+q)^2+i\cuteps$ and $F=(k+p+q)^2+i\cuteps$ and defined $I_\text{E}\equiv I(k,l)$ and $I_\text{F}\equiv I(l,k)$.
After inserting the representation of the form factor (\ref{eq:KLrepre}) we get
\begin{equation}
\frac{i\mathcal{M}_\text{BS}}{\mathcal{F}(0)}
=\left[\frac1{E}+\int_{4m_{\pi}^2}^{\Lambda^2}\diff s\,\mathcal{A}(s)\frac1{E-s+i\cuteps}\right]I_\text{E}
+(E\leftrightarrow F)\,.
\end{equation}
An attentive reader might have noticed that in the previous formula the explicit $i\cuteps$ is not necessary.
At the same time, it might be sensible to explicitly write the infinitesimal imaginary part of the propagator in the following expressions.
To avoid any potential confusion further on, let us define $e\equiv\operatorname{Re}E$ and, similarly, $f\equiv\operatorname{Re}F$.
After the operator $J$ is applied on the matrix element squared and summed over all the spins and polarizations $\overline{|\mathcal{M}_\text{BS}|^2}\equiv\sum_\text{sp., pol.}|\mathcal{M}_\text{BS}|^2$, we can actually use the symmetry $k\leftrightarrow l\Leftrightarrow E\leftrightarrow F$ term by term, which results into
\begin{equation}
\begin{split}
&J\hspace{-.5mm}\left\{\frac{\overline{|\mathcal{M}_\text{BS}|^2}}{\mathcal{F}^2(0)}\right\}
=2\operatorname{Re}J\bigg\{
\left[
\frac1{|E|^2}
+2\,\frac1{E^*}\int\hspace{-1mm}\frac{\mathcal{A}(s)\diff s}{{e}-s+i\cuteps}\right.\\
&\left.+\int\hspace{-2.5mm}\int\hspace{-1mm}\frac{\mathcal{A}(s)\diff s}{{e}-s+i\cuteps}\frac{\mathcal{A}(s^\prime)\diff s^\prime}{{e}^{}-s^\prime-i\cuteps^\prime}
\right]\overline{|I_\text{E}|^2}\\
&+\hspace{-.5mm}\left[
\frac1{EF^*}
+2\,\frac1{F^*}\int\hspace{-1mm}\frac{\mathcal{A}(s)\diff s}{{e}-s+i\cuteps}\right.\\
&+\left.\left.\int\hspace{-2.5mm}\int\hspace{-1mm}\frac{\mathcal{A}(s)\diff s}{{e}-s+i\cuteps}\frac{\mathcal{A}(s^\prime)\diff s^\prime}{{f}^{}-s^\prime-i\cuteps^\prime}
\right]\overline{I_\text{E} I_\text{F}^*}\right\}.
\label{eq:MBSJ}
\end{split}
\end{equation}
In order to perform the integrations of the $J$ operator on the respective terms, we need to do a few fraction-product decompositions; the procedure is described in detail in Appendix~\ref{app:eps}.
Taking also into account that
\begin{equation}
\operatorname{Re}\int\hspace{-2.5mm}\int\diff s\diff s^\prime
\frac{\mathcal{A}(s)\mathcal{A}(s^\prime)}{s-s^\prime-i(\cuteps+\cuteps^\prime)}
=-\pi\operatorname{Im}\int\diff s\,\mathcal{A}^2(s)
=0\,,
\label{eq:ReIntAA}
\end{equation}
we can rewrite (\ref{eq:MBSJ}) into
\begin{equation}
\begin{split}
&J\hspace{-.5mm}\left\{\frac{\overline{|\mathcal{M}_\text{BS}|^2}}{\mathcal{F}^2(0)}\right\}
=4\operatorname{Re}J\hspace{-.5mm}\left\{\left[
\frac12\frac1{|E|^2}
\right.\right.\\
&+\left.\int\diff s\,\mathcal{A}(s)
\left[\frac1s+\mathcal{I}^*(s)\right]\hspace{-1mm}
\left(\frac{1}{{e}-s+i\cuteps}-\frac1{E^*}\right)
\right]\overline{|I_\text{E}|^2}\\
&+\left[
\frac1{V_0}\frac1E
+\mathcal{I}^*(V_0)\frac1{E}
+\int\diff s\,\frac{\mathcal{A}(s)}{V_0-s+i\cuteps}\frac1{{e}-s+i\cuteps}\right.\\
&+\left.\left.
\int\diff s\,\mathcal{A}(s)\operatorname{Re}\{\mathcal{I}(V_0-s)\}\,\frac{1}{{e}-s+i\cuteps}
\right]\overline{I_\text{E} I_\text{F}^*}\right\}.
\label{eq:MBSJresI}
\end{split}
\end{equation}
Following the definition (\ref{eq:KLrepre}), we made use of the fact that
\begin{equation}
\mathcal{I}(s)
=\operatorname{p.v.}\int_{4m_{\pi}^2}^{\Lambda^2}\diff s^\prime\,\frac{\mathcal{A}(s^\prime)}{s-s^\prime}-i\pi\mathcal{A}(s)\,.
\label{eq:Is}
\end{equation}
With the previous simplifications (\ref{eq:A_V}), (\ref{eq:A1fit}) and (\ref{eq:Gamma_rho}), the above integral can be evaluated analytically which speeds up the time-demanding numerics.

For the sake of integrating out the bremsstrahlung photon it is convenient to define the following three rescaled parts of the matrix element squared:
\begin{align}
\frac{e^8}{4}\text{Tr}_{\text{E}^2}&\equiv\frac{\overline{|I_\text{E}|^2}}{|E|^2}\,,\label{eq:TrE2}\\
\frac{e^8}{4}\text{Tr}_{\text{E}}(s)&\equiv\frac{\overline{|I_\text{E}|^2}}{{e}-s+i\cuteps}\,,\label{eq:TrEs}\\
\frac{e^8}{4}\text{Tr}_\text{EF}(s)&\equiv\frac{\overline{I_\text{E} I_\text{F}^*}}{{e}-s+i\cuteps}\label{eq:TrEFs}\,.
\end{align}
One gets the remaining building blocks of (\ref{eq:MBSJresI}) by performing the limit $s\to0$, i.e.\ taking $\text{Tr}_\text{E}(0)$ and $\text{Tr}_\text{EF}(0)$.
Note that it does not matter if one uses $1/E^*$ or simply $1/E$ for the Feynman propagators in the definitions of $\text{Tr}_\text{E}(0)$ and $\text{Tr}_\text{EF}(0)$ since these give real contributions.
The name $\text{Tr}$ is chosen to be in agreement with~\cite{Mikaelian:1972yg}.
Note that the original work of Mikaelian and Smith (M\&S) --- which in some sense corresponds to $\mathcal{A}(s)=0$ since the form factor could have been factorized out --- is connected with the previous definitions through
\begin{equation}
\overline{|\mathcal{M}_\text{BS}^\text{M\&S}|^2}
=\frac{e^8}4\left|\mathcal{F}(M_P^2x)\right|^2\text{Tr}^\text{M\&S}\,,
\end{equation}
where
\begin{equation}
J\hspace{-.5mm}\left\{\text{Tr}^\text{M\&S}\right\}
=4J\hspace{-.5mm}\left\{\frac12\text{Tr}_{\text{E}^2}+\frac{\text{Tr}_\text{EF}(0)}{M_P^2(1+x-x_\gamma)}\right\}.
\label{eq:JTrMS}
\end{equation}
For numerical reasons, the integration of the resulting expression is performed term by term.
We can name these terms for further convenience.
Consequently, we write
\begin{equation}
J\hspace{-.5mm}\left\{\frac{\overline{|\mathcal{M}_\text{BS}|^2}}{\mathcal{F}^2(0)}\right\}
\equiv\frac{e^8}4\operatorname{Re}J\hspace{-.5mm}\left\{\text{Tr}\right\}\,,\quad\text{Tr}=4\sum_i{t_i}\,,
\label{eq:MBSJresTr}
\end{equation}
where
\begin{alignat}{2}
t_{1\text{a}}&=\frac12\text{Tr}_{\text{E}^2}\,,\quad
t_{1\text{b}}=\frac1{V_0}\text{Tr}_\text{EF}(0)\,,\label{eq:t1}\\
t_2&=\int\diff s\,\mathcal{A}(s)\left[\frac1s+\mathcal{I}^*(s)\right]\hspace{-.5mm}\left[\text{Tr}_{\text{E}}(s)-\text{Tr}_{\text{E}}(0)\right],\label{eq:t2}\\
t_{3\text{a}}&=\mathcal{I}^*(V_0)\,\text{Tr}_\text{EF}(0)\,,\quad
t_{3\text{b}}=\int\diff s\,\frac{\mathcal{A}(s)}{V_0-s+i\cuteps}\,\text{Tr}_\text{EF}(s)\,,\label{eq:t3}\\
t_4&=\int\diff s\,\mathcal{A}(s)\operatorname{Re}\{\mathcal{I}(V_0-s)\}\,\text{Tr}_\text{EF}(s)\label{eq:t4}\,.
\end{alignat}
Terms $t_{1\text{a}}$ and $t_{1\text{b}}$ are the only ones present in the M\&S case.
Terms $t_{1\text{a}}$ and $t_{2}$ are dominant.
By construction (cf.~(\ref{eq:MBSJres})), terms $t_{3\text{a}}$ and $t_{3\text{b}}$ belong together since separately they develop peaks which are exactly compensated only in the sum.
Let us mention that for the purpose of (\ref{eq:MBSJresTr}), in $t_{3\text{a}}$ it is sufficient to take only the real part of $\mathcal{I}(V_0)$ since $\text{Tr}_\text{EF}(0)$ gives only a real-part contribution.

For the correction $\delta^\text{BS}(x,y)$ it holds
\begin{equation}
\delta^\text{BS}(x,y)
\sim\int J\hspace{-.5mm}\left\{\frac{\overline{|\mathcal{M}_\text{BS}(x,y;k,l)|^2}}{\mathcal{F}^2(0)}\right\}\diff x_\gamma\,.
\end{equation}
Together with (\ref{eq:MBSJres}) this suggests that we need to perform five subsequent integrations: two nontrivial analytical ones are implicitly hidden in the $J$ operator and at least two need to be inevitably performed numerically.
Let us briefly look at the explicit integral structure --- for instance on the case of $t_4$ --- and how it contributes to the bremsstrahlung correction.
In terms of definitions (\ref{eq:Is}) and (\ref{eq:TrEFs}) we have
\begin{equation}
\begin{split}
&\delta_{t_4}^\text{BS}(x,y)\\
&\sim\operatorname{Re}\int\hspace{-2.5mm}\int\mathcal{A}(s)\,\operatorname{Re}\{\mathcal{I}(V_0-s)\}\,J\{\text{Tr}_\text{EF}(s)\}\diff x_\gamma\diff s\,.
\end{split}
\end{equation}
The term $\operatorname{Re}\{\mathcal{I}(V_0-s)\}$ can be evaluated analytically by taking the fit (\ref{eq:A1fit}) of the spectral function instead of its numerical form (\ref{eq:A1Omnes}) given by the term containing the Omn\`es function.
This procedure significantly speeds up the numerics.
The evaluation of $J\{\text{Tr}_\text{EF}(s)\}$ is performed analytically and the strategy is similar to the one used in~\cite{Husek:2015sma}.
Our goal is to rewrite the expressions under consideration into basic integrals, which are then listed in Appendix D of~\cite{Husek:2015sma} and Appendix~\ref{app:J} of the presented work.
For this purpose, symmetries and properties of operator $J$ are used together with the reduction procedure described in detail in Section V of~\cite{Husek:2015sma} and summarized in Appendix B therein.
Due to the presence of the effective mass $s$ in the denominators, this procedure needs to be slightly modified since now also $E-s$ appears instead of simple $E$.
As a trivial example, we choose the following reduction:
\begin{equation}
\begin{split}
&J\bigg\{\frac1{AB(E-s)}\bigg\}
=\frac1{M_P^2x-s+i\cuteps}\\
&\times\left[J\bigg\{\frac1{AB}\bigg\}-2J\bigg\{\frac1{A(E-s)}\bigg\}-2J\bigg\{\frac1{B(E-s)}\bigg\}\right].
\end{split}
\end{equation}
Note also that the IR-divergent terms (those which diverge after integrating over $x_\gamma$) need to be treated separately.
The current case is discussed in Appendix~\ref{app:JTr}.
For further details and an introduction we refer an interested reader to~\cite{Husek:2015sma}.

To conclude, let us finally write the bremsstrahlung contribution to the radiative corrections (cf.\ (37) in~\cite{Husek:2015sma}):
\begin{equation}
\delta^\text{BS}(x,y)
=\frac{1}{64}\frac{\alpha}{\pi}\frac{4x}{(1-x)^2}\bigg|\frac{\mathcal{F}(0)}{\mathcal{F}\big(M_P^2x\big)}\bigg|^2\frac{4\int \sum_i J\hspace{-.5mm}\left\{{t_i}\right\}\,{\mathrm d}x_\gamma}{\left[1+y^2+\frac{\nu_\ell^2}{x}\right]}\,,
\label{eq:dBSxy}
\end{equation}
where index $i$ runs over all the subscripts appearing in definitions (\ref{eq:t1})-(\ref{eq:t4}): 1a, 1b (the form-factor-independent parts) and 2, 3a, 3b and 4 (the form-factor-dependent parts).


\section{Results}
\label{sec:res}

In the previous sections, we discussed the three main parts of radiative corrections as we referred to them in (\ref{eq:delta_origins}), i.e. the virtual corrections in Section~\ref{sec:virt}, the 1$\gamma$IR correction in Section~\ref{sec:1gIR} and the bremsstrahlung in Section~\ref{sec:BS}.
In order to get the final result, one then simply sums over the partial results: Eqs. (\ref{eq:dvirt}), (\ref{eq:d1gIRVMD}) and (\ref{eq:dBSxy}).
Let us now comment once again on these contributions in detail.
The model-independent expression for the virtual corrections (\ref{eq:dvirt}) includes the photon self-energy contribution (\ref{eq:Pi}), consisting of the exact (at NLO) leptonic (\ref{eq:PiL}) and the experimental-data-based hadronic (\ref{eq:PiH}) parts, and the vertex correction expressed in terms of the $F_1$ and $F_2$ form factors.
The expressions for these form factors can be found in~\cite{Husek:2015sma}; in particular see Eqs. (22) and (23) therein.
The form factor $F_1$ then contains the IR-divergent piece which cancels with the corresponding term stemming from the bremsstrahlung contribution, as shown at the end of Appendix~\ref{app:JTr}.
The bremsstrahlung contribution (\ref{eq:dBSxy}) itself consists of the fully model-independent ingredients (\ref{eq:t1}), already present in the M\&S case, and the pieces (\ref{eq:t2}-\ref{eq:t4}) that still contain some model dependence, though mitigated by our dispersive, data-driven input.
All the necessary definitions are then presented in Appendices~\ref{app:JTr} and \ref{app:J}.
Finally, for the model-dependent 1$\gamma$IR contribution (\ref{eq:d1gIRVMD}) we used the VMD-inspired model discussed in Appendix~\ref{app:etaVMD}.
Let us again stress at this point that we used a rather general approach applicable for a wide family of rational models.
The final result within a particular model can then be found as an appropriate linear combination of the building blocks (\ref{eq:d1gIRh}), with the form factors $A$ and $T$ defined in Appendix~\ref{app:M1gIR}.
As another example of a suitable model we mention the one based on the Resonance Chiral Theory framework~\cite{Guevara:2018rhj}.
There are other --- data-driven --- models~\cite{Escribano:2015vjz,Masjuan:2017tvw}, which are then not directly applicable using the presented approach.
However, we expect that different models would lead to compatible values in the numerical results.

\begin{figure*}[!ht]
\subfloat[][$\eta\to e^+e^-\gamma$]{
\includegraphics[width=\columnwidth]{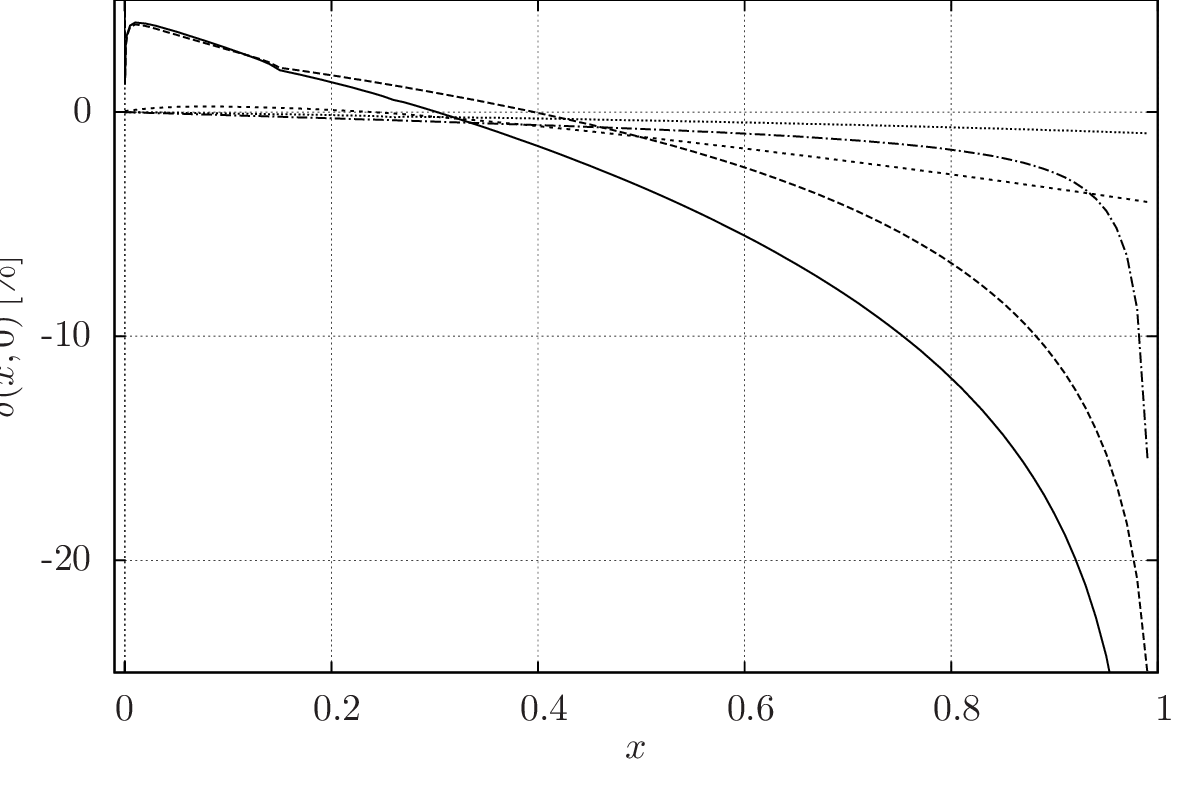}
}
\subfloat[][$\eta\to \mu^+\mu^-\gamma$]{
\includegraphics[width=\columnwidth]{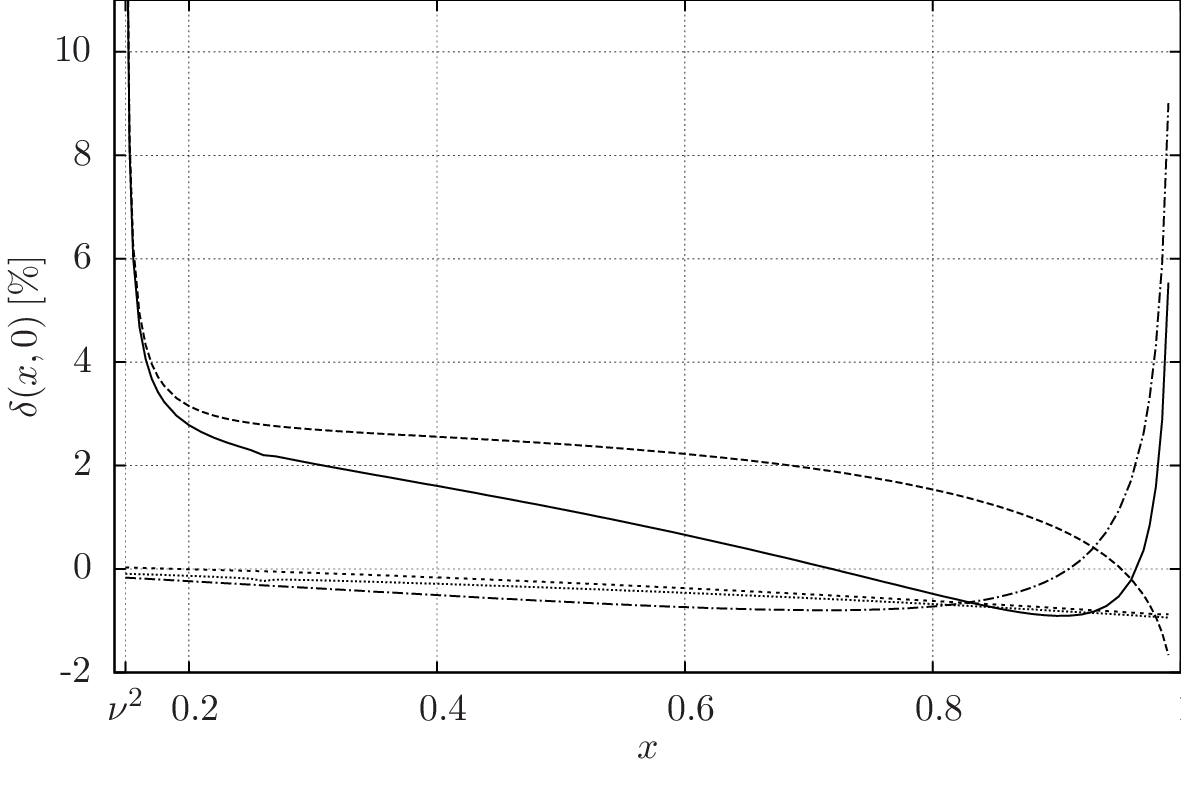}
}

\subfloat[][$\eta^\prime\to e^+e^-\gamma$]{
\includegraphics[width=\columnwidth]{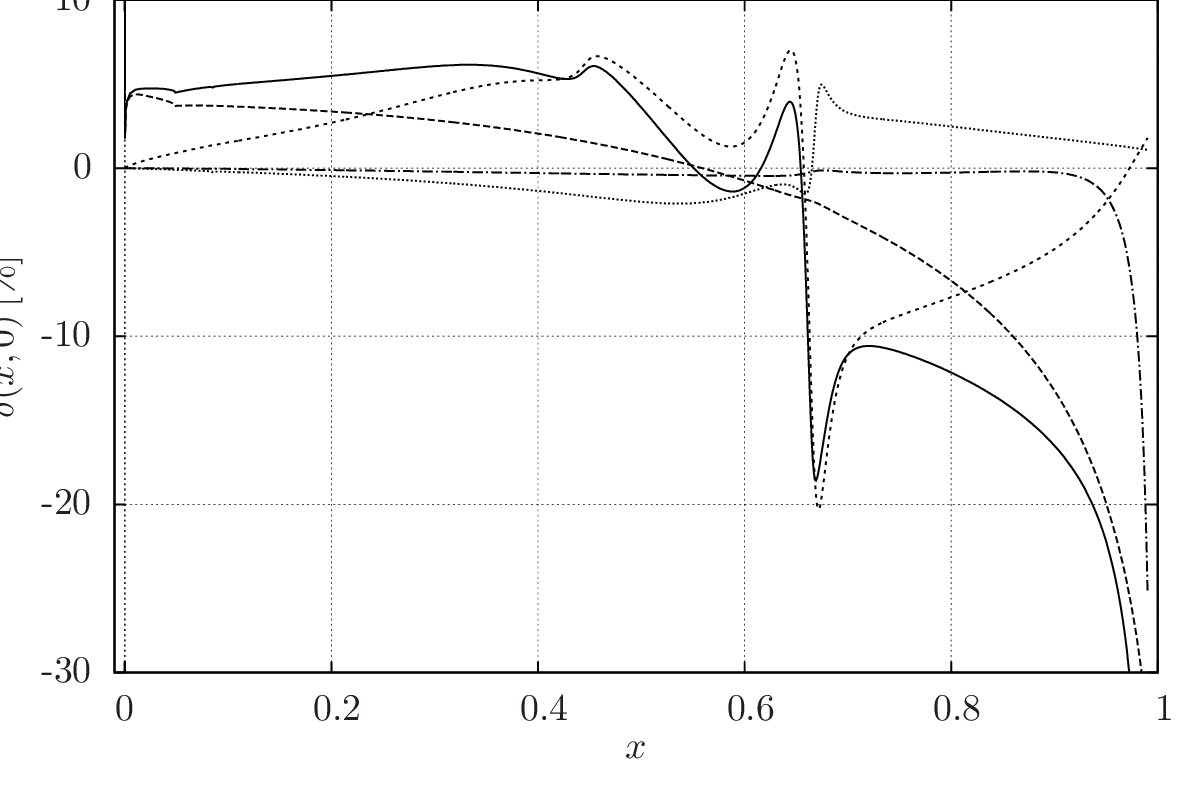}
}
\subfloat[][$\eta^\prime\to \mu^+\mu^-\gamma$]{
\includegraphics[width=\columnwidth]{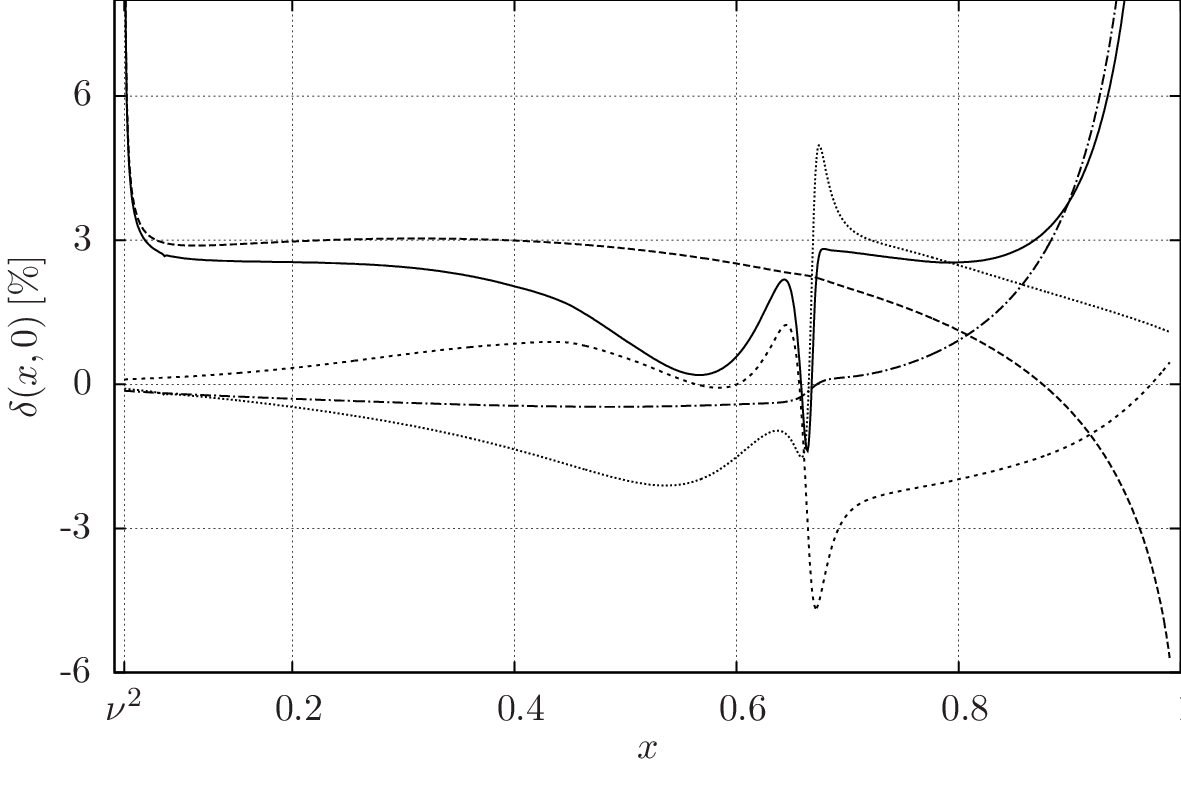}
}
\caption{\label{fig:deltax0}
The overall NLO correction $\delta(x, 0)$ (solid line) in comparison to its constituents for the decays $\eta^{(\prime)}\to \ell^+\ell^-\gamma$.
The sum $\delta^\text{virt}|_{\Pi=\Pi_\text{L}}+\delta_\text{D}^\text{BS}+\delta_{t_{1a}}^\text{BS}+\delta_{t_{1b}}^\text{BS}$ is depicted as a dashed line.
It would have directly corresponded to the M\&S correction~\cite{Mikaelian:1972jn} if muon loops had been included to the vacuum polarization and higher orders in lepton masses ($\mathcal{O}(\nu_\ell^4)$) had not been neglected in~\cite{Mikaelian:1972jn}.
This contribution also includes the IR-divergent part of the bremsstrahlung.
The hadronic contribution $\delta_\text{H}^\text{virt}\equiv\delta^\text{virt}-\delta^\text{virt}|_{\Pi=\Pi_\text{L}}=\delta_\Pi^\text{virt}-\delta_{\Pi_\text{L}}^\text{virt}$ to the virtual radiative corrections is shown as a small-spaced dotted line.
The part of the bremsstrahlung correction which is dependent on the model of the form factor and is nonzero for any nonvanishing spectral function and which corresponds to the sum $\delta_{t_2}^\text{BS} + \delta_{t_{3\text{a}}}^\text{BS} + \delta_{t_{3\text{b}}}^\text{BS} + \delta_{t_4}^\text{BS}$ is shown as a large-spaced dotted line.
The one-photon-irreducible contribution $\delta^{1\gamma\text{IR}}$ is then shown as a dash-dot line.
The divergent behavior of $\delta(x)$ near $x=\nu_\ell^2$ has its origin in the electromagnetic form factor $F_1(x)$ and is connected to the Coulomb self-interaction of the dilepton at threshold.
}
\end{figure*}

\begin{figure*}[!t]
\vspace{-4mm}
\subfloat[][$\eta^\prime\to e^+e^-\gamma$]{
\includegraphics[width=\columnwidth]{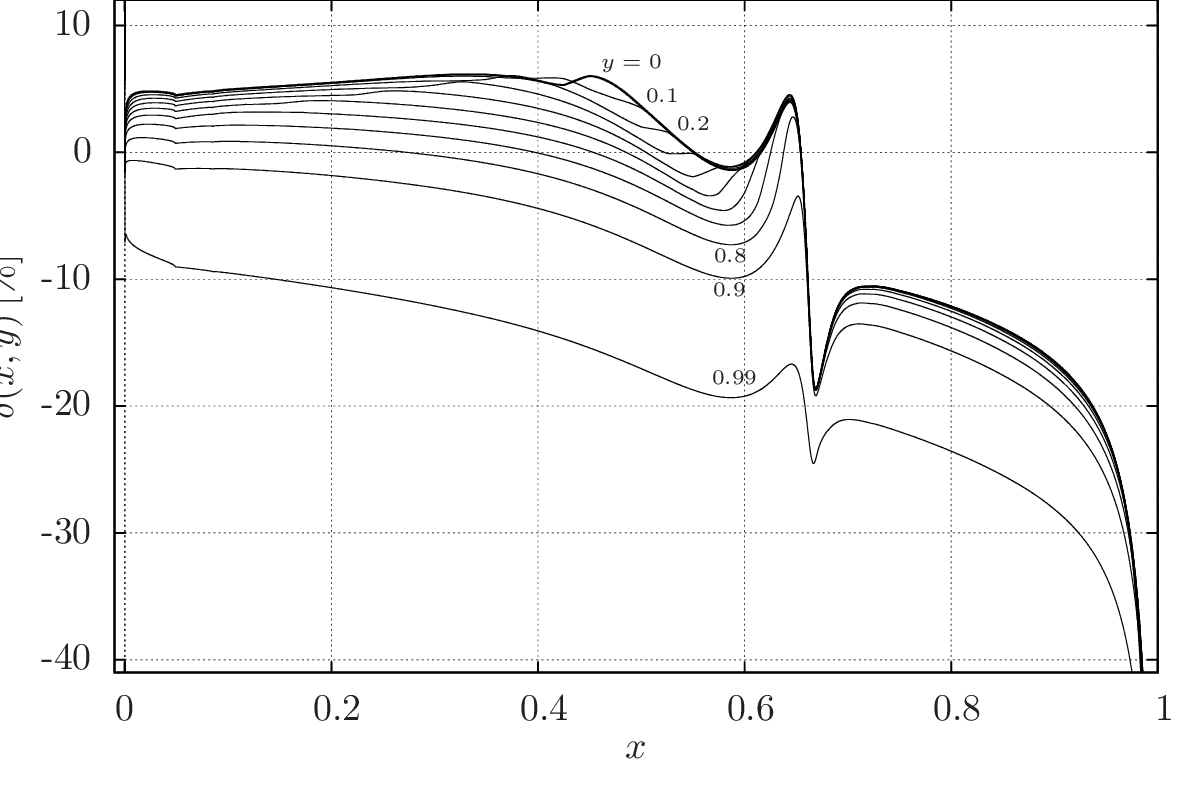}
}
\subfloat[][$\eta^\prime\to \mu^+\mu^-\gamma$]{
\includegraphics[width=\columnwidth]{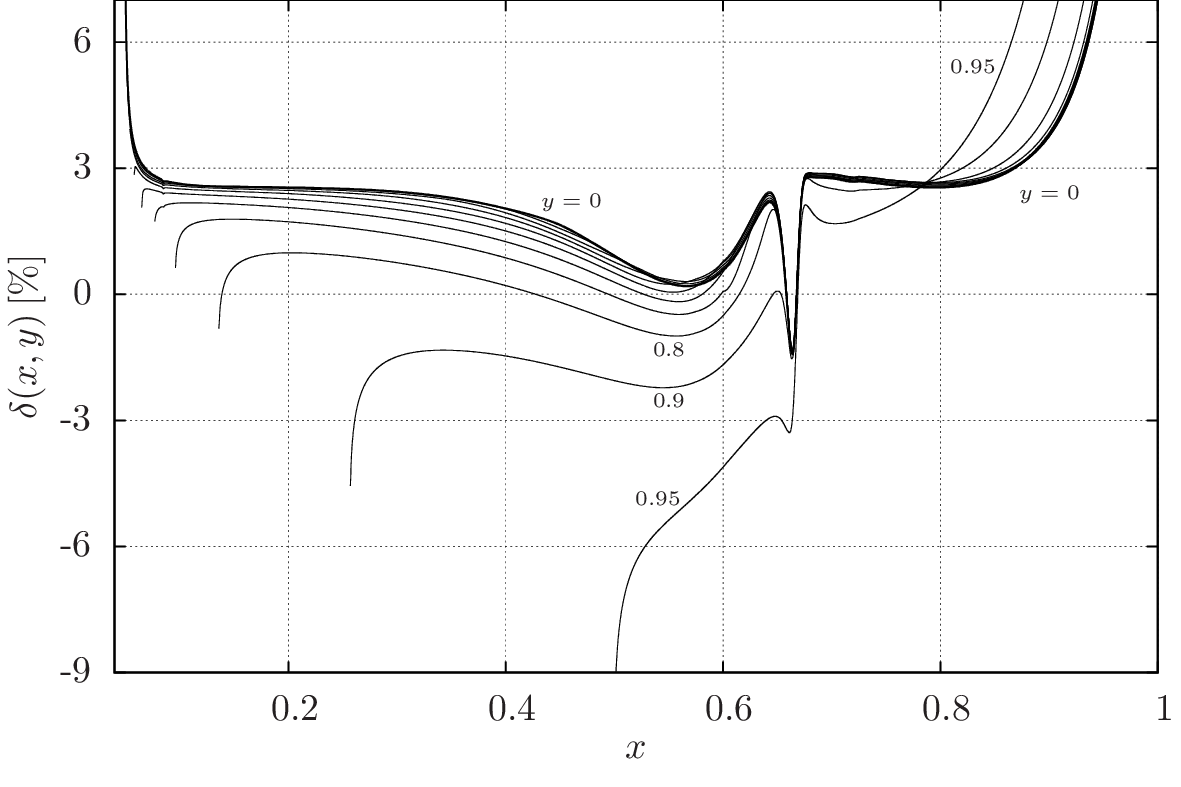}
}
\caption{\label{fig:deltaxy}
The NLO radiative corrections $\delta(x,y)$ given in percent for $\eta^\prime\to\ell^+\ell^-\gamma$ processes.
Different lines correspond to different values of $y$.
These start at $y=0$ and are equidistantly separated by 0.1.
The last value is $y=0.99$ for the electrons and $y=0.95$ for the muons due to kinematical restrictions.
The corrections (integrably) diverge at $x\to\nu_\ell^2$, which happens only for $y=0$ since for $y>0$, $x=\nu_\ell^2$ is not kinematically allowed: indeed, for given $x$, $|y|<\beta(x)$ and $\beta(\nu_\ell^2)=0$.
These plots are complementary to Tabs.~\ref{tab:etap_e} and \ref{tab:etap_mu}, in which the wiggle structure could not be precisely covered.
}
\end{figure*}

The overall NLO radiative corrections to the two-fold differential decay widths and their comparison to the main constituents are shown, for the case $y=0$, in Fig.~\ref{fig:deltax0}.
In light of a direct comparison of our results for the $\eta\to e^+e^-\gamma$ process to the previous work~\cite{Mikaelian:1972jn}, we can mention a significant difference between both approaches.
Taking into account all the discussed contributions, a table of values of corrections $\delta(x,y)$ to the process $\eta\to e^+e^-\gamma$ similar to the one provided in the original work~\cite{Mikaelian:1972jn} can be produced at the very same points; see Table~\ref{tab:eta_e}.
The difference can then be seen even more explicitly when comparing directly the corresponding values in Table~\ref{tab:eta_e} with Table I in~\cite{Mikaelian:1972jn}.
In the same manner, values for the decay $\eta\to \mu^+\mu^-\gamma$ are presented in Table~\ref{tab:eta_mu}.
From the plots in the bottom panels in Fig.~\ref{fig:deltax0} we can see that for the case of an $\eta^\prime$ meson, such a table would need to be much denser to cover the wiggle behavior.
For this reason, the table-like plots were created, which show the overall NLO radiative corrections for various values of $y$; see Fig.~\ref{fig:deltaxy}.
To fully complement these figures, the standard tables of values, where the wiggle region is left out, are provided likewise; see Tables~\ref{tab:etap_e} and \ref{tab:etap_mu}.

In Fig.~\ref{fig:deltax0} we can see that the 1$\gamma$IR correction has an intriguing character.
For the decays $\eta^{(\prime)}\to e^+e^-\gamma$ it is negative for the whole range of values of $x$ and enhances thus the effect of the M\&S-like corrections which are also negative in a wide range of $x$.
On the other hand, for the process $\eta^{(\prime)}\to\mu^+\mu^-\gamma$ the situation is different due to higher masses involved.
In this case, $\delta^{1\gamma\text{IR}}$ is very close to zero, mostly slightly negative, but then it starts to be significantly positive.
Its dominant behavior for large $x$ contributes to the fact that $\delta(x,y)$ is positive over nearly the whole range of $x$.

The form-factor effects are significant and shown in Fig.~\ref{fig:deltax0} as large-spaced dotted lines.
Out of these contributions, $\delta_{t_2}^\text{BS}$ is most prominent and responsible --- together with the hadronic part of the vacuum polarization --- for the wiggles in the case of $\eta^\prime$ decays.
The terms $\delta_{t_{3\text{a}}}^\text{BS} + \delta_{t_{3\text{b}}}^\text{BS}$ and $\delta_{t_4}^\text{BS}$ are on the other hand --- although being most complicated and time consuming to calculate --- rather suppressed.
Together with another numerically small contribution to the bremsstrahlung correction, $\delta_{t_{1\text{b}}}^\text{BS}$, they can be even considered negligible in the case of the $\eta^{(\prime)}\to \mu^+\mu^-\gamma$.
In the case of the decays into electrons, the considered correction is of order 0.1\,\% for small $|y|$, otherwise negligible, too.
Except for the already mentioned form-factor-dependent term $\delta_{t_2}^\text{BS}$, there is then only one significant contribution to the bremsstrahlung correction $\delta_{t_{1\text{a}}}^\text{BS}$, which comes with its separately treated IR-divergent counterpart $\delta_\text{D}^\text{BS}$.

\begin{figure*}[!ht]
\subfloat[][$\eta\to e^+e^-\gamma$]{
\includegraphics[width=\columnwidth]{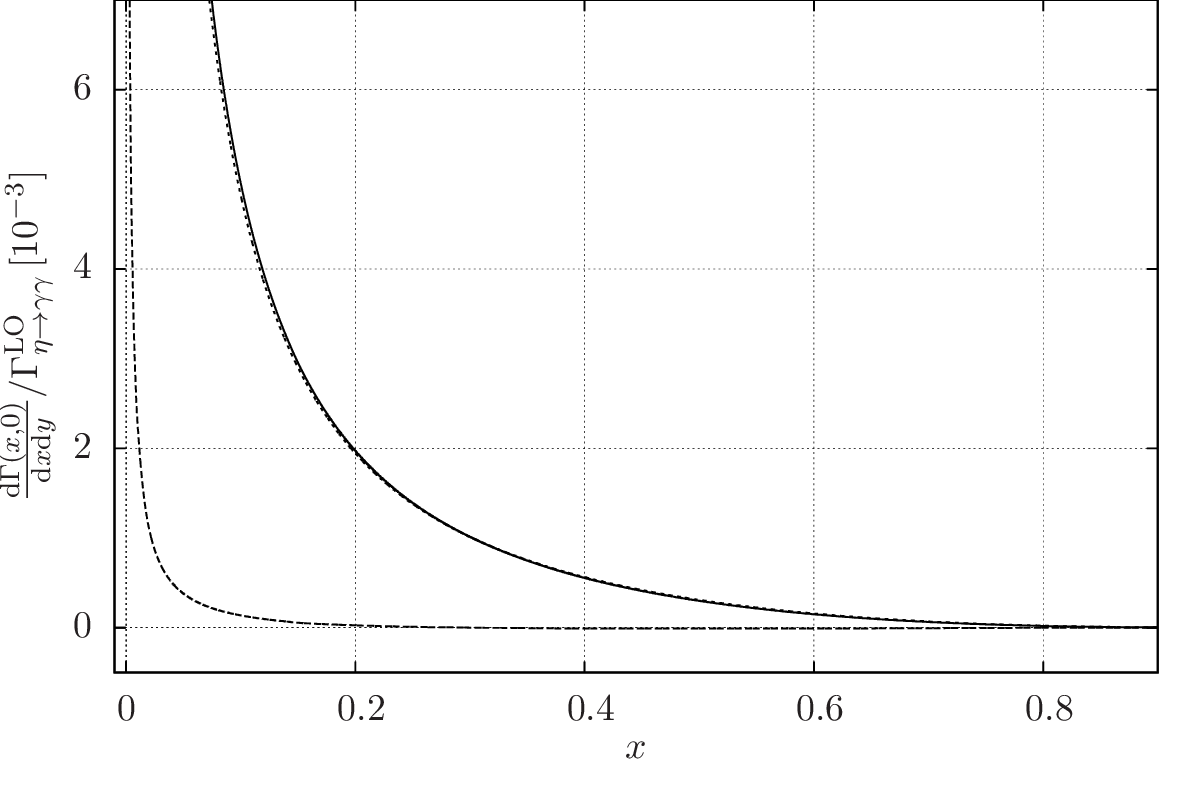}
}
\subfloat[][$\eta\to \mu^+\mu^-\gamma$]{
\includegraphics[width=\columnwidth]{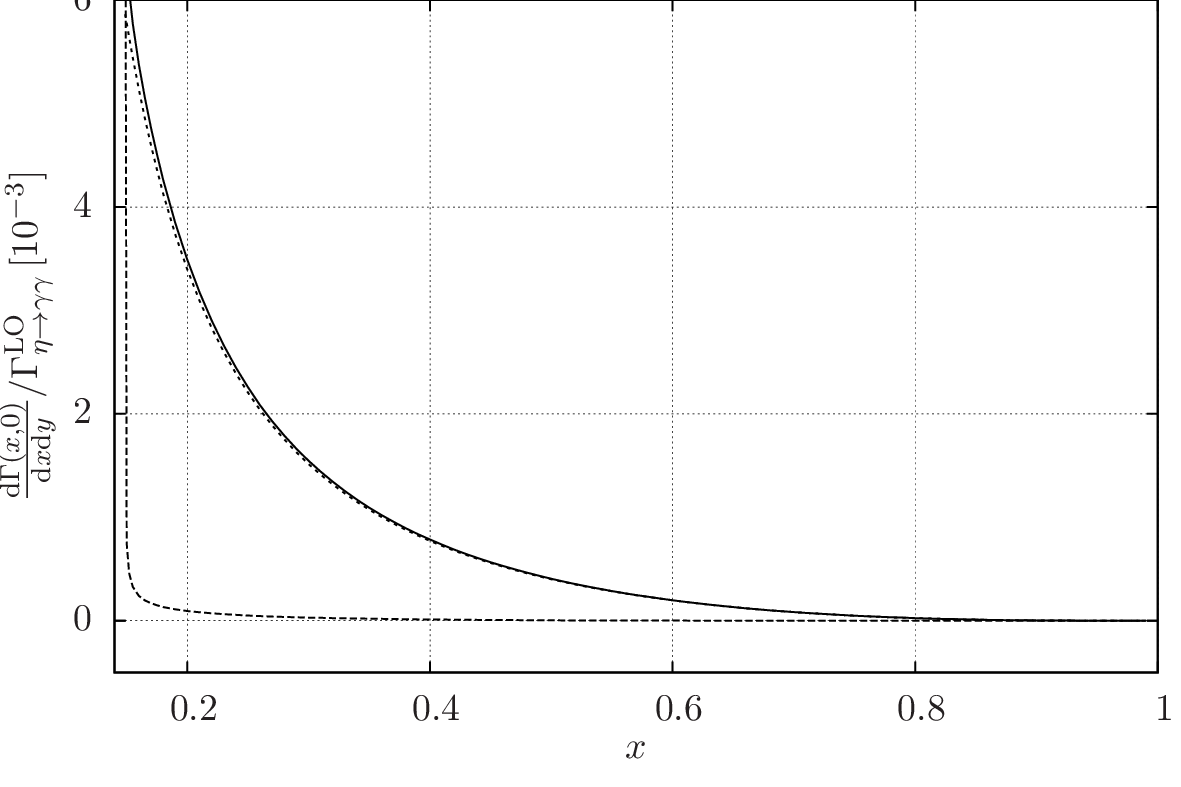}
}

\subfloat[][$\eta^\prime\to e^+e^-\gamma$]{
\includegraphics[width=\columnwidth]{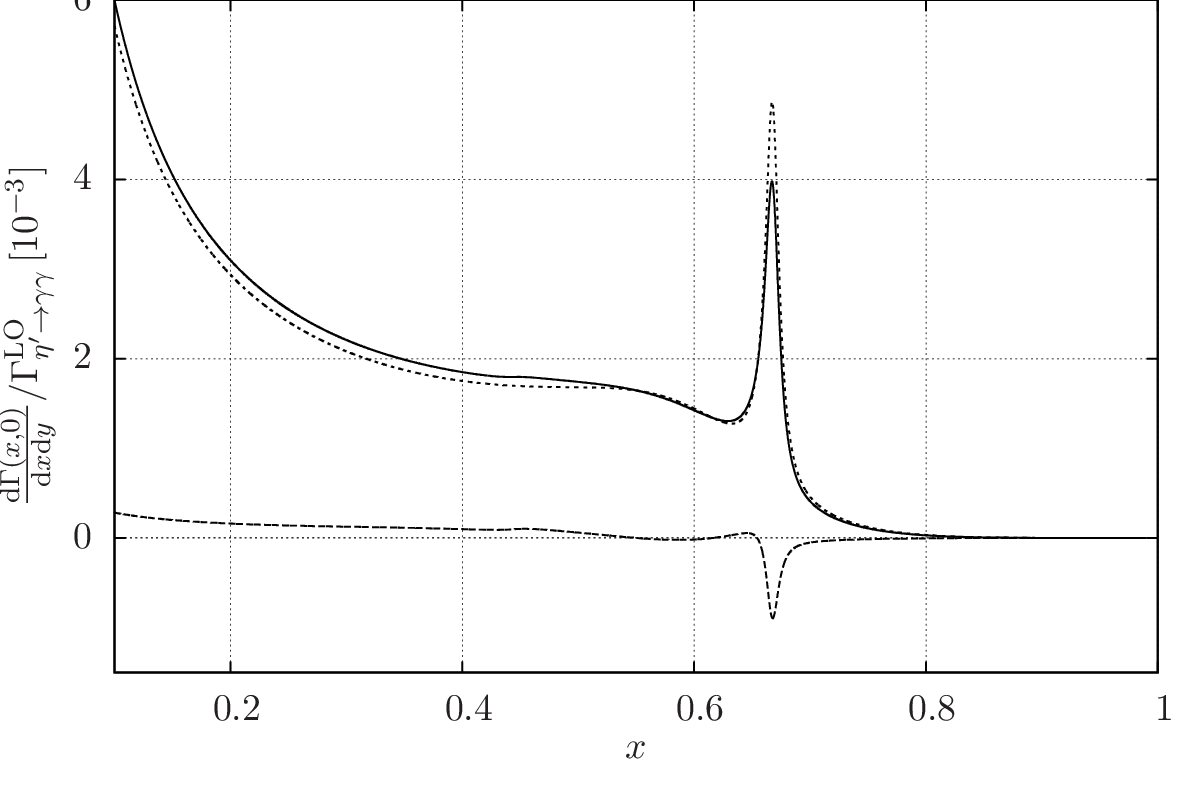}
}
\subfloat[][$\eta^\prime\to \mu^+\mu^-\gamma$]{
\includegraphics[width=\columnwidth]{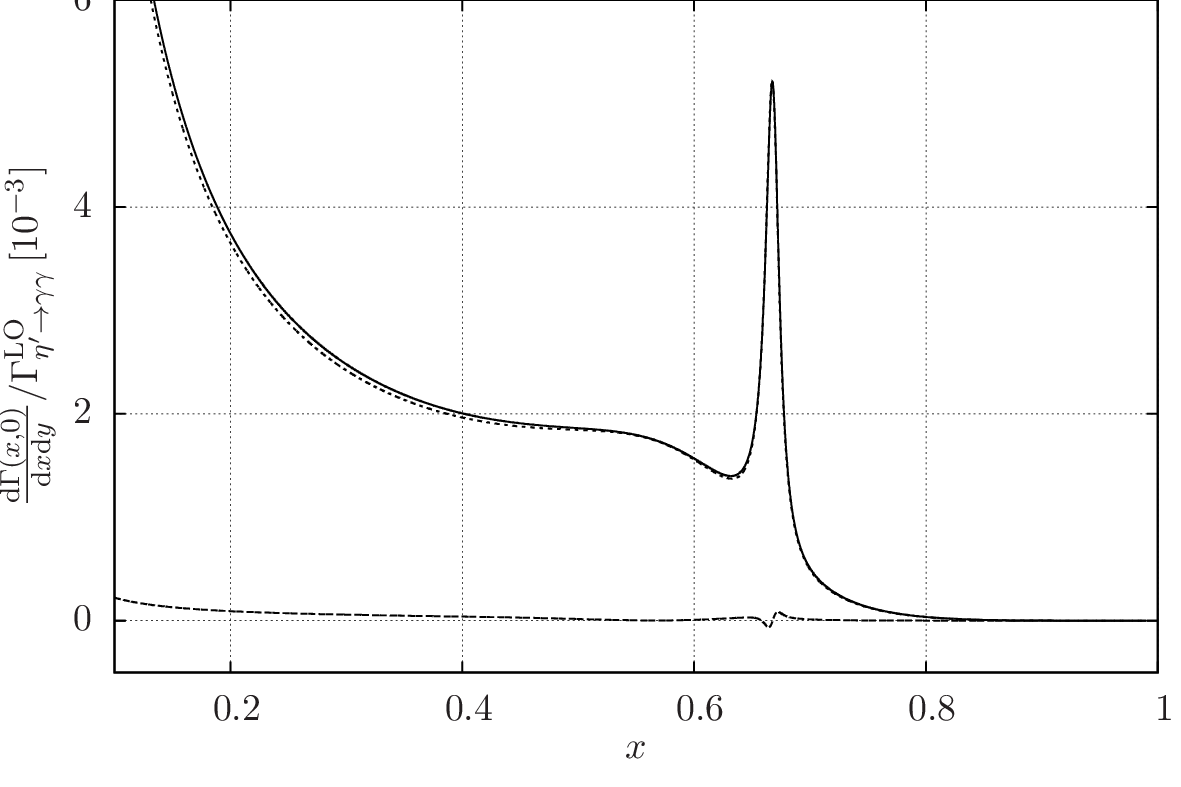}
}
\caption{\label{fig:dGammaNLO}
The two-fold differential decay widths $\mathrm{d}\Gamma(x,0)$ at NLO (solid line) and its constituents for the $\eta^{(\prime)}\to\ell^+\ell^-\gamma$ decays.
The LO differential decay width for $y=0$ is shown as a dotted line.
The corresponding NLO contribution to the differential decay width is represented by a dashed line.
}
\vspace{-1.5cm}
\end{figure*}

In order to fully understand the effect of the amusingly looking wiggles showed in Fig.~\ref{fig:deltax0} and Fig.~\ref{fig:deltaxy} and eventually appreciate the size of the calculated radiative corrections, we present the differential decay widths in Fig.~\ref{fig:dGammaNLO}.
In these plots, we can see a direct comparison of the LO and NLO decay widths, together with their sum.
In general, the total radiative corrections look rather negligible, with an exception of the process $\eta^\prime\to e^+e^-\gamma$.
In this case, the corrections are significant and the NLO differential decay width is sizable compared to the LO width.
The wiggles, which appear only in the $\eta^\prime$ case, then contribute in such a way that they change the shape of the resonance peaks and make them smaller.
In particular, Fig.~\ref{fig:dGammaNLO}c shows that the height of the $\omega$ peak is considerably influenced by the radiative corrections.
This might be interesting for the extraction of $\omega$ properties or of the $\omega$-$\eta^\prime$ interplay.
One might deduce such information from $\eta^\prime\to\omega\gamma\to e^+e^-\gamma$ or from $\eta^\prime\to\omega\gamma\to\pi^+\pi^-\pi^0\gamma$.
It can be expected that the radiative corrections are different for these two decay branches.
Ignoring such radiative corrections in the analyses of these decays might lead to contradictory conclusions.

Finally, let us shortly mention that the hadronic part of the vacuum polarization has, in contrary to the $\eta$ case, a significant impact on the corrections for the $\eta^\prime$ decays, with up to a $\sim$5\,\% effect in the $M_{\eta^\prime}^2x\simeq M_\omega^2$ region.

\section*{Acknowledgment}
T.H. would like to thank Uppsala University and University of Birmingham for their hospitality.
We would also like to thank A. Kup\'s\'c for providing us with the Omn\`es function data used in~\cite{Hanhart:2016pcd}.

This work was supported by the Czech Science Foundation grant GA\v{C}R 18-17224S and by the ERC starting grant 336581 ``KaonLepton''.

\onecolumngrid
\twocolumngrid




\begingroup
\squeezetable
\begin{table*}[!t]
\caption{
\label{tab:eta_e}
The overall NLO correction $\delta(x,y)$ given in percent for a range of values of $x$ and $y$ (i.e.\ the Dalitz-plot corrections) for the process $\eta\to e^+e^-\gamma$.
It is sufficient to show the results only for positive values of $y$ since the corrections are symmetric under $y\to-y$.
The larger values at the edge of the kinematical region (as $x\to1$) are naturally present due to the fact that the correction itself is defined as a ratio of the NLO and LO decay widths which both vanish for $x\to1$.
The LO differential decay width, however, approaches zero much faster than the interference term of the LO graph with the 1$\gamma$IR one.
Let us also note that these larger values are beyond any practical significance in the final spectrum (see Fig.~\ref{fig:dGammaNLO}).
}
\begin{ruledtabular}
\begin{tabular}{l d d d d d d d d d d d}
\, \backslashbox{$x$}{$y$} & 0.00 & 0.10 & 0.20 & 0.30 & 0.40 & 0.50 & 0.60 & 0.70 & 0.80 & 0.90 & 0.99 \\
\hline\\
0.01 & 4.00   & 3.92      & 3.72   & 3.47   & 3.17   & 2.82   & 2.39   & 1.82   & 0.96    & -0.63  & -6.83 \\
0.02 & 4.02   & 3.95      & 3.79   & 3.54   & 3.24   & 2.88   & 2.42   & 1.80   & 0.88    & -0.81  & -7.36 \\
0.03 & 3.90   & 3.84      & 3.69   & 3.46   & 3.16   & 2.79   & 2.32   & 1.69   & 0.74    & -1.02  & -7.77 \\
0.04 & 3.75   & 3.70      & 3.55   & 3.33   & 3.04   & 2.67   & 2.19   & 1.55   & 0.57    & -1.22  & -8.11 \\
0.05 & 3.59   & 3.54      & 3.41   & 3.19   & 2.91   & 2.54   & 2.05   & 1.40   & 0.41    & -1.42  & -8.42 \\
0.06 & 3.43   & 3.39      & 3.26   & 3.05   & 2.77   & 2.40   & 1.91   & 1.25   & 0.25    & -1.60  & -8.70 \\
0.07 & 3.28   & 3.23      & 3.11   & 2.91   & 2.63   & 2.26   & 1.77   & 1.10   & 0.09    & -1.78  & -8.96 \\
0.08 & 3.13   & 3.09      & 2.97   & 2.77   & 2.49   & 2.12   & 1.64   & 0.96   & -0.06   & -1.95  & -9.21 \\
0.09 & 2.98   & 2.94      & 2.83   & 2.63   & 2.36   & 1.99   & 1.50   & 0.82   & -0.21   & -2.12  & -9.44 \\
0.10 & 2.84   & 2.80      & 2.68   & 2.49   & 2.22   & 1.85   & 1.36   & 0.68   & -0.36   & -2.29  & -9.66 \\
\\
0.15 & 1.87   & 1.83      & 1.72   & 1.54   & 1.27   & 0.91   & 0.41   & -0.29  & -1.36   & -3.34  & -10.9 \\
0.20 & 1.34   & 1.30      & 1.19   & 1.01   & 0.74   & 0.38   & -0.13  & -0.84  & -1.93   & -3.96  & -11.7 \\
0.25 & 0.71   & 0.68      & 0.57   & 0.39   & 0.12   & -0.25  & -0.76  & -1.48  & -2.58   & -4.65  & -12.6 \\
0.30 & 0.03   & -0.00     & -0.11  & -0.29  & -0.55  & -0.92  & -1.43  & -2.16  & -3.28   & -5.38  & -13.4 \\
0.35 & -0.71  & -0.74     & -0.85  & -1.03  & -1.29  & -1.66  & -2.17  & -2.91  & -4.04   & -6.16  & -14.3 \\
0.40 & -1.52  & -1.55     & -1.65  & -1.83  & -2.09  & -2.46  & -2.98  & -3.71  & -4.85   & -7.00  & -15.2 \\
0.45 & -2.39  & -2.42     & -2.52  & -2.70  & -2.96  & -3.32  & -3.84  & -4.58  & -5.73   & -7.89  & -16.2 \\
0.50 & -3.34  & -3.37     & -3.47  & -3.64  & -3.90  & -4.26  & -4.77  & -5.51  & -6.67   & -8.84  & -17.2 \\
0.55 & -4.37  & -4.40     & -4.50  & -4.67  & -4.92  & -5.28  & -5.79  & -6.53  & -7.68   & -9.87  & -18.3 \\
0.60 & -5.51  & -5.54     & -5.63  & -5.80  & -6.04  & -6.40  & -6.90  & -7.64  & -8.80   & -11.0  & -19.5 \\
\\
0.65 & -6.78  & -6.81     & -6.90  & -7.05  & -7.29  & -7.64  & -8.14  & -8.87  & -10.0   & -12.2  & -20.8 \\
0.70 & -8.22  & -8.24     & -8.33  & -8.48  & -8.71  & -9.04  & -9.53  & -10.3  & -11.4   & -13.6  & -22.2 \\
0.75 & -9.88  & -9.90     & -9.98  & -10.1  & -10.3  & -10.7  & -11.1  & -11.9  & -13.0   & -15.2  & -23.8 \\
0.80 & -11.9  & -11.9     & -12.0  & -12.1  & -12.3  & -12.6  & -13.1  & -13.8  & -14.9   & -17.2  & -25.8 \\
0.85 & -14.4  & -14.4     & -14.5  & -14.6  & -14.8  & -15.1  & -15.5  & -16.2  & -17.3   & -19.6  & -28.3 \\
0.90 & -17.9  & -17.9     & -18.0  & -18.1  & -18.2  & -18.5  & -18.9  & -19.6  & -20.7   & -22.9  & -31.7 \\
0.95 & -24.3  & -24.3     & -24.3  & -24.3  & -24.4  & -24.6  & -24.9  & -25.5  & -26.6   & -28.7  & -37.6 \\
0.99 & -45.5  & -45.4     & -45.1  & -44.7  & -44.3  & -43.9  & -43.6  & -43.6  & -44.1   & -45.8  & -54.5 \\
\\
\end{tabular}
\end{ruledtabular}
\vspace{\fill}
\end{table*}
\endgroup
\begingroup
\squeezetable
\begin{table*}[!ht]
\caption{
\label{tab:eta_mu}
The overall NLO correction $\delta(x,y)$ given in percent for the process $\eta\to \mu^+\mu^-\gamma$.
The dot-filled entries correspond to kinematically forbidden combinations of $x$ and $y$.
See also the caption of Table~\ref{tab:eta_e} for more details.
}
\begin{ruledtabular}
\begin{tabular}{l d d d d d d d d d d}
\, \backslashbox{$x$}{$y$} & 0.00 & 0.10 & 0.20 & 0.30 & 0.40 & 0.50 & 0.60 & 0.70 & 0.80 & 0.90 \\
\hline\\
0.15 & 12.9   & ...    & ...    & ...    & ...    & ...     & ...    & ...    & ...     & ...   \\
0.16 & 4.69   & 4.67   & 4.65   & ...    & ...    & ...     & ...    & ...    & ...     & ...   \\
0.17 & 3.68   & 3.67   & 3.65   & 3.60   & ...    & ...     & ...    & ...    & ...     & ...   \\
0.18 & 3.23   & 3.22   & 3.21   & 3.18   & 3.01   & ...     & ...    & ...    & ...     & ...   \\
0.19 & 2.97   & 2.96   & 2.96   & 2.93   & 2.85   & ...     & ...    & ...    & ...     & ...   \\
0.20 & 2.78   & 2.78   & 2.78   & 2.76   & 2.71   & 2.26    & ...    & ...    & ...     & ...   \\
0.21 & 2.65   & 2.65   & 2.65   & 2.64   & 2.59   & 2.41    & ...    & ...    & ...     & ...   \\
0.22 & 2.54   & 2.54   & 2.54   & 2.53   & 2.50   & 2.37    & ...    & ...    & ...     & ...   \\
0.23 & 2.45   & 2.45   & 2.45   & 2.45   & 2.42   & 2.32    & ...    & ...    & ...     & ...   \\
0.24 & 2.37   & 2.37   & 2.38   & 2.37   & 2.35   & 2.26    & 1.81   & ...    & ...     & ...   \\
\\
0.25 & 2.30   & 2.30   & 2.30   & 2.30   & 2.28   & 2.21    & 1.90   & ...    & ...     & ...   \\
0.30 & 2.04   & 2.04   & 2.05   & 2.05   & 2.04   & 2.00    & 1.86   & 1.01   & ...     & ...   \\
0.35 & 1.82   & 1.82   & 1.83   & 1.84   & 1.84   & 1.80    & 1.70   & 1.36   & ...     & ...   \\
0.40 & 1.61   & 1.61   & 1.62   & 1.63   & 1.63   & 1.61    & 1.52   & 1.28   & ...     & ...   \\
0.45 & 1.39   & 1.39   & 1.40   & 1.42   & 1.42   & 1.40    & 1.33   & 1.14   & 0.20    & ...   \\
0.50 & 1.16   & 1.16   & 1.18   & 1.19   & 1.20   & 1.19    & 1.13   & 0.96   & 0.35    & ...   \\
0.55 & 0.92   & 0.92   & 0.94   & 0.95   & 0.96   & 0.96    & 0.91   & 0.76   & 0.28    & ...   \\
0.60 & 0.66   & 0.67   & 0.68   & 0.70   & 0.72   & 0.71    & 0.67   & 0.55   & 0.15    & ...   \\
\\
0.65 & 0.39   & 0.40   & 0.42   & 0.44   & 0.46   & 0.46    & 0.43   & 0.32   & -0.02   & ...   \\
0.70 & 0.11   & 0.12   & 0.13   & 0.16   & 0.18   & 0.19    & 0.17   & 0.07   & -0.21   & ...   \\
0.75 & -0.18  & -0.18  & -0.16  & -0.13  & -0.11  & -0.09   & -0.11  & -0.18  & -0.41   & ...   \\
0.80 & -0.48  & -0.47  & -0.45  & -0.43  & -0.40  & -0.38   & -0.38  & -0.44  & -0.61   & -3.20 \\
0.85 & -0.75  & -0.74  & -0.72  & -0.70  & -0.67  & -0.65   & -0.64  & -0.66  & -0.77   & -2.07 \\
0.90 & -0.91  & -0.90  & -0.89  & -0.87  & -0.84  & -0.81   & -0.79  & -0.76  & -0.74   & -1.28 \\
0.95 & -0.53  & -0.53  & -0.52  & -0.50  & -0.47  & -0.41   & -0.32  & -0.14  & 0.24    & 1.07  \\
0.99 & 5.54   & 5.56   & 5.62   & 5.75   & 5.99   & 6.43    & 7.23   & 8.76   & 12.1    & 22.4  \\
\\
\end{tabular}
\end{ruledtabular}
\end{table*}
\endgroup


\begingroup
\squeezetable
\begin{table*}[t]
\vspace{\fill}
\caption{
\label{tab:etap_e}
The overall NLO correction $\delta(x,y)$ given in percent for the process $\eta^\prime\to e^+e^-\gamma$.
See also the caption of Table~\ref{tab:eta_e} for more details.
}
\begin{ruledtabular}
\begin{tabular}{l d d d d d d d d d d d}
\, \backslashbox{$x$}{$y$} & 0.00 & 0.10 & 0.20 & 0.30 & 0.40 & 0.50 & 0.60 & 0.70 & 0.80 & 0.90 & 0.99 \\
\hline\\
0.01 & 4.66   & 4.56   & 4.33  & 4.02    & 3.67   & 3.26   & 2.77   & 2.11    & 1.13   & -0.64  & -7.44 \\
0.05 & 4.51   & 4.45   & 4.28  & 4.02    & 3.68   & 3.23   & 2.67   & 1.89    & 0.72   & -1.31  & -9.04 \\
0.10 & 4.95   & 4.90   & 4.74  & 4.49    & 4.15   & 3.68   & 3.13   & 2.14    & 0.87   & -1.29  & -9.51 \\
0.15 & 5.22   & 5.17   & 5.01  & 4.75    & 4.37   & 3.86   & 3.17   & 2.09    & 0.75   & -1.50  & -10.1 \\
0.20 & 5.49   & 5.43   & 5.26  & 4.95    & 4.48   & 4.06   & 3.03   & 1.91    & 0.51   & -1.83  & -10.7 \\
0.25 & 5.79   & 5.72   & 5.49  & 5.08    & 4.82   & 3.87   & 2.78   & 1.63    & 0.18   & -2.26  & -11.3 \\
0.30 & 6.07   & 5.96   & 5.62  & 5.30    & 4.68   & 3.55   & 2.43   & 1.24    & -0.26  & -2.79  & -12.1 \\
0.35 & 6.12   & 5.97   & 5.75  & 5.37    & 4.21   & 3.07   & 1.94   & 0.71    & -0.86  & -3.49  & -13.0 \\
0.40 & 5.64   & 5.84   & 5.66  & 4.59    & 3.46   & 2.35   & 1.21   & -0.06   & -1.69  & -4.42  & -14.1 \\
0.45 & 6.02   & 4.96   & 4.11  & 3.20    & 2.23   & 1.20   & 0.10   & -1.19   & -2.88  & -5.69  & -15.4 \\
0.50 & 3.60   & 3.49   & 2.00  & 1.00    & 0.22   & -0.62  & -1.61  & -2.85   & -4.53  & -7.35  & -17.1 \\
0.55 & 0.05   & 0.07   & 0.11  & -0.09   & -1.92  & -2.93  & -3.80  & -4.89   & -6.45  & -9.18  & -18.7 \\
0.60 & -1.17  & -1.13  & -1.02 & -0.88   & -0.79  & -0.97  & -3.17  & -5.37   & -7.09  & -9.78  & -19.2 \\
\\
0.70 & -11.1  & -11.1  & -11.1 & -11.0   & -11.0  & -11.1  & -11.2  & -11.5   & -12.2  & -13.7  & -21.1 \\
0.75 & -10.9  & -10.9  & -10.9 & -10.9   & -10.9  & -11.0  & -11.2  & -11.6   & -12.4  & -14.2  & -22.0 \\
0.80 & -12.2  & -12.2  & -12.1 & -12.1   & -12.2  & -12.3  & -12.5  & -13.0   & -13.8  & -15.7  & -23.6 \\
0.85 & -13.9  & -13.9  & -13.9 & -13.9   & -14.0  & -14.1  & -14.4  & -14.8   & -15.7  & -17.6  & -25.5 \\
0.90 & -16.5  & -16.5  & -16.5 & -16.5   & -16.6  & -16.8  & -17.0  & -17.5   & -18.4  & -20.3  & -28.1 \\
0.95 & -22.2  & -22.2  & -22.2 & -22.2   & -22.2  & -22.3  & -22.6  & -23.0   & -23.9  & -25.7  & -33.4 \\
0.99 & -55.8  & -55.6  & -55.0 & -54.0   & -52.9  & -51.8  & -50.7  & -49.9   & -49.5  & -50.2  & -56.6 \\
\\
\end{tabular}
\end{ruledtabular}
\vspace{\fill}
\end{table*}
\endgroup
\begingroup
\squeezetable
\begin{table*}[!ht]
\caption{
\label{tab:etap_mu}
The overall NLO correction $\delta(x,y)$ given in percent for the process $\eta^\prime\to \mu^+\mu^-\gamma$.
The dot-filled entries correspond to kinematically forbidden combinations of $x$ and $y$.
See also the caption of Table~\ref{tab:eta_e} for more details.
}
\begin{ruledtabular}
\begin{tabular}{l d d d d d d d d d d d}
\, \backslashbox{$x$}{$y$} & 0.00 & 0.10 & 0.20 & 0.30 & 0.40 & 0.50 & 0.60 & 0.70 & 0.80 & 0.90 & 0.95 \\
\hline\\
0.05 & 7.47  & 7.35  & ...   & ...   & ...   & ...    & ...    & ...     & ...    & ...    & ...   \\
0.10 & 2.63  & 2.63  & 2.61  & 2.57  & 2.51  & 2.39   & 2.17   & 1.36    & ...    & ...    & ...   \\
0.15 & 2.56  & 2.56  & 2.54  & 2.51  & 2.45  & 2.33   & 2.14   & 1.79    & 0.65   & ...    & ...   \\
0.20 & 2.54  & 2.54  & 2.52  & 2.47  & 2.40  & 2.26   & 2.06   & 1.72    & 0.99   & ...    & ...   \\
0.25 & 2.51  & 2.50  & 2.47  & 2.41  & 2.31  & 2.16   & 1.95   & 1.60    & 0.92   & ...    & ...   \\
0.30 & 2.44  & 2.43  & 2.39  & 2.31  & 2.18  & 2.01   & 1.78   & 1.42    & 0.75   & -1.48  & ...   \\
0.35 & 2.30  & 2.28  & 2.24  & 2.12  & 1.98  & 1.81   & 1.56   & 1.18    & 0.52   & -1.33  & ...   \\
0.40 & 2.04  & 2.03  & 1.94  & 1.82  & 1.69  & 1.51   & 1.25   & 0.87    & 0.21   & -1.47  & ...   \\
0.45 & 1.65  & 1.56  & 1.46  & 1.36  & 1.24  & 1.07   & 0.83   & 0.45    & -0.19  & -1.73  & ...   \\
0.50 & 0.91  & 0.92  & 0.88  & 0.77  & 0.65  & 0.51   & 0.29   & -0.06   & -0.65  & -2.05  & -10.3 \\
0.55 & 0.26  & 0.28  & 0.32  & 0.36  & 0.24  & 0.05   & -0.16  & -0.46   & -0.99  & -2.22  & -5.37 \\
0.60 & 0.58  & 0.59  & 0.64  & 0.70  & 0.76  & 0.79   & 0.54   & 0.07    & -0.51  & -1.68  & -4.11 \\
\\
0.70 & 2.76  & 2.76  & 2.78  & 2.80  & 2.83  & 2.86   & 2.87   & 2.86    & 2.79   & 2.52   & 1.68  \\
0.75 & 2.61  & 2.62  & 2.63  & 2.66  & 2.69  & 2.72   & 2.73   & 2.73    & 2.68   & 2.51   & 2.06  \\
0.80 & 2.54  & 2.55  & 2.56  & 2.58  & 2.60  & 2.63   & 2.65   & 2.66    & 2.67   & 2.77   & 2.96  \\
0.85 & 2.77  & 2.77  & 2.77  & 2.78  & 2.79  & 2.80   & 2.83   & 2.88    & 3.02   & 3.61   & 4.95  \\
0.90 & 3.82  & 3.82  & 3.80  & 3.77  & 3.75  & 3.75   & 3.80   & 3.95    & 4.42   & 6.18   & 10.1  \\
0.95 & 8.11  & 8.08  & 8.01  & 7.91  & 7.83  & 7.83   & 8.02   & 8.62    & 10.3   & 16.1   & 28.7  \\
0.99 & 38.4  & 38.4  & 38.3  & 38.4  & 38.7  & 39.7   & 42.1   & 47.2    & 59.1   & 98.2   & 179.  \\
\\
\end{tabular}
\end{ruledtabular}
\vspace{\fill}
\end{table*}
\endgroup


\clearpage
\appendix


\onecolumngrid

\section{Matrix element of the one-photon-irreducible contribution}
\label{app:M1gIR}

The building-block matrix element for the one-photon-irreducible contribution can be expressed in terms of scalar form factors in the following (manifestly gauge-invariant) way (cf.\ (24) in~\cite{Husek:2015sma}):
\begin{equation}
\begin{split}
i\mathcal{M}_{1\gamma\text{IR}}^h\big[g(M_1^2,M_2^2)\big](p,q,k)
&=-\frac{i e^5}{2}\mathcal{F}_{\pi^0\gamma^*\gamma^*}(0,0)\epsilon^{*\rho}(k)\\
&\hspace{-2mm}\times\Big\{P\big[g(M_1^2,M_2^2)\big](x,y)\left[\left(k\cdot p\right) q_\rho-\left(k\cdot q\right) p_\rho\right]\left[\bar{u}(p,m)\gamma_5 v(q,m)\right]\\
&+A\big[g(M_1^2,M_2^2)\big](x, y)\Big[\bar{u}(p,m)\left[\gamma_\rho\left(k\cdot p\right)-p_\rho\slashed k\right]\gamma_5 v(q,m)\Big]\\
&-A\big[g(M_1^2,M_2^2)\big](x,-y)\Big[\bar{u}(p,m)\left[\gamma_\rho\left(k\cdot q\right)-q_\rho\slashed k\right]\gamma_5 v(q,m)\Big]\\
&+T\big[g(M_1^2,M_2^2)\big](x,y)\left[\bar{u}(p,m)\gamma_\rho\slashed k\gamma_5 v(q,m)\right]\Big\}\,.
\end{split}
\end{equation}
The form factor $P$ does not contribute to $\delta^{1\gamma\text{IR}}$, so we do not include it in this appendix.
Note also that only the box diagram (see Fig.~\ref{fig:diagrams}d) contributes to $A(x,y)$.
Let us also define the two-point (bubble) scalar one-loop integral as a reference point for the used notation:
\begin{equation}
i\pi^2 B_0(0,m^2,m^2)
\equiv(2\pi\mu)^{4-d}\int\frac{\text{d}^dl}
{\left[l^2-m^2+i\epsilon\right]^2}
=i\pi^2\left[\frac1\dimeps-\gamma_E+\log4\pi+\log\frac{\mu^2}{m^2}\right].
\end{equation}
Above, we have introduced $\dimeps=2-d/2$.
Before providing the final expressions, let us define
\begin{align}
\sigma_{+\pm}&\equiv2-(1+x) (1\pm y)\,,\\
\sigma_{-\pm}&\equiv2-(1-x) (1\pm y)\,.
\end{align}
Using the dimensional regularization, the dimensional reduction scheme~\cite{Siegel:1979wq,Novotny:1994yx}) and Passarino--Veltman reduction~\cite{Passarino:1978jh}, the explicit results for the two contributing form factors, $A$ and $T$, in terms of scalar one-loop integrals read (we use again for simplicity $m\equiv m_\ell$, $\nu\equiv\nu_\ell$ and $M\equiv M_P$)
\begin{align*}
&\frac{M}{2\nu}\left\{-16i\pi^2T[g(M_1^2,M_2^2)](x,y)\right\}=\\
&-B_0\left(0,M_1^2,m^2\right)\frac{\left(M_1^2-m^2\right)}{m^2 (1-x) \left(1-y^2\right)}\\
&+B_0\left(0,M_2^2,m^2\right) \left[\frac{1}{2 (1-x)\left(1-y^2\right)}-\frac{1}{2 \left(2-\sigma_{-+}\right)+\nu ^2}\left(1+\frac{M_2^2}{M^2}\frac{2}{2-\sigma_{-+}}\right)\right]
+ (y\to-y,~\sigma_{-+}\to\sigma_{--})\\
&+B_0\left(M^2,M_1^2,M_2^2\right)\left[\frac{M_2^2-M_1^2}{M^2}\frac{4}{\sigma_{-+}^2-4 \nu ^2}\right]
+ (y\to-y,~\sigma_{-+}\to\sigma_{--})\\
&+B_0\left(m^2,M_1^2,m^2\right)\frac{M_1^2-M_2^2}{M^2}{\left[\frac{\sigma_{-+}}{\sigma_{-+}^2-4 \nu ^2}+\frac{1}{(1-x)\left(1-y^2\right)}\right]}
+ (y\to-y,~\sigma_{-+}\to\sigma_{--})\\
&+B_0\left(m^2,M_1^2,m^2\right) \left[\frac{\sigma_{-+}}{\sigma_{-+}^2-4 \nu ^2}+\left(3-y+\frac{M_1^2}{m^2}\right)\frac{1}{(1-x) \left(1-y^2\right)}\right]\\
&-B_0\left(m^2,M_2^2,m^2\right) \left[\frac{\sigma_{-+}}{\sigma_{-+}^2-4 \nu ^2}-\frac{1}{2-\sigma_{--}}\right]\\
&+\left[B_0\left(\frac{M^2}2(1-x)(1+y)+m^2,M_2^2,m^2\right)-B_0\left(\frac{M^2}2(1-x)(1+y)+m^2,M_1^2,m^2\right)\right]\\
&\quad\times\left[\frac{4-\sigma_{-+}}{\sigma_{-+}^2-4 \nu ^2}+\frac{1}{2-\sigma_{-+}}\right]\\
&+B_0\left(\frac{M^2}2(1-x)(1+y)+m^2,M_2^2,m^2\right)\\
&\quad\times\left[
\frac{M_1^2-M_2^2}{M^2}{\left(\frac{4-\sigma_{-+}}{\sigma_{-+}^2-4 \nu ^2}-\frac{1}{2-\sigma_{-+}}\right)}
+\left(1+\frac{M_2^2}{M^2}\frac2{2-\sigma_{-+}}\right)\frac{1}{2(2-\sigma_{-+})+\nu ^2}
\right]
+ (y\to-y,~\sigma_{-+}\to\sigma_{--})\\
&-C_0\left(m^2,0,\frac{M^2}2(1-x)(1+y)+m^2,M_1^2,m^2,m^2\right)\\
&\quad\times\frac{(1+y)}{2 \left[x \left(1-y^2\right)-\nu ^2\right]}
\left[y\left(M_2^2-M_1^2\right)
+\left(2x+(1-x) y^2-2 \nu ^2\right) M^2\right]
+ (y\to-y, M_1\leftrightarrow M_2)\\
%
&+C_0\left(m^2,M^2,\frac{M^2}2(1-x)(1+y)+m^2,m^2,M_1^2,M_2^2\right)
\left\{
\frac{\lambda \left(M^2,M_1^2,M_2^2\right)}{M^2}\left(\frac{2-\sigma_{-+}}{\sigma_{-+}^2-4 \nu ^2}-\frac{1}{2-\sigma_{-+}}\right)\right.\\
&\qquad+\left.4M_1^2\frac{2-\sigma_{-+}}{\sigma_{-+}^2-4 \nu ^2}
-\frac1{M^2}\left[M_2^4-M_1^4-M^2 \left(2 M_2^2-M^2\right)\right]\frac{2}{\sigma_{-+}^2-4 \nu ^2}\right\}\\
&+C_0\left(m^2,M^2,\frac{M^2}2(1-x)(1+y)+m^2,m^2,M_2^2,M_1^2\right)
\left\{
M^2\left[\frac{\sigma_{-+}}{\sigma_{-+}^2-4 \nu ^2}-\frac{y}{1-x}\left(2-\frac{y(1+x)\sigma_{++}}{2 \left[x \left(1-y^2\right)-\nu ^2\right]}\right)\right]
\right.\\
&\qquad\left.
+\left(M_2^2-M_1^2\right) \left(\frac{\sigma_{-+}}{\sigma_{-+}^2-4 \nu ^2}-\frac{1}{1-x}\left(1-\frac{y \sigma_{++}}{2 \left[x \left(1-y^2\right)-\nu ^2\right]}\right)\right)
-M_2^2\frac{4}{\sigma_{-+}^2-4 \nu ^2}
\right\}\\
&+C_0\left(m^2,M^2,\frac{M^2}2(1-x)(1-y)+m^2,m^2,M_1^2,M_2^2\right)\\
&\quad\times\left\{
\frac{\left(M_2^2-M_1^2\right)^2}{M^2} \left(\frac{2-\sigma_{--}}{\sigma_{--}^2-4 \nu ^2}-\frac{1}{2-\sigma_{--}}\right)
-\frac{M_2^4-M_1^4}{M^2}\left(\frac{2}{\sigma_{--}^2-4\nu ^2}\right)
+\frac{4 M_2^2-M^2}{2-\sigma_{--}}
\right.\\
&\qquad+\left.\left(M_2^2-M_1^2\right)
\left(\frac{\sigma_{--}}{\sigma_{--}^2-4 \nu ^2}-\frac{1+y}{2-\sigma_{--}}+\frac{y \sigma_{+-}}{2 (1-x) \left[x \left(1-y^2\right)-\nu ^2\right]}\right)
+\frac{M^2y}{1-x}\left(2+\frac{y(1+x) \sigma_{+-}}{2 \left[x \left(1-y^2\right)-\nu ^2\right]}\right)\right\}\\
&+D_0\left(m^2,M^2,m^2,0,\frac{M^2}2(1-x)(1-y)+m^2,\frac{M^2}2(1-x)(1+y)+m^2,m^2,M_1^2,M_2^2,m^2\right)\\
&\quad\times\frac{1}{4}\left[\frac{y}{2 \left(x \left(1-y^2\right)-\nu ^2\right)}\left\{y\left(1-y^2\right)(1+x)^2 M^4
+2 M_1^2 \left[2(1-y)M_1^2+2yM_2^2-(1+x) (1-y) (1+3y) M^2\right]\right\}\right.\\
&\qquad-\left.\vphantom{\frac12}M^2 \left\{2 M_1^2 (2-3 y)-M^2 \left(1-2 y^2\right)\right\}\right]
+ (y\to-y, M_1\leftrightarrow M_2)\,,
\addtocounter{equation}{1}\tag{\theequation}
\end{align*}
\begin{align*}
&-\frac{M^2}{4}(1-x)\left\{-16i\pi^2A[g(M_1^2,M_2^2)](x,y)\right\}=
\frac1{1-y}\\
&+B_0\left(0,M_2^2,m^2\right)
\frac{4\left(M_2^2-m^2\right)}{M^2}
\frac1{ (1-y) \left[2 \left(2-\sigma _{--}\right)+\nu ^2\right]}\\
&+B_0\left(M^2,M_1^2,M_2^2\right)
\frac{2 \left(M_2^2-M_1^2\right)}{M^2}
\left(
\frac{ \sigma _{-+}\left(4-\sigma_{-+}\right)}{\sigma _{++} \left(\sigma _{-+}^2-4 \nu ^2\right)}
-\frac{\sigma _{--}^2}{\sigma _{+-} \left(\sigma _{--}^2-4 \nu ^2\right)}
-\frac{4 x y (1+y)}{\sigma _{+-} \sigma _{++} \left(x \left(1-y^2\right)-\nu ^2\right)}
\right)\\
&+B_0\left(m^2, M_1^2, m^2 \right)
\frac{2 \nu^2\left(M_2^2 - M_1^2 \right)} {M^2 (1 - x)}
\frac {x\sigma _ {--} - (1+x)\nu^2} {\left(\sigma _{--}^2 - 4 \nu^2 \right)\left[x\left (1 - y^2 \right) - \nu^2 \right]}\\
&+B_0\left(m^2,M_2^2,m^2\right)
\left\{
\frac1{(1-x)(1-y)^2}\left(\frac{2 M_2^2}{M^2}-\nu ^2\right)\right.\\
&\qquad-\left.\frac{M_2^2-M_1^2}{M^2(1-x)}
\left[
\frac{1+x}{2}-\frac{2}{1-y}
-\frac{(1-x)^2}{2 (1+x)} \left(1-\frac{4}{(1-x) \sigma _{++}}\right) \left(1+\frac{4 \nu ^2}{\sigma _{-+}^2-4 \nu ^2}\right)
-\frac{2 y x^2(1+y)^2}{\sigma _{++} \left[x \left(1-y^2\right)-\nu ^2\right]}
\right]\right\}\\
&+B_0\left(\frac{M^2}{2} (1-x) (1+y)+m^2,M_1^2,m^2\right)\\
&\quad\times\left\{
\frac{M_2^2-M_1^2}{2 M^2}
\left[
\frac{1+x}{1-x}
-\frac1{\sigma _{++}}\left(\frac{\sigma _{-+}\left(4-\sigma _{-+}\right)^2}{\sigma _{-+}^2-4 \nu ^2}
-\frac{4 x y (1+y) (2-x (1+y))}{(1-x) \left[x \left(1-y^2\right)-\nu ^2\right]}\right)
\right]\right\}\\
&+B_0\left(\frac{M^2}{2} (1-x) (1-y)+m^2,M_2^2,m^2\right)
\frac{\nu ^2}{2 (1-x) (1-y)^2}
\left\{
3
-\frac{\nu ^2}{2 \left(2-\sigma _{--}\right)+\nu ^2}\right.\\
&\qquad+\left.\frac{4\left(M_2^2-M_1^2\right)}{M^2}\frac{(1-y)}{\sigma_{+-}}
  \left[\frac{\left(2-\sigma _{--}\right) \left(4-\sigma_{--}\right)}{\sigma _{--}^2-4 \nu ^2}
  +\frac{y [2-x (1-y)]}{x \left(1-y^2\right)-\nu ^2}\right]
-\frac{2 M_2^2}{m^2}\left(1-\frac{\nu ^2}{4 \left(2-\sigma _{--}\right)+2 \nu ^2}\right)
\right\}\\
&+C_0\left(m^2,0,\frac{M^2}{2} (1-x) (1-y)+m^2,M_2^2,m^2,m^2\right)
\frac{M^2 (1-y)}{2 \left(x \left(1-y^2\right)-\nu ^2\right)^2}
\left\{
2 y \nu^2\frac{\left(M_2^2-M_1^2\right)^2}{M^4}\right.\\
&\qquad
-\frac{2 M_2^2 \left(M_2^2-M_1^2\right)}{M^4}\left[x (1+y)^2-\nu ^2\right]
+\frac{\nu ^2 \left(M_2^2-M_1^2\right)}{M^2}\left[(1+y) (x+y)-\nu ^2\right]\\
&\qquad-\left.\left[x \left(1-y^2\right)-\nu ^2\right] \left[\frac{2-\sigma _{-+}}{2}\left(
2-\sigma _{--}-\frac{4M_2^2}{M^2}\right)
   +\nu ^2 \left(1-\frac{2 x}{1-y}\right)+\frac{\nu ^4}{(1-y)^2}\right]
\right\}\\
&+C_0\left(m^2,0,\frac{M^2}{2} (1-x) (1+y)+m^2,M_1^2,m^2,m^2\right)
\frac{M^2 (1+y)}{2 \left(x \left(1-y^2\right)-\nu ^2\right)^2}
\left\{
2 y \nu^2\frac{\left(M_2^2-M_1^2\right)^2}{M^4}\right.\\
&\qquad
-\frac{2 M_2^2 \left(M_2^2-M_1^2\right)}{M^4}\left[x (1+y)^2-\nu ^2\right]
+\frac{\nu ^2 \left(M_2^2-M_1^2\right)}{M^2}\left[(1+y) (x+y)-\nu ^2\right]\\
&\qquad-\left.\left[x \left(1-y^2\right)-\nu ^2\right] \left[\frac{2-\sigma _{-+}}{2}\left(
2-\sigma _{--}-\frac{4M_2^2}{M^2} +\frac{2\nu ^2}{1+y}\right)
\right]
\right\}\\
&+C_0\left(m^2,M^2,\frac{M^2}{2} (1-x) (1-y)+m^2,m^2,M_1^2,M_2^2\right)
\frac{M^2}{2}
\left\{
\frac{(1+y) \left(1-(x+y-x y)^2\right)+4y\nu ^2}{2 \left[x \left(1-y^2\right)-\nu ^2\right]}\right.\\
&\qquad
+\frac{\left(M_2^2-M_1^2\right)^2}{M^4}\left(1+\frac{2}{1-x}\right)
-\frac{2 M_1^2}{M^2}\left(2+\frac{(1+y) \sigma _{+-}}{x \left(1-y^2\right)-\nu ^2}\right)\\
&\qquad
-\frac{M_2^2-M_1^2}{M^2}
\frac{\sigma _{--}^2}{\sigma _{+-} \left(\sigma _{--}^2-4 \nu ^2\right)}
\left(
\left(2-\sigma _{--}\right)\frac{\left(M_2^2-M_1^2-M^2\right)}{M^2}-\frac{2
   \left(M_1^2+M_2^2-M^2\right)}{M^2}
\right)\\
&\qquad
-\frac{M_2^2-M_1^2}{M^2 (1-x)}
\left[
2 - 3 y - x (4 + y)
+\frac{x y (1+y) \sigma _{+-}}{\left(x \left(1-y^2\right)-\nu ^2\right)^2}
\left((1+x) \left(1-y^2\right)-2 \left((1-y)\frac{M_1^2}{M^2}+(1+y)\frac{M_2^2}{M^2}\right)\right)\right.\\
&\qquad\qquad
+\left.(1 + y)\frac{(2-x)(1-x)^2+\left(3+x^2\right) y-x (1+x) (1-3x) y^2-(1+x) [1+x (5+2 x)] y^3}{\sigma _{+-} \left[x\left(1-y^2\right)-\nu ^2\right]}
\right]\\
&\qquad
+\frac{2 \left(M_2^2-M_1^2\right)}{M^2\sigma _{+-}(1-x) \left[x \left(1-y^2\right)-\nu ^2\right]}\\
&\qquad\quad\times
\left.\left[
\left((1+y)\frac{M_2^2}{M^2}-y\frac{M_1^2}{M^2}\right) \left[x^2 (1-y)^2-4 x \left(1-y^2\right)+(1+y)^2\right]
+2x (1 + y)^2\frac{M_2^2}{M^2}
\right]
\right\}\\
&+C_0\left(m^2,M^2,\frac{M^2}{2} (1-x) (1+y)+m^2,m^2,M_2^2,M_1^2\right)
\frac{M^2}{2}
\left\{
\frac{(1-y) \left(1-(x-y+x y)^2\right)-4y\nu ^2}{2 \left[x \left(1-y^2\right)-\nu ^2\right]}\right.\\
&\qquad
-\frac{\left(M_2^2-M_1^2\right)^2}{M^4}\left(1+\frac{2}{1-x}\right)
+\frac{2 M_2^2}{M^2}\left(2-\frac{(1+y) \sigma _{++}}{x \left(1-y^2\right)-\nu ^2}\right)\\
&\qquad
-\frac{M_2^2-M_1^2}{M^2}
\frac{\sigma _{-+}(4-\sigma _{-+})}{\sigma _{++} \left(\sigma _{-+}^2-4 \nu ^2\right)}
\left(
\left(2-\sigma _{-+}\right)\frac{\left(M_2^2-M_1^2+M^2\right)}{M^2}+\frac{2\left(M_1^2+M_2^2-M^2\right)}{M^2}
\right)\\
&\qquad
-\frac{M_2^2-M_1^2}{M^2 (1-x)}
\left[
2 + 3 y + x y
+\frac{x y (1+y) \sigma _{++}}{\left(x \left(1-y^2\right)-\nu ^2\right)^2}
\left((1+x) \left(1-y^2\right)-2 \left((1-y)\frac{M_1^2}{M^2}+(1+y)\frac{M_2^2}{M^2}\right)\right)\right.\\
&\qquad\qquad
+\left.(1 + y)\frac{x(1-x)^2-\left(1-5x^2\right) y+(2+5x-3x^3) y^2-(1+x) [1+x (5+2 x)] y^3}{\sigma _{++} \left[x\left(1-y^2\right)-\nu ^2\right]}
\right]\\
&\qquad
+\frac{2 \left(M_2^2-M_1^2\right)}{M^2\sigma _{++}(1-x) \left[x \left(1-y^2\right)-\nu ^2\right]}\\
&\qquad\quad\times
\left.\left[
\left((1+y)\frac{M_2^2}{M^2}-y\frac{M_1^2}{M^2}\right) \left[x^2 (1+y)^2-4 x \left(1-y^2\right)+(1-y)^2\right]
+2x (1 + y)^2\frac{M_2^2}{M^2}
\right]
\right\}\\
&+D_0\left(m^2,M^2,m^2,0,\frac{M^2}{2} (1-x) (1-y)+m^2,\frac{M^2}{2} (1-x) (1+y)+m^2,m^2,M_1^2,M_2^2,m^2\right)\\
&\quad\times\frac{M^4}{4 \left[x \left(1-y^2\right)-\nu ^2\right]}
\left\{
\frac{1}{2} (1-x) \left(1-y^2\right) \left[4 \nu ^2+\left(2-\sigma _{--}\right) \left(2-\sigma_{-+}\right)\right]
+\frac{8 M_2^4}{M^4} \left(2-\sigma _{-+}\right)\right.\\
&\qquad
-\frac{2 M_1^2}{M^2} (1-x) (2+y) \left[(1-x) \left(1-y^2\right)+2 \nu ^2\right]
-\frac{2 \left(M_2^2-M_1^2\right)^2}{M^4}
\left(\frac{\nu ^2 \left(1-y^2\right) [x (1-y)+2 y]-\nu ^4 (1-3 y)}{x \left(1-y^2\right)-\nu ^2}\right)\\
&\qquad
-\frac{2 M_2^2 \left(M_2^2-M_1^2\right)}{M^4}
\left(\frac{x (1-3 y) (1+y)^2 \left(2-\sigma _{--}\right)-\nu ^2 (1+3x) (1+y)^2+4 \nu ^4}{x \left(1-y^2\right)-\nu ^2}\right)\\
&\qquad
-\frac{4 \left(M_2^2-M_1^2\right)}{M^2}
\left[
\frac{\left(M_1^2 (1-y)+M_2^2 (1+y)\right)^2}{M^4}
\left(
1+\frac{x y (1+y)}{x \left(1-y^2\right)-\nu ^2}
\right)
-\frac{M_1^2}{M^4}\left[M_1^2 (1-y)+M_2^2 (1+y)\right]
\right]\\
&\qquad
-(1+y)\left.\frac{M_2^2-M_1^2}{M^2}
\left(\frac{3x(1-x)^2\left(1-y^2\right)^2-\nu ^2 \left(1-y^2\right) (3-13x+8x^2-(1-x)y)+\nu ^4 [5x+y(3+x)-7]}{x\left(1-y^2\right)-\nu ^2}\right)
\right\}.
\addtocounter{equation}{1}\tag{\theequation}
\end{align*}


\section{Bremsstrahlung matrix element squared}
\label{app:JTr}

In this appendix we provide explicit formulae for the building blocks of the bremsstrahlung invariant matrix element squared, as described in Eqs.~(\ref{eq:MBSJresI}-\ref{eq:t4}).
If we generalize (\ref{eq:TrE2}) to
\begin{equation}
\frac{e^8}{4}\text{Tr}_{\text{E}^2}(s)\equiv\frac{\overline{|I_\text{E}|^2}}{(E-s)^2}\,,
\label{eq:TrE2s}
\end{equation}
we only need to provide its explicit form in order to cover $\text{Tr}_{\text{E}^2}$ and $\text{Tr}_\text{E}(s)$ defined in (\ref{eq:TrE2}) and (\ref{eq:TrEs}).
Indeed, one can obtain $\text{Tr}_\text{E}(s)$ by means of multiplying $\text{Tr}_{\text{E}^2}(s)$ by the appropriate propagator: $\text{Tr}_\text{E}(s)=(E-s)\text{Tr}_{\text{E}^2}(s)$.
For completeness, $\text{Tr}_{\text{E}^2}=\text{Tr}_{\text{E}^2}(0)$.

In the following expressions, as denoted below, the symmetries of the $J$ operator were already taken into account and the reduction procedure (in the sense of Appendix B of~\cite{Husek:2015sma}) into the basic integrals was performed as well.
Before we get to the desired expressions, we need to define a complete set of the Feynman denominators (while suppressing the `+$i\cuteps$' part):
\begin{equation}
A=l\cdot q\,,\quad
B=l\cdot p\,,\quad
C=k\cdot q\,,\quad
D=k\cdot p\,,\quad
E=(p+q+l)^2\,,\quad
F=(p+q+k)^2\,.
\label{eq:ABCDEF}
\end{equation}
For simplicity, in what follows we use $m\equiv m_\ell$, $\nu\equiv\nu_\ell$ and $M\equiv M_P$.
The term which includes squares of the bremsstrahlung diagrams reads
\begin{equation}
\begin{split}
&J\hspace{-.5mm}\left\{\text{Tr}_{\text{E}^2}(s)\right\}
=J\bigg\{G^2(s)\text{Tr}_\text{D}\\
&+4M^2\left[\frac{2s}{M^2}+x_\gamma-(1-x)(1-y)-\frac{\nu^2}2\right]\frac1A
+8\,\frac BA
-2\nu^2M^2\frac{B}{A^2}
-16s\,\frac1{E-s}
-8(s^2+M^4)\,\frac1{(E-s)^2}\\
&+M^4\,G^2(s)\left[\left(1+\frac{\nu^2}{2x}\right)\frac{x_\gamma^2}{AB}
-\left[x_\gamma+2y(1-x)\right]\frac{\nu^2}{2x}\frac{x_\gamma}{A^2}\right]\\
&-\frac{\nu^2}{2x}\left[6s^2-4M^2s\,g_++M^4\,[1+(1-g_+)^2]+2\nu^2M^2(s-M^2)\right]\hspace{-.5mm}\left[\frac{M^2x}{A^2(E-s)}-G(s)\frac1{A^2}\right]\\
&-\frac{\nu^2}{2x^2}\left[\frac{2s^3}{M^2}-2s^2g_++M^2s\,[1+(1-g_+)^2]+\nu^2(s-M^2)^2\right]\hspace{-.5mm}\left[\frac{M^4x^2}{A^2(E-s)^2}-G^2(s)\frac1{A^2}\right]\\
&+4M^4\left\{\frac{3s^2}{M^4}-\frac{2s}{M^2}\,(g_+-x)+h_0-h_+\,G^2(s)+\nu^2\left[(1+x_\gamma)-h_-\,\frac{G^2(s)}{2x}\right]-\frac{\nu^4}2\left[1-(1-x)^2\frac{G^2(s)}{x^2}\right]\right\}\frac1{A(E-s)}\\
&+4M^6\left\{\frac{s^3}{M^6}-\frac{s^2}{M^4}\,(g_+-x)+\frac{s}{M^2}\,h_0-x\{1-\,h_+\left[1-G(s)\right]\}+\nu^2\left[\frac{s}{M^2}\,(1+x_\gamma)+\frac{h_-}2\left[1-G(s)\right]-1\right]\right.\\
&\hspace{15mm}-\left.\frac{\nu^4}2\left[\frac{s}{M^2}+\frac{(1-x)^2}x\left[1-G(s)\right]-\frac1x\right]\right\}\frac1{A(E-s)^2}\\
&+(y\to-y)\bigg\}\,,
\label{eq:JTrE2}
\end{split}
\end{equation}
the IR-divergent part of which, up to $G^2(s)$ and the $y\to-y$ part, is given as
\begin{equation}
\text{Tr}_\text{D}
=M^4(1-x)^2\left(1+y^2+\frac{\nu^2}x\right)\left[\left(1-\frac{\nu^2}{2x}\right)\frac1{AB}-\frac{\nu^2}{2x}\frac1{A^2}\right].
\end{equation}
This is in agreement with Appendix A of~\cite{Husek:2015sma} up to an additional factor of 1/2 visible in matching (\ref{eq:JTrMS}).
Above, we have used the following definitions:
\begin{align}
G(s)&\equiv\frac{M^2x}{M^2x-s+i\cuteps}\,,\label{eq:Gs}\\
v_0&\equiv\frac12(1+x-x_\gamma)\;,\quad V_0=2M^2v_0\,,\\
g_+&\equiv2-x_\gamma-(1-x)(1+y)=2v_0-y(1-x)\,,\\
h_0&\equiv1-(1-x)(x-y)-x_\gamma(1-x_\gamma)+y^2(1-x)^2\,,\\
h_\pm&\equiv x_\gamma^2+(1-x)^2\big(1\pm y^2\big)\,.
\end{align}

The second building block which arises from the interference of bremsstrahlung diagrams under consideration can be expressed as
\begin{equation}
\begin{split}
&J\hspace{-.5mm}\left\{\text{Tr}_{\text{EF}}(s)\right\}
=J\bigg\{4(V_0-2s)
-8\left(s^2-V_0s+M^4\right)\frac1{(E-s)}\\
&+M^2(V_0-2s)\left[x_\gamma-(1-x)(1+y)+\frac{2\nu^2[1+y(1-x)]}{x_\gamma-(1-x)(1+y)}\right]\frac1A\\
&+\frac1{M^2g_+-2s}\frac{M^8}2\left[4v_0^2(v_0^2-x)-(1-x)^2y^2(2v_0^2-x)+\frac14(1-x)^4y^4\right.\\
&\hspace{10mm}+\left.\frac{\nu^2}2\left\{x_\gamma\left[12v_0^2-2(1+x)+2(1-x)^2+4x_\gamma-3(1-x)^2y^2\right]-xh_-\right\}\right]\frac1{BC}\\
&+\text{Tr}_{X(E-s)}(s;A,x,1)+\text{Tr}_{X(E-s)}(s;D,1-x_\gamma,-1)\\
&+(y\to-y)\bigg\}\,,
\end{split}
\label{eq:JTrEF}
\end{equation}
where we have defined the function
\begin{equation}
\begin{split}
&\text{Tr}_{X(E-s)}(s;X,\tilde x,\xi)\\
&\equiv2\left\{\xi(s-M^2\tilde x)\left\{s^2-V_0s+\frac{M^4}2\big[1+(x-x_\gamma)^2\big]\right\}-M^6xx_\gamma(1+x+x_\gamma)+M^4y(1-x)[s-M^2\tilde x-\xi M^2xx_\gamma]\right.\\
&+\left.\frac12\xi\big(s-M^2\tilde x\big)M^4y^2(1-x)^2+\frac{2M^6xx_\gamma(x^2+x_\gamma^2)}{x_\gamma-(1-x)(1+\xi y)}+\xi\frac{2s(s-V_0)(s^2-V_0s+M^4x)}{M^2g_+-2s}\right.\\
&+\left.M^2\nu^2\left[\big[s^2-V_0s+M^4(x+x_\gamma)\big]\left(1-\frac{x+x_\gamma}{x_\gamma-(1-x)(1+\xi y)}+\xi\frac{M^2(x-3x_\gamma)}{M^2g_+-2s}\right)+\xi\frac{4M^6x_\gamma^2}{M^2g_+-2s}\right]\right\}\frac1{X(E-s)}\,.
\end{split}
\end{equation}
Needless to say, whenever $(y\to-y)$ appears above it applies for entire expressions.
Let us note that after inserting the expressions (\ref{eq:JTrE2}) and (\ref{eq:JTrEF}) into formula (\ref{eq:JTrMS}), one indeed recovers expressions presented in Appendix A of~\cite{Husek:2015sma}.

In (\ref{eq:MBSJresTr}), which we can reformulate into
\begin{equation}
J\hspace{-.5mm}\left\{\frac{\overline{|\mathcal{M}_\text{BS}|^2}}{\mathcal{F}^2(0)}\right\}
=\frac{e^8}4\operatorname{Re}J\hspace{-.5mm}\left\{\text{Tr}\right\}\,,\quad\text{Tr}=4\sum{t_i}\,,
\end{equation}
only $t_{1\text{a}}$ and $t_2$ contribute to the overall IR-divergent part, which can be rewritten as
\begin{equation}
\text{Tr}^\text{div}
=4(t_{1\text{a}}^\text{div}+t_2^\text{div})
=4\left\{
\frac12\text{Tr}_{\text{E}^2}^\text{div}
+\int\diff s\,\mathcal{A}(s)\left[\frac1s+\mathcal{I}^*(s)\right]\hspace{-.5mm}\left[\text{Tr}_{\text{E}}^\text{div}(s)-\text{Tr}_{\text{E}}^\text{div}(0)\right]
\right\}.
\end{equation}
We already know that $J\{\text{Tr}_{\text{E}^2}^\text{div}\}=J\{\text{Tr}_\text{D}\}+(y\to-y)$; note that $G(0)=1$.
In addition, by definition, the IR-divergent part of $\text{Tr}_\text{E}(s)$ is given by the divergent part of $(E-s)\text{Tr}_{\text{E}^2}(s)$.
To this, only $\text{Tr}_{\text{E}^2}^\text{div}$ can contribute.
Moreover, we realize that $E-s=M^2x-s+2A+2B$ and that only its $M^2x-s+i\cuteps$ part does not suppress the essentially divergent terms $1/A^2, 1/B^2$ and $1/AB$ appearing in $\text{Tr}_{\text{E}^2}^\text{div}$.
Consequently, $J\{\text{Tr}_{\text{E}}^\text{div}(s)\}=(M^2x-s+i\cuteps)G^2(s)J\{\text{Tr}_{\text{E}^2}^\text{div}\}$.
We thus find
\begin{equation}
J\big\{\text{Tr}^\text{div}\big\}
=2J\big\{\text{Tr}_{\text{E}^2}^\text{div}\big\}\left\{
1+2\int\diff s\,\mathcal{A}(s)\left[\frac1s+\mathcal{I}^*(s)\right]\hspace{-.5mm}\left[(M^2x-s+i\cuteps)G^2(s)-M^2x\right]
\right\},
\end{equation}
which, using (\ref{eq:Gs}) and (\ref{eq:KLrepre}), translates into
\begin{equation}
J\big\{\text{Tr}^\text{div}\big\}
=2J\big\{\text{Tr}_{\text{E}^2}^\text{div}\big\}\left\{
1+2M^2x\int\diff s\,\frac{\mathcal{A}(s)}{M^2x-s+i\cuteps}
+\int\diff s\int\diff s^\prime\,\frac{\mathcal{A}(s)\mathcal{A}(s^\prime)}{s-s^\prime-i\cuteps}\left[\frac{2sM^2x}{M^2x-s+i\cuteps}\right]
\right\}.
\end{equation}
Due to the symmetries of the double integral and the fact that in the end we are nevertheless interested only in the real part, the term in the square brackets can be further recast into
\begin{equation}
\frac{2sM^2x}{M^2x-s+i\cuteps}
\stackrel{\text{eff}}{=}
-2M^2x
+\frac{(M^2x)^2}{M^2x-s+i\cuteps}
-\frac{(M^2x)^2}{M^2x-s^\prime-i\cuteps}
\stackrel{\text{eff}}{=}
\frac{(M^2x)^2(s-s^\prime-i\cuteps)}{(M^2x-s+i\cuteps)(M^2x-s^\prime-i\cuteps)}\,.
\end{equation}
Taking the real part of the previous result, we finally get (cf.\ definition (\ref{eq:KLrepre}))
\begin{equation}
J\hspace{-.5mm}\left\{{\overline{|\mathcal{M}_\text{BS}^\text{div}|^2}}\right\}
=\frac{e^8}4\mathcal{F}^2(0)\operatorname{Re}J\big\{\text{Tr}^\text{div}\big\}
=2|{\mathcal{F}(M^2x)}|^2\big[J\{\text{Tr}_\text{D}\}+(y\to-y)\big]
\end{equation}
The form factor squared cancels after normalizing on the LO differential decay width and, subsequently, we get the same $\delta_\text{D}^\text{BS}(x,y)$ as in~\cite{Husek:2015sma}, Appendix A.
This is the desired result, since $\delta_\text{D}^\text{BS}(x,y)$ contains the terms which exactly cancel the IR divergences stemming from the virtual corrections: $\delta_\text{D}^\text{BS}(x,y)+2\operatorname{Re}\,\{F_1(x)\}$ is then IR-safe.

Let us recall that the integration of $J\{\text{Tr}_\text{D}\}$ over $x_\gamma$ needs to be performed analytically.
Note that it was only kept as a part of $J\{\text{Tr}_{\text{E}^2}(s)\}$ since it also contributes to the IR-convergent part of $J\{\text{Tr}_\text{E}(s)\}$.


\vspace{5mm}
\twocolumngrid
\section{Basic \texorpdfstring{$J$}{J} terms}
\label{app:J}

In this appendix we provide a list of integrals generated by acting of operator $J$, which are not covered by Appandix~D in~\cite{Husek:2015sma}.
These appear as a consequence of the generalization of the bremsstrahlung matrix element, which was necessary in our current approach.
This is connected to the fact that we took the effect of vector-meson resonances into account.
First, let us recall that
\begin{align}
A+C=\frac{M^2}4[(1-x)(1+y)-x_\gamma]\,,\\
B+D=\frac{M^2}4[(1-x)(1-y)-x_\gamma]\,,
\end{align}
and define
\begin{equation}
W_0
\equiv(A+C)+(B+D)
=\frac{M^2}2(1-x-x_\gamma)\,.
\end{equation}
Next, we define functions $K$, $P$ and $Q$ in the following manner:
\begin{equation}
K(a,b)
\equiv\frac1{2b}\left[\log(a+b)-\log(a-b)\right],
\end{equation}
\begin{equation}
\begin{split}
&P[g(s)]
\equiv\frac2{w_2(\alpha_1-\alpha_2)}\\
&\times\big[\log(1-\alpha_1)-\log(1-\alpha_2)-\log(-\alpha_1)+\log(-\alpha_2)\big]\,,
\end{split}
\end{equation}
where $\alpha_{1,2}=\frac1{2w_2}(-w_1\pm\sqrt{w_1^2-4w_0w_2})$ with
\begin{align}
w_2&=2(B+D)g(s)+g^2(s)+m^2M^2x_\gamma\,,\\
w_1&=2(A+C)g(s)+M^2x_\gamma(M^2x-2m^2)\,,\\
w_0&=m^2M^2x_\gamma\,,
\end{align}
and finally
\begin{equation}
\begin{split}
&Q(u_0,u_1,u_2;g(s))
=-\frac4{w_1^2-4w_0w_2}\\
&\times\left\{
u_0+u_1-u_2+\frac{(u_0+u_1+u_2)(w_2-w_0)}{w_0+w_1+w_2}-\frac{u_0w_1}{w_0}\right.\\
&\qquad+\left.\frac12\left(2u_2w_0-u_1w_1+2u_0w_2\right)P[g(s)]
\right\}.
\end{split}
\end{equation}
Furthermore, with the definitions
\begin{align}
g(s)&\equiv M^2x-s+i\cuteps\,,\\
\tilde g(s)&\equiv M^2(1-x_\gamma)-s+i\cuteps
\end{align}
at hand, we can express the missing basic $J$ integrals as
\begin{align}
J\bigg[\frac1{E-s}\bigg]&=K\Big(W_0+g(s),\sqrt{W_0^2-M^4xx_\gamma}\Big)\,,\\
J\bigg[\frac1{A(E-s)}\bigg]&=P[g(s)]\,,\\
J\bigg[\frac1{C(E-s)}\bigg]&=-P[-\tilde g(s)]\,,\label{eq:CEs}\\
J\bigg[\frac1{D(E-s)}\bigg]&=J\bigg[\frac1{C(E-s)}\bigg]_{y\to-y}\,,\\
J\bigg[\frac1{A^2(E-s)}\bigg]&=2\,Q(v_1,v_2-v_1,-v_2;g(s))\,,\label{eq:A2Es}
\end{align}
where we have used $v_1=(A+C)$ and $v_2=(B+D)+g(s)$.
Note that (\ref{eq:A2Es}) includes an IR-divergent part $J[1/A^2]/g(s)$.
Let us also mention that even though it might seem like that after inspecting definition (\ref{eq:JTrE2}) of $J\{\text{Tr}_{\text{E}^2}(s)\}$, it is not necessary to define other integrals, e.g.\
\begin{align}
J\bigg[\frac1{A(E-s)^2}\bigg]&=Q(0,v_1,v_2;g(s))\,,
\end{align}
since only $\text{Tr}_{\text{E}^2}(0)$ and $(E-s)\text{Tr}_{\text{E}^2}(s)$ appear in the final expression for $\delta^\text{BS}(x,y)$.
For completeness, let us take advantage of our notation and add that $J[1/{(C(E-s)^2)}]=Q(0,v_1,v_2;-\tilde g(s))$\,.
All the other terms can be found in Appendix~D of~\cite{Husek:2015sma}.
At this point, let us mention that in~\cite{Husek:2015sma}, terms (D20) and (D21) were not listed properly in light of definition (D1).
It was not stated explicitly that the second version of (D1) was used for evaluating all the terms.
This (second) version, however, equals to the original definition
\begin{equation}
L(a,b)
\equiv\frac1{\sqrt{a^2-b}}\log\Bigg|\frac{a+\sqrt{a^2-b}}{a-\sqrt{a^2-b}}\Bigg|
\label{eq:Lab}
\end{equation}
only for $a>0$, as mentioned in (D1).
When $a<0$, an additional overall minus sign appears.
This was not emphasized and shall at least be clarified at this point.
Taking thus into account (\ref{eq:Lab}), which is universal for any $a$, we find
\begin{align}
J\bigg[\frac1{CE}\bigg]&=-\frac2{M^4}L(\rho^\prime,\nu^2xx_\gamma)\,,\label{eq:CE}\\
J\bigg[\frac1{BC}\bigg]&=-\frac8{M^4}L(\omega,\nu^2x_\gamma\omega)\,.
\end{align}
Other expressions in Appendix~D of~\cite{Husek:2015sma} require no changes.
Let us also mention that one should indeed get now (\ref{eq:CE}) by putting $s\to0$ in (\ref{eq:CEs}).


\section{Partial fraction decompositions for the bremsstrahlung matrix element squared}
\label{app:eps}

In this appendix we show how to rewrite the bremsstrahlung matrix element squared (\ref{eq:MBSJ}) into a form which will allow us to perform the integrations of the $J$ operator on the respective terms.
To achieve this we need to use a few fraction-product decompositions.
First, we take the simplest case:
\begin{equation}
\frac1{{e}-s+i\cuteps}\frac1{E^*}
=\frac1s\left[\frac1{{e}-s+i\cuteps}-\frac1{E^*}\right].
\end{equation}
Note that $E$ represents a Feynman denominator corresponding to the virtual photon and in this and following applications the difference between $E$ and $E^*$ plays no role.
We also write $s$ instead of $s-i(\cuteps+\cuteps^\prime)$ since $s\ne0$ due to the positive limits of the integration.
Similarly,
\begin{equation}
\begin{split}
&\frac1{{e}-s+i\cuteps}\frac1{{e}^{}-s^\prime-i\cuteps^\prime}\\
&=\frac1{s-s^\prime-i(\cuteps+\cuteps^\prime)}\left[\frac1{{e}-s+i\cuteps}-\frac1{{e}^{}-s^\prime-i\cuteps^\prime}\right].
\end{split}
\label{eq:s-sp}
\end{equation}
In the denominators, $\cuteps$ and $\cuteps^\prime$ represent positive and infinitesimally small independent numbers and the numerical result remains the same even when we assume for simplicity that $\cuteps^\prime\to\cuteps$.
Now, we can use the fact that this term is multiplied by two spectral functions $\mathcal{A}(s)$ and $\mathcal{A}(s^\prime)$ and integrated symmetrically over $s$ and $s^\prime$.
After we rename $s\leftrightarrow s^\prime$ in the second term on the right-hand side of (\ref{eq:s-sp}), we realize that due to the symmetric integration we obtain the complex conjugate of the first term.
Since we are anyway interested in the real part only, we just get the factor of 2.
In the following we use the knowledge of the fact that ${e}+{f}=M_P^2(1+x-x_\gamma)\equiv V_0$ is a significant $J$-invariant combination of kinematical variables:
\begin{equation}
\frac1{{e}-s+i\cuteps}\frac1{F^*}
=\frac1{V_0-s+i(\cuteps-\cuteps^\prime)}\left[\frac1{{e}-s+i\cuteps}+\frac1{F^*}\right].
\end{equation}
Note that due to the presence of the $J$ operator, we can substitute $1/F^*$ by $1/E^*$ on the right-hand side of the previous equation.
Above, $\epsilon^\prime$ is actually artificial and can be safely sent to zero.
This can be also seen due to some other reasons.
First, on the left-hand side we could have already written $f$ instead of $F^*$.
Another, more general, reasoning is similar to the following case.
The last term which needs to be treated can be decomposed as
\begin{equation}
\begin{split}
&\frac1{{e}-s+i\cuteps}\frac1{{f}^{}-s^\prime-i\cuteps^\prime}
=\frac1{V_0-s-s^\prime+i(\cuteps-\cuteps^\prime)}\\
&\times\left[\frac1{{e}-s+i\cuteps}+\frac1{{f}^{}-s^\prime-i\cuteps^\prime}\right].
\end{split}
\label{eq:EFss}
\end{equation}
Again, $\cuteps$ and $\cuteps^\prime$ are infinitesimally small and independent.
For a moment we can assume these to be finite though small.
The left-hand side does obviously not depend numerically on the fact if $\cuteps$ is greater or smaller than $\cuteps^\prime$ so the apparent dependence of this type needs to be canceled on the right-hand side.
For this purpose both of the terms in the brackets are necessary.
Without loss of generality we can assume that $\cuteps>\cuteps^\prime$ and put $\cuteps=\cuteps^\prime+\delta$ with $\delta>0$ small.
Since $\cuteps$ and $\cuteps^\prime$ can be arbitrary for (\ref{eq:EFss}) to be valid, we can set $\cuteps\simeq\cuteps^\prime$.
In the corresponding limit $\delta\to0$ we can apply the Sochocki's formula to find
\begin{equation}
\begin{split}
&\frac1{{e}-s+i\cuteps}\frac1{{f}^{}-s^\prime-i\cuteps^\prime}
\simeq\left[\frac1{{e}-s+i\cuteps}+\frac1{{f}^{}-s^\prime-i\cuteps}\right]\\
&\times\bigg[\operatorname{p.v.}\frac1{V_0-s-s^\prime}
-i\pi\operatorname{sgn}(\cuteps-\cuteps^\prime)\,\delta\big(V_0-s-s^\prime\big)\bigg],
\end{split}
\label{eq:EFss2}
\end{equation}
where $\operatorname{p.v.}$ stands for the principal value.
The part containing the delta function then trivially vanishes (due to the first bracket) together with the dependence on the sign of the difference $\cuteps-\cuteps^\prime$.
Again, using the fact that the integration is symmetric in $s$ and $s^\prime$ and that we can on the right change $F^*\to E^*$, we obtain the sum of a term and its complex conjugate, which results in a factor of 2 under the real-part operator; note that $\overline{I_\text{E} I_\text{F}^*}$ has no (nonvanishing) imaginary part since it is related to the tree diagram.

Taking into account the previous decompositions, Eqs.~(\ref{eq:ReIntAA}) and (\ref{eq:Is}), we can rewrite (\ref{eq:MBSJ}) as
\begin{equation}
\begin{split}
&J\hspace{-.5mm}\left\{\frac{\overline{|\mathcal{M}_\text{BS}|^2}}{\mathcal{F}^2(0)}\right\}
=4\operatorname{Re}J\hspace{-.5mm}\left\{\left[
\frac12\frac1{|E|^2}
+\int\diff s\,\mathcal{A}(s)
\right.\right.\\
&\left.
\times\left(\frac1s+\int\hspace{-1mm}\frac{\mathcal{A}(s^\prime)\diff s^\prime}{s-s^\prime-i\cuteps}\right)
\hspace{-1mm}\left(\frac{1}{{e}-s+i\cuteps}-\frac1{E^*}\right)
\right]\overline{|I_\text{E}|^2}\\
&+\left[
\frac1{V_0}\frac1E
+\int\diff s\,\frac{\mathcal{A}(s)}{V_0-s+i\cuteps}\left(\frac1{F^*}+\frac1{{e}-s+i\cuteps}\right)
\right.\\
&+\left.\left.
\int\hspace{-2.5mm}\int\diff s\diff s^\prime\operatorname{p.v.}\frac{\mathcal{A}(s)\mathcal{A}(s^\prime)}{V_0-s-s^\prime}\frac{1}{{e}-s+i\cuteps}
\right]\overline{I_\text{E} I_\text{F}^*}\right\}.
\label{eq:MBSJres}
\end{split}
\end{equation}
and consequently into (\ref{eq:MBSJresI}).


\section{VMD-inspired model for the \texorpdfstring{$\eta^{(\prime)}$}{eta(')} electromagnetic transition form factors}
\label{app:etaVMD}

\subsection{Introduction}

For a phenomenological model of a transition between the $\eta^{(\prime)}$ meson and virtual photons we need to take into account a strange-quark content of $\eta^{(\prime)}$.
It is more convenient to work in the quark-flavor basis than in the octet-singlet one~\cite{Feldmann:1998sh,Escribano:2005qq}.
In such a basis, the vector currents related to physical states of $\omega$, $\rho^0$ and $\phi$ mesons are identical to the basis currents.
Having a standard definition of vector currents and pseudoscalar densities in the octet-singlet basis%
\footnote{Our convention is $\gamma_5 = i \gamma^0\gamma^1\gamma^2\gamma^3$.}%
(note that $a=0,1,\dots,8$)
\begin{equation}
j^a_\mu(x)
\equiv\bar q(x)\gamma_\mu T^a q(x)\,,\quad
j^a(x)
\equiv\bar q(x)i\gamma_5 T^a q(x)\,,
\label{VPcur}
\end{equation}
we can write for the currents of our interest
\begin{alignat}{4}
j^\omega_\mu
&=\frac12\big[\bar u\gamma_\mu u+\bar d\gamma_\mu d\big]
&&=\sqrt\frac{2}{3}j^0_\mu+\frac{1}{\sqrt3}j^8_\mu
&&\equiv j^{\ell 0}_\mu\,,\label{eq:jl0}\\
j^{\rho^0}_\mu
&=\frac12\big[\bar u\gamma_\mu u-\bar d\gamma_\mu d\big]
&&=j^3_\mu
&&\equiv j^{\ell 1}_\mu\,,\label{eq:jl1}\\
j^\phi_\mu
&=\frac1{\sqrt2}\big[\bar s\gamma_\mu s\big]
&&=\frac{1}{\sqrt3}j^0_\mu-\sqrt\frac{2}{3}j^8_\mu
&&\equiv j^\text{s}_\mu\,.\label{eq:js}
\end{alignat}
Note that for simplicity we have left out the spacetime coordinates $x$ of the currents and quark fields.
Above, we see the relations between neutral-meson-related vector currents, appropriate combinations of quark-flavor-diagonal vector currents, their octet-singlet basis decomposition and finally the quark-flavor basis definition.
The electromagnetic current reads
\begin{equation}
\begin{split}
\frac1ej_\mu^\text{em}
&=\frac23\bar u\gamma_\mu u-\frac 13\bar d\gamma_\mu d-\frac 13\bar s\gamma_\mu s\\
&=\frac 13j_\mu^{\ell0}+j_\mu^{\ell1}-\frac{\sqrt2}3j_\mu^\text{s}\,.
\end{split}
\label{eq:jem}
\end{equation}

In the chiral limit, the $PVV$ correlator $\Pi(r^2;p^2,q^2)$ is defined in the octet-singlet basis by 
\begin{equation}
\begin{split}
&d^{abc}\epsilon_{\mu\nu\alpha\beta}p^{\alpha}q^{\beta}\Pi(r^2;p^2,q^2)\\
&\equiv\int\text{d}^4x\,\text{d}^4y\,e^{ip\cdot x+iq\cdot y}\langle 0|T[j^a(0)j_\mu^b(x)j_\nu^c(y)]|0\rangle
\end{split}
\label{eq:PVVdef}
\end{equation}
with $r=p+q$.
In the above formulae we have used
\begin{equation}
\text{Tr}\big[T^a,T^b\big]
=\frac12\delta^{ab}\,,\quad
d^{abc}
\equiv2\,\text{Tr}\big[\{T^a,T^b\}T^c\big]  \,.
\label{eq:normdef}
\end{equation}
As it is common, $T^a\equiv{\lambda^a}/{2}$ where $\lambda^a$ denote the Gell-Mann matrices in the flavor space and $d^{abc}$ are the $\text{U}(3)$ symmetric symbols.

If, for simplicity, we rewrite (\ref{eq:PVVdef}) schematically as
\begin{equation}
\text{corr}(j^a,j_\mu^b,j_\nu^c)
=d^{abc}\Pi\,,
\end{equation}
then --- using linearity and definitions (\ref{eq:jl0}), (\ref{eq:jl1}) and (\ref{eq:js}) --- we get only three nontrivial combinations of currents in the quark-flavor basis:
\begin{align}
\text{corr}(j^{\ell},j_\mu^{\ell 0},j_\nu^{\ell 0})
&=\Pi^\ell\label{eq:Pi_l0}\\
\text{corr}(j^{\ell},j_\mu^{\ell 1},j_\nu^{\ell 1})
&=\Pi^\ell\label{eq:Pi_l1}\\
\text{corr}(j^\text{s},j_\mu^\text{s},j_\nu^\text{s})
&=\sqrt2\,\Pi^\text{s}\,.
\label{eq:corr_sqrt2}
\end{align}
In this way we have found the normalization relation among bases (there is an additional factor $\sqrt{2}$ in the case of the strange correlator).
Note that $j^{\ell}\equiv\frac i2\big[\bar u\gamma_5 u+\bar d\gamma_5 d\big]$ and $j^\text{s}\equiv\frac i{\sqrt2}\big[\bar s\gamma_5 s\big]$\,.
In Eqs.~(\ref{eq:Pi_l0})-(\ref{eq:corr_sqrt2}) we have gone beyond the chiral limit: from now on we will distinguish between the light and strange correlators.

Since the quark content of the $\eta^{(\prime)}$ physical states is not equal to the $\text{U}(3)$ isoscalar states, there is a mixing between $\eta$ and $\eta^\prime$ mesons.
In the quark-flavor basis, this mixing occurs (for $A\in\{\ell,\text{s}\}$) among the states $|\eta^A\rangle$ defined as $\langle 0|j^A|\eta^B\rangle=\delta^{AB}\mathcal{Z}_{\eta^A}$ together with the orthonormality relation $\langle\eta^A|\eta^B\rangle=\delta^{AB}$\,.
The resulting mixing (in the quark-flavor basis) can be written as
\begin{align}
|\eta\rangle&=\cos\phi\,|\eta^\ell\rangle-\sin\phi\,|\eta^\text{s}\rangle\,,\label{eq:etamixing}\\
|\eta^{\prime}\rangle&=\sin\phi\,|\eta^\ell\rangle+\cos\phi\,|\eta^\text{s}\rangle\,.
\label{eq:etaprimemixing}
\end{align}

Next we define the correlator $\eta^A VV$ for each basis state (again $A\in\{\ell,\text{s}\}$):
\begin{equation}
\Pi_{\eta^A VV}(p^2,q^2)
\equiv \frac1{\mathcal{Z}_{\eta^A}}\lim_{r^2\to M_\eta^2}(r^2-M_\eta^2)\Pi^A(r^2;p^2,q^2)\,,
\label{eq:FFp2q2}
\end{equation}
The factors $\mathcal{Z}_{\eta^A}=\langle 0|j^A|\eta^A\rangle$ are related to the pion case $\mathcal{Z}_{\pi}=B_0F_\pi$ by $\mathcal{Z}_{\eta^A}=\mathcal{Z}_{\pi}f_A$: we have introduced the ratio of the decay constants $f_A\equiv F_A/F_\pi$.
For the $\eta VV$ correlator we can then finally write
\begin{equation}
\Pi_{\eta VV}(p^2,q^2)
=\cos\phi\,\Pi_{\eta^\ell VV}-\sqrt2\sin\phi\,\Pi_{\eta^\text{s} VV}\,.
\end{equation}
The $\sqrt2$ factor comes from (\ref{eq:corr_sqrt2}) and the mixing factors from (\ref{eq:etamixing}).
To avoid difficulties connected with calculation of the decay constant values and defining the final $\eta VV$ correlator all the way through (\ref{eq:FFp2q2}), we will use the VMD ansatz
\begin{equation}
\Pi_{\eta^A VV}^\text{VMD}(p^2,q^2)
=-\frac{N_\text{c}}{8\pi^2 F_A}\frac{M_A^4}{(p^2-M_A^2)(q^2-M_A^2)}\,,
\label{eq:VMDansatz}
\end{equation}
where the light and strange channels are saturated by associated resonances: $M_\ell=M_{\rho^0/\omega}$ and $M_\text{s}=M_{\phi}$.

Numerical values for the mixing angle $\phi$ and decay constants $f_\ell$ and $f_\text{s}$ are obtained from various experimental data (for details see next part of this appendix) in terms of a global fit with the result
\begin{equation}
\phi=41(2)^\circ\,,\quad f_\ell=1.07(6)\,,\quad f_s=1.74(3)\,.
\end{equation}
These values are used throughout the paper.

\subsection{Model relevancy check}

Now we have to check if such a VMD-inspired model is phenomenologically successful and if it is suitable for calculating 1$\gamma$IR contribution to the NLO virtual radiative corrections.

\subsubsection{Decays including vector mesons}

We may start with processes containing vector mesons.
To investigate this we need to take into account few more quantities.
We now introduce the overlap between a $\mathcal{V}$ meson ($\mathcal{V}\in\{\omega,\rho^0,\phi\}$) and the vector current $j_\mu^\mathcal{V}$, i.e.
\begin{equation}
\mathcal{Z}_\mathcal{V}\epsilon_\mu(\vec{p},\lambda_\mathcal{V})
\equiv\langle 0|j_\mu^\mathcal{V}|\phi(\vec{p},\lambda_\mathcal{V})\rangle\,.
\end{equation}
Here we assume that $\rho^0$ and $\omega$ mesons contain only the light quarks (no hidden strangeness) and the $\phi$ meson contains only the strange quarks.
This is a fairly good approximation to the real world \cite{Olive:2016xmw}.
For later convenience we also define the $\gamma$-$\mathcal{V}$ coupling strength $F_\mathcal{V}\equiv {\mathcal{Z}_\mathcal{V}}/{M_\mathcal{V}}$.
With this at hand we can obtain $F_\mathcal{V}$ from the $\mathcal{V}\to e^+e^-$ decay processes.
The direct calculation from the Lorentz invariant matrix element
\begin{equation}
\begin{split}
&i\mathcal{M}_{\mathcal{V}\to e^+e^-}\\
&=i\kappa_\mathcal{V}\mathcal{Z}_\mathcal{V}\bar{u}(\vec{q}_1)(-ie\gamma^\mu)v(\vec{q}_2)\frac{(-i)}{M_\mathcal{V}^2}\epsilon_\mu(\vec{p},\lambda_\omega)\,,
\end{split}
\end{equation}
yields in the limit $m_e^2\ll M_\mathcal{V}^2$ after averaging and summing over polarizations
\begin{equation}
F_\mathcal{V}
\simeq
\frac{1}{e|\kappa_\mathcal{V}|}\sqrt{{12\pi M_\mathcal{V}{\Gamma({\mathcal{V}\to e^+e^-})}}}\,.
\label{eq:FV}
\end{equation}
Above, $\kappa_\mathcal{V}$ is the overlap of the electromagnetic current $j_\mu^\text{em}$ with the meson-related current $j_\mu^\mathcal{V}$\,,
For a little more details concerning the $\omega$ case see~\cite{Husek:2015wta}.
Necessary numerical inputs and results are shown in Table~\ref{tab:GVee}.

\begin{table}[!ht]
\begin{ruledtabular}
\begin{tabular}{c | c c c }
$\mathcal{V}$ & $\omega$ & $\rho^0$ & $\phi$\\
\hline
$M_\mathcal{V}$\,[MeV] & 782.65(12) & 775.26(25) & 1019.46(2)\\
$\Gamma(\mathcal{V})$\,[MeV] & 8.49(8) & 147.8(9) & $4.266(31)$\\
$B({\mathcal{V}\to e^+e^-})$\,[$10^{-5}$] & 7.28(14) & 4.72(5) & 29.54(30)\\
\hline
$\kappa_\mathcal{V}$ & $e$/3 & $e$ & $-e\sqrt{2}$/3\\
$\Gamma(\mathcal{V}\to e^+e^-)$\,[keV] & 0.62(2) & 7.0(1) & 1.26(2)\\
$F_\mathcal{V}$\,[MeV] & 140(2) & 156(2) & 161(1)\\
\end{tabular}
\end{ruledtabular}
\caption{
Photon to meson coupling strength $F_\mathcal{V}$ together with used experimental values~\cite{Olive:2016xmw}.
In the upper part of this table we see masses, decay widths and branching ratios of the $\mathcal{V}\to e^+e^-$ processes for the vector mesons under consideration.
In the lower part we then see the overlaps $\kappa_\mathcal{V}$ (coefficients in (\ref{eq:jem})), decay widths of $\mathcal{V}\to e^+e^-$ calculated from values provided in the upper part and finally the resulting coupling strengths $F_\mathcal{V}$ evaluated according to (\ref{eq:FV}).
}
\label{tab:GVee}
\end{table}

For the following, it is convenient to introduce the $\eta^{(\prime)}\mathcal{V}V$ correlator
\begin{equation}
\Pi_{\eta^{(\prime)}\mathcal{V}V}(q^2)
=\frac1{\mathcal{Z}_\mathcal{V}}\lim_{p^2\to M_A^2}(p^2- M_A^2)\Pi_{\eta^{(\prime)} VV}(p^2,q^2)\,.
\label{eq:FFq2}
\end{equation}
Consequently, the $\eta^{(\prime)}\mathcal{V}\gamma^*$ form factor within the VMD model can be written as
\begin{equation}
\begin{split}
e\mathcal{F}_{\eta^{(\prime)}\mathcal{V}\gamma^*}^\text{VMD}(q^2)
=
-\frac{N_\text{c}}{8\pi^2 F_\pi}
\frac{M_A^2}{\mathcal{Z}_\mathcal{V}}\,
\kappa_\mathcal{V}\kappa_{\eta^{(\prime)}\mathcal{V}}\,\frac{M_A^2}{q^2-M_A^2}
\,.
\end{split}
\label{eq:FFVMDphi}
\end{equation}
The additional factors $\kappa_{\eta^{(\prime)}\mathcal{V}}$, which come from the $\eta$-$\eta^\prime$ mixing, can be found in Tabs.~\ref{tab:GVetag}-\ref{tab:GVetaprimeg}.
Using these form factors we can calculate the two-body decay widths containing one pseudoscalar meson ($\eta^{(\prime)}$), one vector meson ($\omega$, $\rho^0$ or $\phi$) and one photon.
Depending on the masses of the particles involved, we use one of the following two prescriptions:
\begin{equation}
\Gamma_{\mathcal{V}\to\eta^{(\prime)}\gamma}
=\frac{\alpha M_\mathcal{V}^3}{24}|\mathcal{F}_{\eta^{(\prime)}\mathcal{V}\gamma^*}(0)|^2\bigg(1-\frac{M_{\eta^{(\prime)}}^2}{M_\mathcal{V}^2}\bigg)^3\,,
\label{eq:Gweg}
\end{equation}
\begin{equation}
\Gamma_{\eta^{\prime}\to\mathcal{V}\gamma}
=\frac{\alpha M_{\eta^\prime}^3}{8}|\mathcal{F}_{\eta^{\prime}\mathcal{V}\gamma^*}(0)|^2\bigg(1-\frac{M_\mathcal{V}^2}{M_{\eta^{\prime}}^2}\bigg)^3\,.
\label{eq:Gewg}
\end{equation}
The results together with a comparison with the experimental data can be found, for the $\eta$ case, in Table~\ref{tab:GVetag}, and for the $\eta^\prime$ case in Table~\ref{tab:GVetaprimeg}.
\begin{table}[!ht]
\begin{ruledtabular}
\begin{tabular}{c | c c c }
$\mathcal{V}$ & $\omega$ & $\rho^0$ & $\phi$\\
\hline
$B(\mathcal{V}\to\eta\gamma)\,[10^{-4}]$ & 6.6(1.7) & 3.0(2) & 130.9(2.4)\\
\hline
$\kappa_{\eta \mathcal{V}}$ & $\cos(\phi)/f_\ell$ & $\cos(\phi)/f_\ell$ & $-\sqrt{2}\sin(\phi)/f_\text{s}$\\
$\Gamma_{\mathcal{V}\to\eta\gamma}^\text{exp}$\,[keV] & 5.6(1.4) & 44(3) & 56(1)\\
$\Gamma_{\mathcal{V}\to\eta\gamma}^\text{VMD}$\,[keV] & 5.7(7) & 37(5) & 50(5)\\
\end{tabular}
\end{ruledtabular}
\caption{
Comparison of the VMD model with the experimental data~\cite{Olive:2016xmw} for the decays $\mathcal{V}\to\eta\gamma$.
The $\sqrt2$ factor in $\kappa_{\eta\phi}$ comes from (\ref{eq:corr_sqrt2}).
}
\label{tab:GVetag}
\end{table}
\begin{table}[!ht]
\begin{ruledtabular}
\begin{tabular}{c | c c c }
$\mathcal{V}$ & $\omega$ & $\rho^0$ & $\phi$\\
\hline
$B(\eta^\prime \mathcal{V}\gamma)\,[\%]$ & 2.75(23) & 29.1(5) & $6.25(21)\times10^{-3}$\\
\hline
$\kappa_{\eta^\prime \mathcal{V}}$ & $\sin(\phi)/f_\ell$ & $\sin(\phi)/f_\ell$ & $\sqrt{2}\cos(\phi)/f_\text{s}$\\
$\Gamma_{\eta^\prime \mathcal{V}\gamma}^\text{exp}$\,[keV] & 5.4(5) & 58(3) & 0.27(1)\\
$\Gamma_{\eta^\prime \mathcal{V}\gamma}^\text{VMD}$\,[keV] & 6.5(9) & 52(7) & 0.30(2) \\
\end{tabular}
\end{ruledtabular}
\caption{
Comparison of the VMD model with the experimental data~\cite{Olive:2016xmw} for the $\eta^\prime \mathcal{V}\gamma$ decays.
}
\label{tab:GVetaprimeg}
\end{table}
We see that the VMD model and the experimental data are in agreement.

The last class of the processes containing vector mesons which we will investigate is a counterpart of the previous decays, when the photon in the final state is virtual and turns into the lepton pair.
The decay width can then be expressed as a form-factor-dependent integral over the dilepton invariant mass
\begin{equation}
\begin{split}
&\Gamma(\mathcal{V}\to\eta\ell^+\ell^-)
=\frac{\alpha^2}{72\pi M_\mathcal{V}^3}
\hspace{-3mm}
\int\limits_{4m_\ell^2}^{(M_\mathcal{V}-M_\eta)^2}
\hspace{-2mm}
\;
\frac{|\mathcal{F}_{\eta \mathcal{V}\gamma^*}(q^2)|^2}{q^2}\\
&\times\sqrt{1-\frac{4m_\ell^2}{q^2}}
\bigg(1+\frac{2m_\ell^2}{q^2}\bigg)
\lambda^{\frac32}\big(M_\mathcal{V}^2,M_\eta^2,q^2\big)\,
{\text{d}q^2}\,,
\end{split}
\end{equation}
where $\lambda$ denotes the K\"all\'en triangle function defined as
\begin{equation}
\lambda(a,b,c)
\equiv a^2+b^2+c^2-2ab-2ac-2bc\,.
\end{equation}
After inserting numerical values we get
\begin{equation}
\Gamma_{\phi\to\eta e^+e^-}^\text{VMD}
=(0.42\pm0.05)\,\text{keV}\,,
\end{equation}
which should be compared with $\Gamma_{\phi\to\eta e^+e^-}^\text{exp}=(0.49\pm0.05)\,\text{keV}$.
For the above evaluation we have used $B(\phi\to\eta e^+e^-)=(1.15\pm0.10)\times10^{-4}$~\cite{Olive:2016xmw}.

Data for $\omega/\rho^0\to\eta\ell^+\ell^-$ and $\eta^\prime\to\omega/\rho^0 e^+e^-$ decays as well as for $\phi\to\eta\mu^+\mu^-$ and $\phi\to\eta^\prime e^+e^-$ decays are not available.

\subsubsection{Electromagnetic transition form factor}

Using previously defined meson-specific factors, we can finally define the doubly virtual electromagnetic transition form factor of an eta meson in the quark-flavor basis:
\begin{equation}
\mathcal{F}_{\eta\gamma^*\gamma^*}(p^2,q^2)
=\sum_\mathcal{V}\kappa_\mathcal{V}^2(\kappa_{\eta\mathcal{V}}f_{A(\mathcal{V})})\Pi_{\eta^{A(\mathcal{V})} VV}(p^2,q^2)\,.
\end{equation}
For $A(\mathcal{V})$ above we simply substitute $A(\rho^0)=A(\omega)=\ell$ and $A(\phi)=\text{s}$\,.
In the VMD case (inserting ansatz (\ref{eq:VMDansatz})) the form factor becomes
\begin{equation}
\begin{split}
e^2\mathcal{F}_{\eta\gamma^*\gamma^*}^\text{VMD}(p^2,q^2)\quad&\\
=
-\frac{N_\text{c}}{8\pi^2 F_\pi}
\frac{2e^2}{3}
\bigg[
\frac53&\frac{\cos\phi}{f_\ell}\,\frac{M_{\rho^0/\omega}^4}{(p^2-M_{\rho^0/\omega}^2)(q^2-M_{\rho^0/\omega}^2)}\\
-\frac{\sqrt2}3&\frac{\sin\phi}{f_\text{s}}\,\frac{M_\phi^4}{(p^2-M_\phi^2)(q^2-M_\phi^2)}
\bigg]\,.
\end{split}
\label{eq:FFeggVMD}
\end{equation}
To get the $\eta^\prime$ form factor, it is only necessary to perform the following substitution:
\begin{equation}
\mathcal{F}_{\eta^\prime\gamma^*\gamma^*}^\text{VMD}(p^2,q^2)
=
\mathcal{F}_{\eta\gamma^*\gamma^*}^\text{VMD}(p^2,q^2)\big|_{\substack{\cos\phi\to\sin\phi\\ \sin\phi\to-\cos\phi}}\,.
\label{eq:FFVMDprime}
\end{equation}

As a simple application, we can have a look at the two-photon decay of a pseudoscalar $P\in\{\pi^0,\eta^{(\prime)}\}$. 
The decay with of such a process can be expressed as follows:
\begin{equation}
\Gamma_{P\to\gamma\gamma}^\text{VMD}
=
\frac12\frac1{16\pi M_P}
\left(\frac{\alpha}{\pi F_\pi}\right)^2\frac{M_P^4}{2}\kappa_P^2\,.
\label{eq:GPgg}
\end{equation}

Of course, for a neutral pion we would have $\kappa_{\pi^0}=1$\,.
In the $\eta^{(\prime)}$ case we can write
\begin{equation}
\kappa_{\eta^{(\prime)}}
=\frac3{2e^2}\sum_\mathcal{V}\kappa_\mathcal{V}^2\kappa_{\eta^{(\prime)}\mathcal{V}}\,,
\end{equation}
which becomes (cf.~(\ref{eq:FFeggVMD}) for $p^2=q^2=0$)
\begin{align}
\kappa_{\eta}&=\frac53\frac{\cos\phi}{f_\ell}-\frac{\sqrt2}3\frac{\sin\phi}{f_\text{s}}=1.0(1)\,,\\
\kappa_{\eta^{\prime}}&=\frac53\frac{\sin\phi}{f_\ell}+\frac{\sqrt2}3\frac{\cos\phi}{f_\text{s}}\simeq1.23(10)\,.
\end{align}
Note that e.g.\ $\kappa_\eta\simeq1$ is consistent with experiment, although it differs significantly from a na\"ive WZW-based calculation~\cite{Savage:1992ac} for which $\kappa_\eta=1/\sqrt{3}$. 
The results are shown in Table~\ref{tab:GVgg}.
\begin{table}[!h]
\begin{ruledtabular}
\begin{tabular}{c c | c c c}
$P$ && $\eta$ & $\eta^{\prime}$ \\
\hline
$M_P$\,[MeV] && 547.862(17) & 957.78(6)\\
$\Gamma_P$\,[keV] && 1.31(5) & 198(9)\\
$B(P\to\gamma\gamma)$\,[\%] && 39.41(20) & 2.20(8)\\
\hline
$\Gamma_{P\to\gamma\gamma}^\text{exp}$\,[keV] && 0.52(2) & 4.4(2)\\
$\Gamma_{P\to\gamma\gamma}^\text{VMD}$\,[keV] && 0.52(8) & 4.2(5)\\
\end{tabular}
\end{ruledtabular}
\caption{
The two-photon decay of $\eta^{(\prime)}$ within the VMD model and its comparison with the data.
As usual, in the upper part of the table we state the experimental inputs~\cite{Olive:2016xmw} and in the lower part we can see the derived quantities.
}
\label{tab:GVgg}
\end{table}
We see that there is a very good agreement between the prediction of the VMD model and the experimental results.

Next, we should investigate the singly virtual transition form factor.
One possibility is to compare the decay widths of the $\eta^{(\prime)}$ Dalitz decays.
These can be written as the following integral:
\begin{equation}
\begin{split}
&\Gamma_{P\to\ell^+\ell^-\gamma}^\text{LO}
=\left(\frac{\alpha}{\pi}\right)\Gamma_{P\to\gamma\gamma}^\text{LO}\\
&\times\int_{\nu^2}^1\text{d}x\,\bigg|\frac{\mathcal{F}_{\eta\gamma^*\gamma^*}(0,xM_P^2)}{\mathcal{F}_{\eta\gamma^*\gamma^*}(0,0)}\bigg|^2\frac{8\beta(x)}{3}\frac{(1-x)^3}{4x}\bigg[1+\frac{\nu^2}{2x}\bigg]\,.
\end{split}
\end{equation}
A simple model we discuss here is however (in the singly virtual mode) not suitable for the Dalitz decays of $\eta^\prime$ due to the unregulated pole which needs to be integrated over.
We thus present results only for the $\eta$ case.
These are shown in Table~\ref{tab:GPllg}.
\begin{table}[!h]
\begin{ruledtabular}
\begin{tabular}{c c | c c c}
process && $\eta\to e^+e^-\gamma$ & $\eta\to \mu^+\mu^-\gamma$\\
\hline
$B\,[10^{-4}]$ && 69(4) & 3.1(4)\\
\hline
$\Gamma^\text{exp}$\,[eV] && 9.0(6) & 0.41(6)\\
$\Gamma^\text{VMD}$\,[keV] && 8.6(2.0) & 0.41(8)\\
\end{tabular}
\end{ruledtabular}
\caption{
Dalitz decays within the VMD model and the comparison with the data~\cite{Olive:2016xmw}.
}
\label{tab:GPllg}
\end{table}

\begin{figure}[!b]
\resizebox{\columnwidth}{!}{\includegraphics{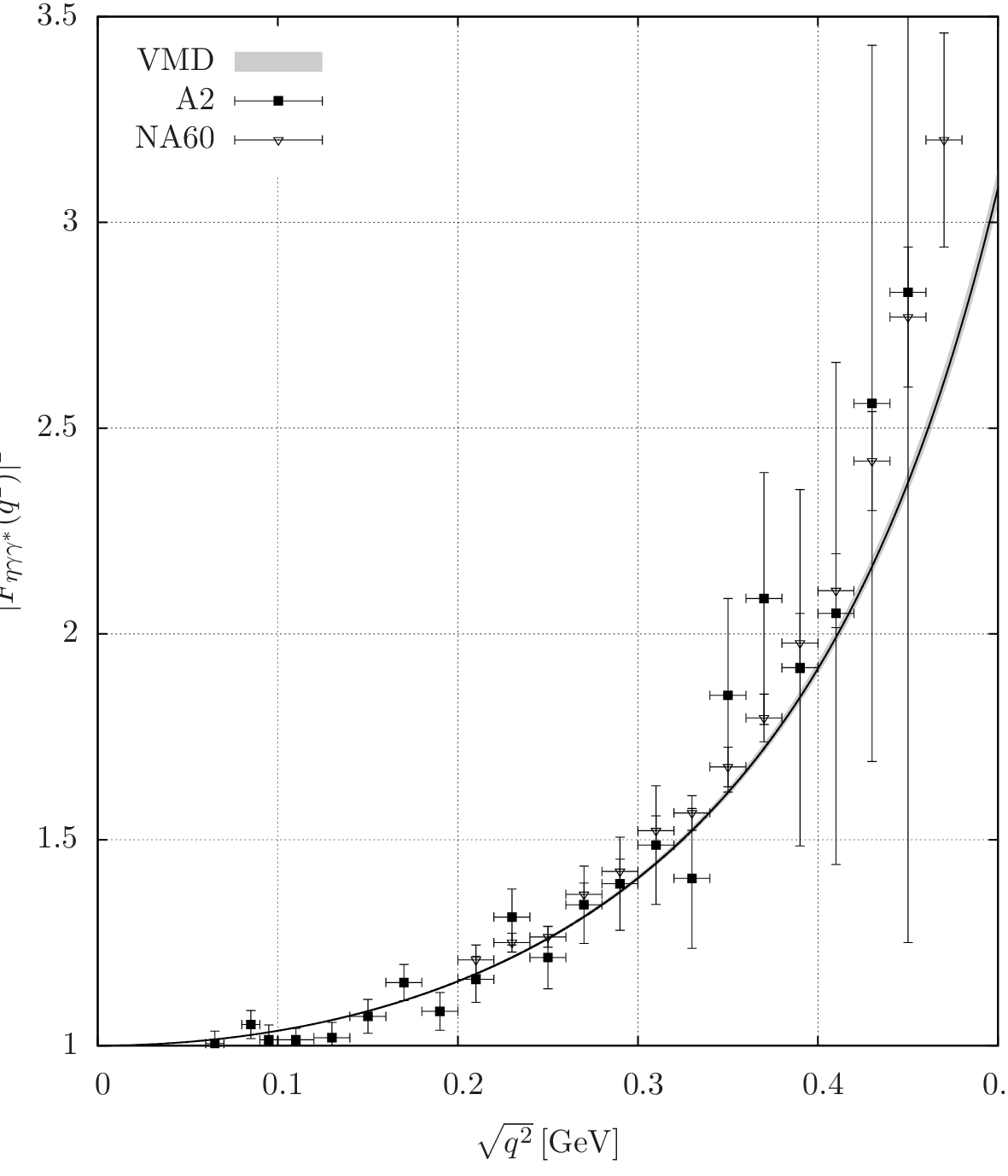}}
\caption{
The normalized singly virtual $\eta$ electromagnetic transition form factor squared within the VMD-inspired model compared to data: time-like region.
The shaded area corresponds to the uncertainty of our fit for the VMD model.
Data are taken from~\cite{Aguar-Bartolome:2013vpw,Arnaldi:2016pzu}.
}
\label{fig:eta_time}
\vspace{-5mm}
\end{figure}

\begin{figure}[!t]
\resizebox{\columnwidth}{!}{\includegraphics{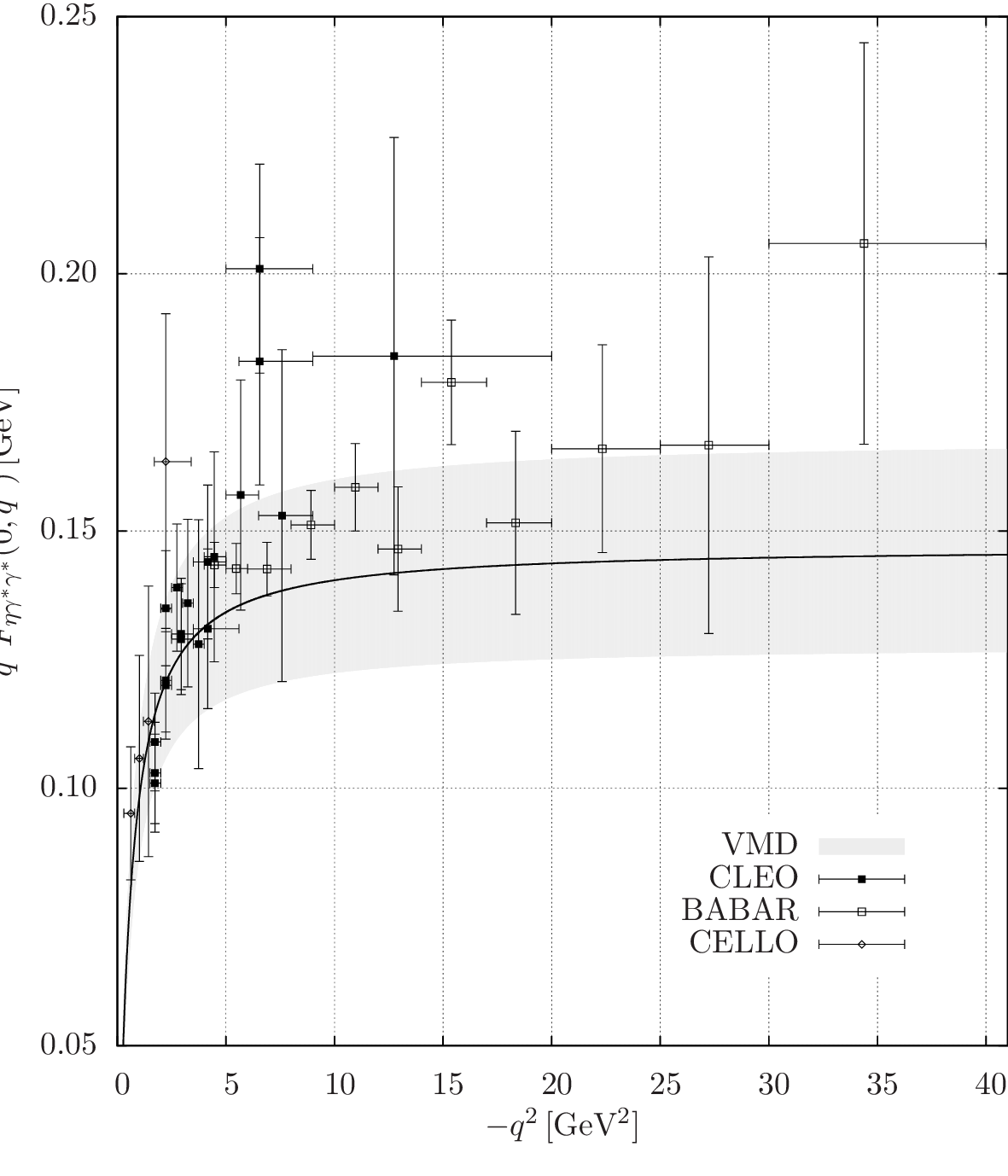}}
\vskip4mm
\resizebox{\columnwidth}{!}{\includegraphics{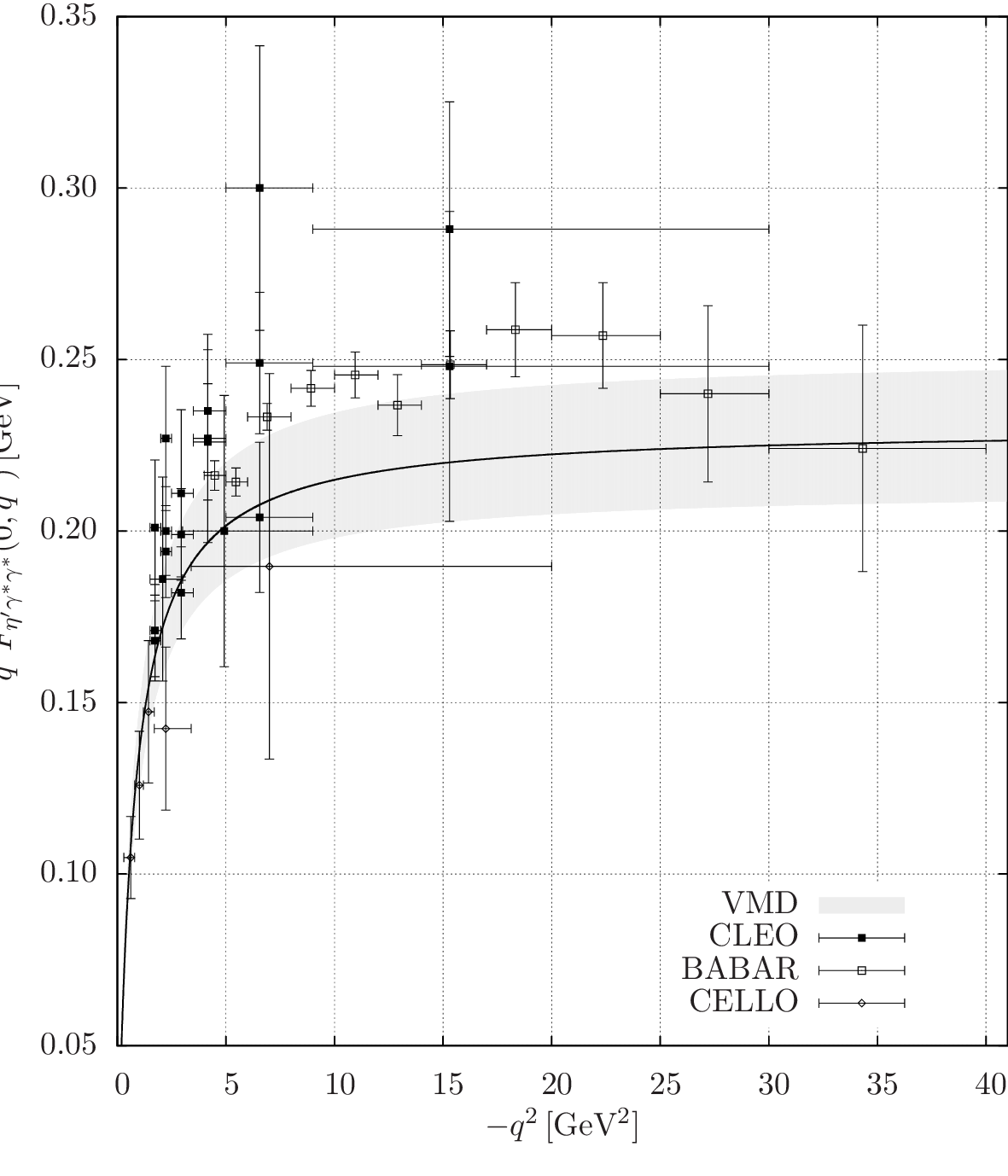}}
\caption{
The singly virtual electromagnetic transition form factors within the VMD-inspired model compared to data: space-like region.
In the first panel we see the $\eta$ form factor, in the second panel there is a plot for the $\eta^\prime$ case.
Data are taken from~\cite{Gronberg:1997fj,BABAR:2011ad,Behrend:1990sr}.
}
\label{fig:eta_space}
\end{figure}

The other way is to directly plot the transition form factor over the respective data.
The measurements of the $\eta$ transition form factor in time-like region are shown in Fig.~\ref{fig:eta_time}.
In the space-like region the results can be found in Fig.~\ref{fig:eta_space}.

Finally, we can test the transition form factor in its whole power, in the doubly virtual mode.
Using the VMD model (\ref{eq:FFeggVMD}), we can calculate the decay width of one of the rare processes $\eta\to\mu^+\mu^-$\,.
Note that data for the rare processes $\eta\to e^+e^-$ and $\eta^\prime\to\ell^+\ell^-$ are not available yet; for predictions for these decays within the model under discussion see Table~\ref{tab:Pll} in Appendix~\ref{app:Pll}.
Experimentally, the branching ratio was found to be $B(\eta\to\mu^+\mu^-)=(5.8\pm0.8)\times10^{-6}$\,, which translates into $\Gamma_{\eta\to\mu^+\mu^-}^\text{exp}=(7.6\pm1.1)\times10^{-6}$\,keV.
Within the VMD model we find $\Gamma_{\eta\to\mu^+\mu^-}^\text{VMD}=(6.0\pm1.4)\times10^{-6}$\,keV, which is in agreement with the experimental value.
Note that we have calculated this value using the convenience of the approach shown in the following appendix.
It is similar to the one we have used to calculate the 1$\gamma$IR contribution to the radiative corrections in Section~\ref{sec:1gIR}.
Finally, let us also mention that for the decays under consideration, more advanced models were developed \cite{Masjuan:2015cjl}, which could also mimic $\pi\pi$ rescattering effects.


\section{Form factors in \texorpdfstring{$P\to\ell^+\ell^-$}{P->l+l-} decays}
\label{app:Pll}

In this appendix we apply the approach explained in Section~\ref{sec:1gIR} to the $P\to\ell^+\ell^-$ decays.
We would like to show how the building block for the matrix element looks like in this case and calculate the coefficients for the specific transition form-factor models.

On account of the Lorentz symmetry and parity conservation, the on-shell matrix element of the $P\to \ell^+\ell^-$ process can be written in terms of just one pseudoscalar form factor in the following form:
\begin{equation}
i\mathcal{M}_{P\to \ell^+\ell^-}
=P_{P\to \ell^+\ell^-}\left[\bar{u}(\vec{q}_1)\gamma_5 v(\vec{q}_2)\right].
\end{equation}
Subsequently, the decay width reads
\begin{equation}
\Gamma(P\to \ell^+\ell^-)
=\frac{2M_P^2|P_{P\to \ell^+\ell^-}|^2}{16\pi M_P}\sqrt{1-\frac{4m_\ell^2}{M_P^2}}\,.
\end{equation}
Taking into account only the LO contribution in the QED expansion, we find for the pseudoscalar form factor
\begin{equation}
P_{P\to \ell^+\ell^-}^\text{LO}
=\frac{ie^4 m_\ell}{M_P^2}
\int\frac{\text{d}^4l}{(2\pi)^4}
\frac{\mathcal{F}_{P\gamma^*\gamma^*}(p^2,q^2)\lambda(M_P^2,p^2,q^2)}{p^2q^2(l^2-m_\ell^2)}\,.
\label{eq:PLO}
\end{equation}
Here, $p=l-q_1$ and $q=l+q_2$, where $q_1$ and $q_2$ are the lepton momenta and $\lambda$ is the triangle K\"all\'en function.
For the rational resonance-saturation models, we will use in agreement with substitution (\ref{eq:FFsubsth}) the following definition:
\begin{equation}
\begin{split}
&P_{P\to\ell^+\ell^-}^h\big[h(c_1,c_2,M_{V_1}^2,M_{V_2}^2)\big]
=
\frac{ie^4 m_\ell}{M_P^2}
\left(-\frac{N_\text{c}}{12\pi^2F_\pi}\right)\\
&\times
\int\frac{\text{d}^4l}{(2\pi)^4}
\frac{h(c_1,c_2,M_{V_1}^2,M_{V_2}^2)\lambda(M_P^2,p^2,q^2)}{(l^2-m_\ell^2)}\,.
\end{split}
\end{equation}
In the case of the process $\eta\to\mu^+\mu^-$ within the VMD model discussed in Appendix~\ref{app:etaVMD}, we can write (cf.~(\ref{eq:FFeggVMD}))
\begin{equation}
\begin{split}
P_{\eta\to \mu^+\mu^-}^\text{VMD}
=\bigg\{
\frac53\frac{\cos\phi}{f_\ell}&P_{\eta\to\mu^+\mu^-}^h\big[h(1,1,M_{\rho/\omega}^2,M_{\rho/\omega}^2)\big]\\
-\frac{\sqrt2}3\frac{\sin\phi}{f_\text{s}}&P_{\eta\to\mu^+\mu^-}^h\big[h(1,1,M_\phi^2,M_\phi^2)\big]
\bigg\}\,.
\end{split}
\label{eq:PLOVMD}
\end{equation}
Note that in the pion case we would simply have
\begin{equation}
P_{\pi^0\to e^+e^-}^\text{VMD}
=P_{\pi^0\to e^+e^-}^h\big[h(1,1,M_{\rho/\omega}^2,M_{\rho/\omega}^2)\big]\,.
\end{equation}
After recalling (\ref{eq:Mhing}) we know that the previous expressions might be obtained in terms of linear combinations of the building blocks $P_{P\to\ell^+\ell^-}^h\big[g(M_{V_1}^2,M_{V_2}^2)\big]$.
Using the dimensional regularization, the dimensional reduction scheme~\cite{Siegel:1979wq,Novotny:1994yx} and Passarino--Veltman reduction~\cite{Passarino:1978jh}, the explicit result of the necessary loop integration in terms of scalar one-loop integrals reads
\begin{equation}
\begin{split}
&P_{P\to\ell^+\ell^-}^h\big[g(M_{V_1}^2,M_{V_2}^2)\big]
=-\frac{e^4 m_\ell}{16\pi^2}
\left(-\frac{N_\text{c}}{12\pi^2F_\pi}\right)\\
&\times\left\{
\frac{M_{V_1}^2}{2m_\ell^2}\Big[B_0(0,M_{V_1}^2,M_{V_1}^2)-B_0(m_\ell^2,m_\ell^2,M_{V_1}^2)+1\Big]\right.\\
&+\frac{M_{V_1}^2}{M_P^2}\Big[B_0(m_\ell^2,m_\ell^2,M_{V_2}^2)-B_0(m_\ell^2,m_\ell^2,M_{V_1}^2)\Big]\\
&-B_0(m_\ell^2,m_\ell^2,M_{V_1}^2)
-\frac12\Big[1+B_0(0,m_\ell^2,m_\ell^2)\\
&-\tilde{C}_0(m_\ell^2,m_\ell^2,M_P^2, M_{V_1}^2,m_\ell^2, M_{V_2}^2)\Big]\bigg\}+\big\{M_{V_1}^2\leftrightarrow M_{V_2}^2\big\}\,.
\end{split}
\label{eq:PllBB}
\end{equation}
Above, it was convenient to introduce the following combination of the three-point scalar one-loop function $C_0$ and the K\"all\'en triangle function $\lambda$:
\begin{equation}
\begin{split}
&\tilde{C}_0(m^2,m^2,M_1^2,M_2^2,m^2,M_3^2)\\
&\equiv\frac{1}{M_1^2}\lambda(M_1^2,M_2^2,M_3^2)\,C_0(m^2,m^2,M_1^2,M_2^2,m^2,M_3^2) \,.
\end{split}
\end{equation}
For completeness, we list the predictions for the branching ratios of the $\eta^{(\prime)}\to\ell^+\ell^-$ decays in Table~\ref{tab:Pll}.
\begingroup
\squeezetable
\begin{table}[!t]
\begin{ruledtabular}
\begin{tabular}{c | c c c c}
 & $\eta\to e^+e^-$ & $\eta\to \mu^+\mu^-$ & $\eta^\prime\to e^+e^-$ & $\eta^\prime\to \mu^+\mu^-$\\
\hline\\
$B$ & $5.4(1.2)\times10^{-9}$ & $4.6(1.0)\times10^{-6}$ & $1.8(3)\times10^{-10}$ & $1.3(2)\times10^{-7}$
\end{tabular}
\end{ruledtabular}
\caption{
Branching ratios for the $\eta^{(\prime)}\to\ell^+\ell^-$ decays within the VMD-inspired model.
For instance, the value for the $\eta^\prime\to\mu^+\mu^-$ process is in good agreement with the value $B(\eta^\prime\to\mu^+\mu^-)=1.4(2)\times10^{-7}$ calculated in~\cite{Silagadze:2006rt}.
\label{tab:Pll}
}
\end{table}
\endgroup

In what follows, we would like to provide some basic examples of the decomposition of the loop integrals containing various models for transition form factors in the case of the rare decay of a neutral pion.
Lets start with some definitions.
The VMD ansatz for the electromagnetic transition form factor of a neutral pion takes a simple form
\begin{equation}
\mathcal{F}_{\pi^0\gamma^*\gamma^*}^\text{VMD}(p^2,q^2)
=-\frac{N_\text{c}}{12\pi^2 F_\pi}
\frac{M_V^4}{(p^2-M_V^2)(q^2-M_V^2)}\,.
\label{eq:piFFVMD}
\end{equation}
The lowest-meson dominance (LMD) model~\cite{Knecht:1999gb}, where also the lowest-lying pseudoscalar multiplet was taken into account, gives the following result:
\begin{equation}
\mathcal{F}_{\pi^0\gamma^*\gamma^*}^\text{LMD}(p^2,q^2)
=\mathcal{F}_{\pi^0\gamma^*\gamma^*}^\text{VMD}(p^2,q^2)
\bigg[1-\frac{4\pi^2F_\pi^2}{N_\text{c} M_V^4}(p^2+q^2)\bigg]\,.
\label{eq:piFFLMD}
\end{equation}
As the last example we introduce the two-hadron saturation (THS) model proposed in~\cite{Husek:2015wta}, which for the $PVV$ correlator takes into account two meson multiplets in both vector and pseudoscalar channels:
\begin{equation}
\begin{split}
&\mathcal{F}_{\pi^0\gamma^*\gamma^*}^\text{THS}(p^2,q^2)
=-\frac{N_\text{c}}{12\pi^2F_\pi}\\
&\times\bigg\{1+\frac{\kappa}{2N_\text{c}}\frac{p^2q^2}{(4\pi F_\pi)^4}
-\frac{4\pi^2F_\pi^2(p^2+q^2)}{N_\text{c} M_{V_1}^2 M_{V_2}^2}\bigg[6+\frac{p^2q^2}{M_{V_1}^2 M_{V_2}^2}\bigg]\bigg\}\\
&\times\frac{M_{V_1}^4 M_{V_2}^4}{(p^2-M_{V_1}^2)(p^2-M_{V_2}^2)(q^2-M_{V_1}^2)(q^2-M_{V_2}^2)}\,.
\end{split}
\label{eq:piFFfinal2}
\end{equation}
In terms of decomposition (\ref{eq:FFsubsth}) we can write for the amplitudes
\begin{align}
\hspace{-2mm}\mathcal{M}_{1\gamma\text{IR}}^\text{VMD}
&=\mathcal{M}_{1\gamma\text{IR}}^h\big[h(1,1,M_V^2,M_V^2)\big]\,,
\label{eq:hVMD}\\
\hspace{-2mm}\mathcal{M}_{1\gamma\text{IR}}^\text{LMD}
&=\mathcal{M}_{1\gamma\text{IR}}^h\Big[h\Big(c_1^\text{LMD},2c_1^\text{LMD}-1,M_V^2,M_V^2\Big)\Big]\,,
\label{eq:hLMD}\\
\begin{split}
\hspace{-2mm}\mathcal{M}_{1\gamma\text{IR}}^\text{THS}
&=\frac14\mathcal{M}_{1\gamma\text{IR}}^h\Big[h\Big(4c_{1,1}^\text{THS},4c_{2,1}^\text{THS},M_{V_1}^2,M_{V_2}^2\Big)\Big]\\
&+\frac14\mathcal{M}_{1\gamma\text{IR}}^h\Big[h\Big(0,4c_{2,2}^\text{THS},M_{V_1}^2,M_{V_1}^2\Big)\Big]\\
&+(M_{V_1}^2\leftrightarrow M_{V_2}^2)\,,
\end{split}
\label{eq:hTHS}
\end{align}
where the coefficients $c_{1,i}^\text{model}$ and $c_{2,i}^\text{model}$ have the following form:
\begin{align}
c_1^\text{LMD}
&=1-\frac{4\pi^2F_\pi^2}{N_\text{c}M_V^2}\,,\\
c_{1,1}^\text{THS}
&=\frac{M_{V_2}^2}{M_{V_2}^2-M_{V_1}^2}\left(1-\frac{24\pi^2F_\pi^2}{N_\text{c}M_{V_2}^2}\right),\\
\begin{split}
c_{2,1}^{\text{THS}}
&=-\frac{M_{V_1}^2M_{V_2}^2}{(M_{V_2}^2-M_{V_1}^2)^2}\\
&\times\bigg[1+\frac{\kappa M_{V_1}^2M_{V_2}^2}{2N_\text{c}(4\pi F_\pi)^4}
-\frac{7(2\pi F_\pi)^2}{N_\text{c} M_{V_1}^2}\bigg(1+\frac{M_{V_1}^2}{M_{V_2}^2}\bigg)\bigg],
\end{split}\\
\begin{split}
c_{2,2}^\text{THS}
&=\frac{M_{V_2}^4}{(M_{V_2}^2-M_{V_1}^2)^2}\\
&\times\bigg[1+\frac{\kappa M_{V_1}^4}{2N_\text{c}(4\pi F_\pi)^4}
-\frac{(4\pi F_\pi)^2}{2N_\text{c} M_{V_2}^2}\bigg(6+\frac{M_{V_1}^2}{M_{V_2}^2}\bigg)\bigg]\,.
\end{split}
\end{align}
We can find the above-listed constants from projecting on the product of the normalized form factor and the photon propagators: for instance we have
\begin{equation}
c_{2,2}^\text{THS}
=\hspace{-2mm}\lim_{p^2,q^2\to M_{V_1}^2}\hspace{-2mm}\frac{\mathcal{F}_{\pi^0\gamma^*\gamma^*}^\text{THS}(p^2,q^2)}{\mathcal{F}_{\pi^0\gamma^*\gamma^*}(0,0)}\frac{(p^2-M_{V_1}^2)(q^2-M_{V_1}^2)}{p^2q^2}\,.
\end{equation}
Taking into account decomposition (\ref{eq:FFsubstg}) into building blocks (\ref{eq:PllBB}), one recovers formulae (A.5-A.7) from~\cite{Husek:2015wta}.

\onecolumngrid

\end{document}